\documentclass[12pt]{article}
\pdfoutput=1
\usepackage{putex}
\usepackage{comment}
\usepackage{graphicx}
\usepackage{caption}
\usepackage{amsmath}
\usepackage{array}
\usepackage{subcaption}
\usepackage{epstopdf}
\usepackage{enumerate}
\usepackage{cite}
\usepackage{youngtab}
\usepackage{tensor}
\usepackage{slashed}
\usepackage[export]{adjustbox}
\usepackage[aligntableaux=center]{ytableau}
\usepackage[utf8]{inputenc}
\usepackage[
      colorlinks=true,
      linkcolor=blue,
      urlcolor=blue,
      filecolor=black,
      citecolor=red,
      ]{hyperref}
\newcommand{\RNum}[1]{\uppercase\expandafter{\romannumeral #1\relax}}

\newcommand {\be} {\begin {equation}}
\newcommand {\ee} {\end {equation}}

\newcommand {\bes} {\begin {equation*}}
\newcommand {\ees} {\end {equation*}}

\newcommand{\cG}{{\mathcal G}}

\newcommand{\cL}{{\mathcal L}}

\newcommand{\beq}{\begin{equation}}
\newcommand{\eeq}{\end{equation}}

\def\ie{\begin{equation}\begin{aligned}}
\def\fe{\end{aligned}\end{equation}}

\numberwithin{equation}{section}

\def\<{\langle}
\def\>{\rangle}

\begin{document}

\preprint{PUPT-2606}

\institution{PU}{Department of Physics, Princeton University, Princeton, NJ 08544, USA}

\title{{\LARGE $O(N)$ Models with Boundary Interactions \\ and their Long Range Generalizations}}

\authors{Simone Giombi\worksat{\PU} and Himanshu Khanchandani\worksat{\PU}
}

\abstract{We study the critical properties of scalar field theories in $d+1$ dimensions with $O(N)$ invariant interactions localized on a $d$-dimensional boundary. By a combination of large $N$ and epsilon expansions, we provide evidence for the existence of non-trivial $O(N)$ BCFTs in $1<d<4$. Due to having free fields in the bulk, these models possess bulk higher-spin currents which are conserved up to terms localized on the boundary. We suggest that this should lead to a set of protected spinning operators on the boundary, and give evidence that their anomalous dimensions vanish. We also discuss the closely related long-range $O(N)$ models in $d$ dimensions, and in particular study a weakly coupled description of the $d=1$ long range $O(N)$ model near the 
upper critical value of the long range parameter, which is given in terms of a non-local non-linear sigma model. By combining the known perturbative descriptions, we provide some estimates of critical exponents in $d=1$.}


\maketitle

\tableofcontents

\section{Introduction and Summary}

Conformal field theories with a boundary have been studied for a long time \cite{Cardy:1984bb, Cardy:1991tv, Diehl:1981zz, McAvity:1993ue, McAvity:1995zd} and have a variety of physical applications, from statistical physics and condensed matter to string theory and holography (for a recent review, see \cite{Andrei:2018die}). A renewed interest in the subject has also taken place in light of the progress in conformal bootstrap methods \cite{Liendo:2012hy, Gliozzi:2015qsa, Mazac:2018biw, Kaviraj:2018tfd, Bissi:2018mcq}. Recently, boundary conformal field theories have also been proposed to play a role as holographic duals of certain single sided black hole microstates \cite{Almheiri:2018ijj, Rozali:2019day}. In this paper, we study a special type of boundary conformal field theory (BCFT) which is obtained by taking free fields in a $(d+1)$-dimensional bulk and adding interactions localized on a $d$-dimensional boundary. Free field theories with localized boundary interactions have been considered before in several different contexts including applications to dissipative quantum mechanics, open string theory and edge states in quantum hall effect \cite{Callan:1993mw, Callan:1994ub, Callan:1995em, Fendley:1994rh, Lukyanov:2003nj, Lukyanov:2012wq}. More recently, several examples of BCFT with non-interacting bulk fields were considered in \cite{Herzog:2017xha, Herzog:2018lqz}. A particularly interesting model, with possible applications to graphene, is obtained by taking a free Maxwell field in four dimensions coupled to fermions localized on a three-dimensional boundary (or ``brane") \cite{Gorbar:2001qt, Son:2007ja, Kaplan:2009kr, Teber:2012de, Teber:2014hna, Kotikov:2013eha, Son:2015xqa, Kotikov:2016yrn, Hsiao:2017lch, Herzog:2017xha, Herzog:2018lqz, DiPietro:2019hqe}. 

\begin{figure}[!ht]
\centering
\includegraphics[scale=0.7]{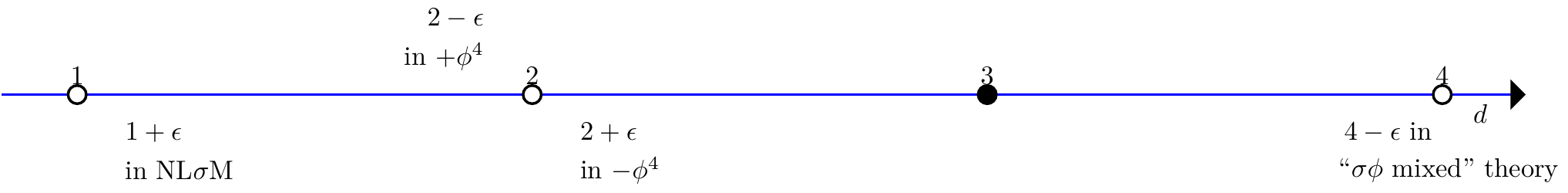}
\caption{$O(N)$ BCFT in $1 < d < 4$}
\label{BCFTFigure}
\end{figure}

In the present paper, we focus on the case of scalar field theory with $O(N)$ invariant boundary interactions. In particular, we investigate the critical properties of the model defined by $N$ real scalar fields $\phi^I$ with the standard quartic interaction restricted to the boundary
\begin{equation}
S= \int d^{d + 1} x \frac{1}{2} \partial_{\mu} \phi^I \partial^{\mu} \phi^I + \int d^{d} x \frac{g}{4} (\phi^I \phi^I)^2.
\end{equation}
With (generalized) Neumann boundary conditions $\partial_n \phi \sim g \phi^3$, the quartic interaction is marginal in $d=2$ and relevant in $d<2$, and hence one may have a non-trivial IR fixed point. As we show below, working in the framework of the $\epsilon$-expansion one indeed finds a weakly coupled Wilson-Fisher fixed point in $d=2-\epsilon$, with real and positive coupling constant (here and below, we shall always assume that relevant quadratic terms have been tuned to criticality). This model was analyzed before in \cite{DE87, PhysRevB.37.5257} with an additional $\phi^6$ coupling in the bulk. Here we will not turn on this bulk coupling. As in the well-known case of the standard critical $O(N)$ models, one may also develop a large $N$ expansion for any $d$ by introducing a Hubbard-Stratonovich field, which in the present case is localized on the $d$-dimensional boundary. This yields a large $N$ BCFT which appears to be unitary in $1/N$ perturbation theory in the range $1 < d < 4$. We perform explicit calculations of various physical quantities in this BCFT, and show that the large $N$ expansion precisely matches onto the $\epsilon$-expansion in the quartic model in $d=2-\epsilon$. On the other hand, in $d=1+\epsilon$ we show that it matches onto the UV fixed point of a non-local non-linear $O(N)$ sigma model with the sphere constraint localized on the boundary. The action of this sigma model is given by 
\begin{equation}
S = \int d^{d + 1} x  \frac{1}{2} \partial_{\mu} \phi^I \partial^{\mu} \phi^I + \int d^{d} x \ \sigma (\phi^I \phi^I - \frac{1}{t^2})\,,
\label{nonlinear-sigma}
\end{equation} 
where $t$ is the boundary coupling constant for which we compute the beta function to order $t^5$. The large $N$ expansion can be formally continued above the upper critical dimension $d=2$, where it remains perturbatively unitary for $d<4$. In $d=4-\epsilon$, we provide strong evidence that the large $N$ expansion matches onto the IR fixed point of a metastable (for sufficiently large $N$ and small $\epsilon$) mixed ``$\sigma \phi$'' theory
\begin{equation}
S = \int d^{d + 1} x \frac{1}{2} (\partial_{\mu} \phi^I)^2 + \int d^{d} x \bigg( \frac{1}{2} (\partial \sigma )^2 + \frac{g_1}{2} \sigma \phi^I \phi^I  + \frac{g_2}{4!} \sigma^4 \bigg)\,.
\end{equation}
The instability arises because at the fixed point the quartic self-interaction of the $\sigma$ field is negative, as we will show below by explicitly computing the beta functions of the model. Correspondingly, one finds real instanton solutions localized on the boundary, which are expected to produce imaginary parts in the scaling dimensions of boundary operators and other observables, as is well-known for the standard $\phi^4$ theory with negative coupling. A summary of the various descriptions of the boundary $O(N)$ BCFTs in $1<d<4$ is given in Figure \ref{BCFTFigure}. The picture we find is a close analogue of the one found for the standard critical $O(N)$ models as a function of $d$. The large $N$ expansion in those models can be developed for any $d$ and it is perturbatively unitary in $2<d<6$. It matches onto the UV fixed points of the non-linear sigma model near $d=2$, and onto the Wilson-Fisher fixed point of the $\phi^4$ theory near $d=4$. As one approaches $d=6$, one finds instead a cubic $O(N)$ symmetric theory \cite{Fei:2014yja, Fei:2014xta} that has perturbative fixed points in $d=6-\epsilon$; non-perturbatively, these are unstable due to instanton effects, which produce small imaginary parts of physical observables \cite{Giombi:2019upv}. 

The fact that the BCFTs we study contain fields which are non-interacting in the bulk has interesting consequences. In particular, it implies  that the boundary operator spectrum has several operators with protected scaling dimensions, as we elaborate on in Section \ref{GeneralRemarks}. The simplest protected boundary operator is just the one induced by the free bulk field $\phi^I$, and has protected dimension $\Delta=(d-1)/2$. While our prime example in this paper are the scalar $O(N)$ models, similar properties are expected to hold in other similar models with free fields in the bulk.    


Recall that a flat boundary in $d + 1$ Euclidean dimensions breaks the conformal symmetry from $SO(d + 2,1)$ to $SO(d+1,1)$, which is the conformal group on the $d$ dimensional boundary. In particular, translational invariance perpendicular to the boundary is broken, which results in a delta-function localized source for the divergence of stress-tensor
\begin{equation}
\partial_{\mu}T^{\mu y} = D(\textbf{x}) \delta(y).
\label{T-non-cons}
\end{equation} 
In most of the paper we assume flat space with a flat boundary, and we will use $\textbf{x}$ for the $d$ coordinates on the boundary and $y$ for the transverse direction with $x^{\mu} = (\textbf{x}, y)$. The above equation is to be understood as an operator equation and it defines the displacement operator denoted by $D (\textbf{x})$. This relation also fixes the dimension of displacement operator to be same as that of stress tensor, $\Delta=d+1$. Since the stress tensor is conserved in the bulk, the displacement operator remains protected even in the presence of interactions and its scaling dimension is not renormalized. This holds in any BCFT. If the bulk theory is free, as in the models we study in this paper, then we also have a set of higher spin currents (see e.g. \cite{Giombi:2016ejx} for a review) which in the scalar field theory take the schematic form
\begin{equation}
J^{\mu_1 \mu_2 ... \mu_s} = \sum_{k = 0}^s c_{sk} \partial_{\{ \mu_1 .... \mu_k} \phi \partial_{\mu_{k + 1} ... \mu_s \}} \phi. 
\end{equation}
If the bulk fields are free, the divergence of these currents vanishes in the bulk. Then, as we explain in section \ref{HSDisplacementGeneral} below, one expects an equation similar to (\ref{T-non-cons}) with a delta-function localized source, defining a set of spinning operators on the boundary with spin ranging from 0 to $s-2$, which we call higher spin displacement operators.\footnote{These operators were also considered in the context of replica twist defect in \cite{Balakrishnan:2017bjg} but they are not protected in that case.} Since the higher-spin currents are conserved in the bulk, we expect that the scaling dimensions of these higher-spin displacement operators should be non-renormalized, despite the presence of interactions at the boundary. We obtain several perturbative checks of this expectation in Section \ref{HSDisplacement}. It would be nice to further study the consequences of having such protected operators in the spectrum, and also study the analogous operators in other examples of BCFT with free fields in the bulk. 

\begin{figure}[!ht]
\centering
\includegraphics[scale=0.7]{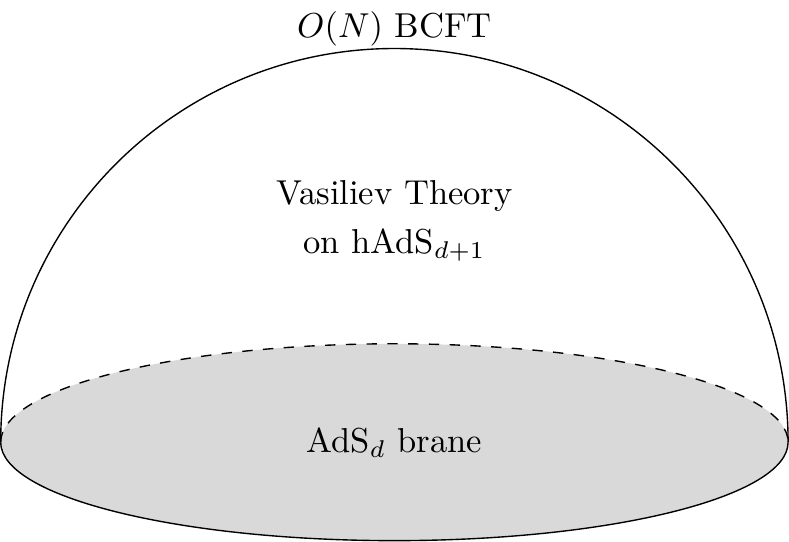}
\caption{AdS/ BCFT setup for $O(N)$ BCFT}
\label{AdSBCFT}
\end{figure}

In light of our $O(N)$ BCFT results, it would be interesting to extend the higher-spin versions of AdS/CFT (see \cite{Giombi:2012ms, Giombi:2016ejx} for reviews) to the case of  AdS/BCFT \cite{Fujita:2011fp}. Type A Vasiliev theory in AdS$_{d + 1}$ space \cite{Vasiliev:1990en , Vasiliev:1992av, Vasiliev:2003ev} is conjectured to be dual to a $d$ dimensional $O(N)$ model, free or interacting depending on the boundary conditions of a bulk scalar field \cite{Klebanov:2002ja}. Similarly, the $O(N)$ BCFT we study should be dual to Vasiliev theory on hAdS$_{d + 1}$, where we have half of AdS$_{d + 1}$ space ending on a AdS$_d$ brane as shown in figure \ref{AdSBCFT}. In such a setup, boundary conditions of  AdS$_{d + 1}$ fields on the AdS$_d$ brane should be determined by the boundary conditions of $O(N)$ BCFT, while as usual, the boundary condition on the asymptotic AdS$_{d + 1}$ boundary will be determined by whether the $O(N)$ model is free or interacting in the bulk of the BCFT (in this paper, we turn off interactions in the bulk, but one could more generally allow for a bulk coupling constant in addition to the boundary one, and study the RG flow of both couplings). 

From the point of view of perturbative calculations of purely boundary observables in the models we study, one essentially computes boundary Feynman diagrams where the scalar fields has a $1/|\textbf{p}|$ propagator, which is induced by the free kinetic term in the bulk (recall that we focus on Neumann boundary conditions). This may be thought of as a particular kind of non-local scalar field theory in $d$ dimensions. A natural generalization is to consider more general non-local propagator parametrized by an arbitrary power $s$, with a propagator $1/|\textbf{p}|^s$ in momentum space. This corresponds to a non-local kinetic term proportional to 
\begin{equation}
\int d^d x d^d y \frac{\phi^I(x) \phi^I (y)}{|x - y|^{d + s}}.
\end{equation}
as can be checked by a Fourier transform to momentum space. Adding $O(N)$ invariant quartic interactions to such a non-local model, one finds fixed points which are expected to describe second order phase transition in a system of $N$-component unit spins interacting with a long range Hamiltonian
\begin{equation}
H = - J \sum_{i,j} \frac{\textbf{s}_i \cdot \textbf{s}_j}{|i - j|^{d + s}}. 
\end{equation}
Critical exponents for the long range interactions fall in three categories \cite{PhysRevLett.29.917, PhysRevB.8.281, PhysRevB.15.4344, Honkonen_1989, Honkonen:1990mr, Luijten_2002, 2014arXiv1401.6805C, Paulos:2015jfa, Behan:2017dwr, Behan:2017emf} : 1) For $s < d/2$, critical exponents are the same as the ones for Gaussian fixed point, 2) for $d/2 < s < s_*$ there is a non trivial long range fixed point and critical exponents can be calculated and 3) for $s > s_*$, the critical exponents take the same value as the corresponding  short range fixed point. The value of $s_*$ is such that the conformal dimension of $\phi$ is continuous at the long range to short range crossover. In the long range fixed point, $\phi$ has no anomalous dimension and its scaling dimension is fixed to be  $(d - s)/2$ (an argument for this is that $\phi$ can be formally thought of as a free field satisfying Laplace equation in a higher dimensional bulk, where $p=2-s$ is the co-dimension). On the other hand, at the short range fixed point, $\phi$ has an anomalous dimension and its scaling dimension is $\Delta_{SR}=(d - 2 + 2 \gamma_{\phi}^{SR})/2$. This fixes $s_* = 2 - 2 \gamma_{\phi}^{SR}$.    

\begin{figure}[!ht]
\centering
\includegraphics[scale=1]{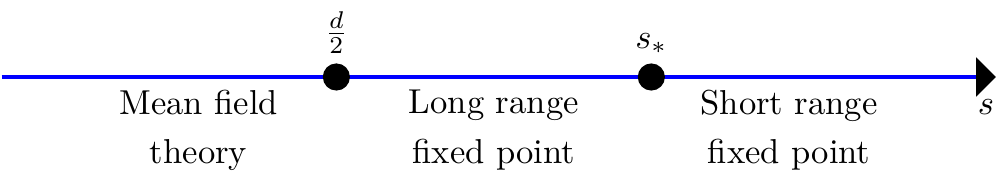}
\caption{Continuum models for various values of $s$.}
\label{LongRangeFigure}
\end{figure}

\begin{figure}[!ht]
\centering
\includegraphics[scale=1]{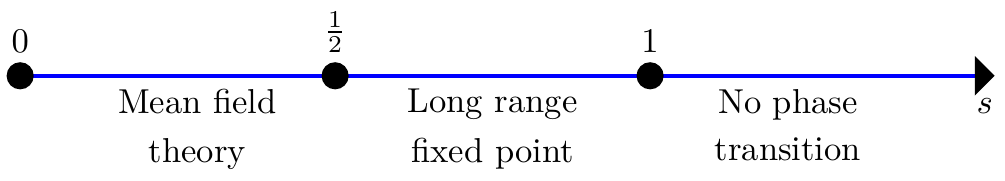}
\caption{Continuum picture for one dimensional $O(N)$ model for various $s$.}
\label{LongRangeFigure1d}
\end{figure}

The crossover from mean field theory to long-range fixed point is relatively under control and perturbation theory can be developed since the usual $\phi^4$ interaction is weakly coupled. An alternative scaling theory was proposed in \cite{Behan:2017dwr, Behan:2017emf}, which is weakly coupled near short range to long range crossover and can be used to do perturbation theory. However, in $d = 1$, there is no short range fixed point, since there is no phase transition in $d =1$ $O(N)$ model, except at zero temperature. At zero temperature, all correlation functions are constant, and hence the anomalous dimension of $\phi$ is commonly assigned an exact value $\gamma_{\phi}^{SR} = 1/2$ which makes $\Delta_{\phi}^{SR} = 0$ and $s_* = 1$. In the long range model, there is a phase transition for $0 < s < 1$ as was shown by Dyson in \cite{dyson1969} and further studied in \cite{PhysRevLett.37.1577, Aizenman1988DiscontinuityOT, Aizenman1988, 2015CMaPh.334..719A}. So $s = 1$ is the upper critical value for the long range universality class in $ d = 1$, which is what we would have naively expected by extrapolating the crossover region from higher dimensions. Hence for $d = 1$, the picture in figure \ref{LongRangeFigure} is modified to figure \ref{LongRangeFigure1d}. Below we will study a non-local non-linear sigma model which becomes weakly coupled in $s = d - \epsilon$ for all $d$, and is a natural generalization of the boundary model (\ref{nonlinear-sigma}). Precisely in $d = 1$, it is weakly coupled near the upper critical value of $s$ for the long range model, and is well suited to do perturbation theory in the vicinity of $s=1$. Unlike the usual local non-linear sigma model, the $\beta$ function for this model is proportional to $N - 1$ instead of $N -2$, hence the description is only valid for $N > 1$. This is in agreement with what was found long ago in \cite{PhysRevLett.37.1577}. Combining results from non-linear sigma model and the quartic model, we give some Pad\'e estimates for critical exponents in the $d = 1$ long range $O(N)$ model. They are in good agreement with the Monte Carlo results of \cite{2014arXiv1401.6805C} for the values of $s$ given there.  It would be interesting to bootstrap this model using techniques similar to the one used for $d = 3$ long range Ising in \cite{Behan:2018hfx}, and compare the results with our estimates.      


This paper is organized as follows: In Section \ref{GeneralRemarks}, we discuss some general aspects of free field theories with interactions localized on the boundary. In Section \ref{Models}, we introduce the boundary $O(N)$ models in $1 < d < 4$ and its various descriptions as a function of dimension, and present various calculations of physical quantities at the fixed points. We explicitly construct a set of spinning operators induced on the boundary by bulk higher spin currents and provide evidence for the vanishing of their anomalous dimension in section \ref{HSDisplacement}. We end by describing long range generalizations of our models and give some estimates for $d = 1$ long range $O(N)$ model in section \ref{LongRange}. Appendices contain some other interesting examples of BCFT with free fields in the bulk and some technical details. 

{\bf Note added:} After completion of this paper, we became aware of \cite{Prochazka:2019fah} which has some overlap with parts of our work.

\section{Free fields with boundary interactions: some general remarks} \label{GeneralRemarks}
The models we consider in this paper have an action of the following general form 
\begin{equation}
S = \int d^{d + 1} x \ \cL_{free} + \int d^d x 
\ \cL_{int}.
\end{equation}
To be concrete, let us consider the case of scalar fields, so that $\mathcal{L}_{free} = (\partial_{\mu} \phi)^2/2$, but most of what we discuss below should have a generalization to the case of other fields. The usual variational principle gives the equation of motion $\partial_{\mu} \partial^{\mu} \phi = 0$, and we have to satisfy either Dirichlet or generalized Neumann boundary condition 
\begin{equation}
\phi(\textbf{x},0)= 0, \ \ \ \textrm{or} \ \ \ \partial_{y}\phi (\textbf{x}, 0) - \frac{\delta \cL_{int}}{\delta \phi} = 0. 
\label{bc}
\end{equation} 
We will be focusing on generalized Neumann in this paper, which allows for the possibility of interesting critical behavior for the boundary $O(N)$ models in $1<d<4$. 

In a CFT with a boundary, in addition to the usual bulk OPE, we also have the boundary OPE where we expand the bulk field $\phi$ into a set of boundary primary operators 
\begin{equation}
\phi (\textbf{x},y) = \sum_{\hat{O}} \frac{B_{\phi}^{\hat{O}}}{(2 y)^{\Delta - \hat{\Delta}}} D^{\hat{\Delta}} (y^2 \vec{\partial}^2) \hat{O}(\textbf{x})
\end{equation}
The differential operator $D^{\hat{\Delta}} (y^2 \vec{\partial}^2)$ can be fixed using conformal invariance as we now review \cite{McAvity:1995zd}. We know by conformal invariance that 
\begin{equation}
\langle \phi(\textbf{x},y) \hat{O} (\textbf{x}') \rangle = \frac{B_{\phi \hat{O}}}{(2 y)^{\Delta -\hat{\Delta}} ((\textbf{x} - \textbf{x}')^2 + y^2)^{\hat{\Delta}}}, \ \ \ \langle \hat{O}(\textbf{x}) \hat{O} (\textbf{x}') \rangle = \frac{C_{\hat{O}}}{(\textbf{x} - \textbf{x}')^{2 \hat{\Delta}}}.
\end{equation} 
Using $B_{\phi \hat{O}} = C_{\hat{O}} B_{\phi}^{\hat{O}}  $, this is satisfied if (here and elsewhere the symbol $(x)_m$ refers to the Pochhammer symbol and is defined by $ (x)_m = \Gamma (x + m )/\Gamma(x)$)
\begin{equation}
D^{\hat{\Delta}} (y^2 \vec{\partial}^2) \frac{1}{(\textbf{x} - \textbf{x}')^{2 \hat{\Delta}}} = \frac{1}{ ((\textbf{x} - \textbf{x}')^2 + y^2)^{\hat{\Delta}}} = \sum_{m = 0}^{\infty} \frac{(\hat{\Delta})_m }{m!} \frac{(- y^2)^m}{(\textbf{x} - \textbf{x}')^{2 \hat{\Delta} + 2 m }}
\end{equation}  
which implies 
\begin{equation}
D^{\hat{\Delta}} (y^2 \vec{\partial}^2) = \sum_{m = 0}^{\infty} \frac{1}{m!} \frac{1}{(\hat{\Delta} + 1 - \frac{d}{2})_m} \left(- \frac{1}{4}y^2 \vec{\partial}^2 \right)^m
\end{equation}
Applying the bulk equation of motion $\partial_{\mu} \partial^{\mu} \phi = 0$ to this OPE, one finds
\begin{equation}
\begin{split}
\partial_{\mu} \partial^{\mu} \phi &= \sum_{\hat{O}} \frac{B_{\phi}^{\hat{O}}}{(2 y)^{\Delta - \hat{\Delta}}}  \sum_{m = 0}^{\infty} \frac{1}{m!} \frac{1}{(\hat{\Delta} + 1 - \frac{d}{2})_m} \bigg( (- \frac{1}{4}y^2)^m (\vec{\partial}^2)^{m + 1} \hat{O} (\textbf{x})  \\
&+ (2 m - \Delta + \hat{\Delta})(2 m - 1 - \Delta + \hat{\Delta}) (- \frac{1}{4}\vec{\partial}^2)^m (y^2)^{m - 1}  \hat{O} (\textbf{x}) \bigg) \\
&= \sum_{\hat{O}} \frac{B_{\phi}^{\hat{O}}}{(2 y)^{\Delta - \hat{\Delta}}}  \sum_{m = 0}^{\infty} \frac{1}{m!} \frac{1}{(\hat{\Delta} + 1 - \frac{d}{2})_m} \bigg(1 - \frac{(2 m + 2 - \Delta + \hat{\Delta})(2 m + 1 - \Delta + \hat{\Delta})}{4 (m + 1) (m + 1 + \hat{\Delta} - \frac{d}{2})} \bigg) \\
& \times \left(- \frac{1}{4}y^2 \vec{\partial}^2 \right)^m  \hat{O} (\textbf{x}). 
\end{split}
\end{equation}
The only allowed operators will be the ones for which the above coefficient vanishes for all integer $m$, because different descendants with different $m$ are independent. Plugging in $\Delta = (d -1)/2$, it is easy to see that the coefficient vanishes only for $\hat{\Delta} = (d - 1)/2$ and $\hat{\Delta} = (d + 1)/2$, so these are the only two operators allowed in the boundary OPE of a free scalar field. In the case where there are no interactions at the boundary, one has either one or the other of these operators, corresponding to Neumann and Dirichlet boundary conditions respectively. For the generalized Neumann boundary conditions in the presence of boundary interactions, as we show below one has both of these operators present in the boundary spectrum. Their dimensions are protected and add to $d$, satisfying a kind of ``shadow relation". Intuitively, the reason for this is clear from the structure of the generalized Neumann boundary condition in (\ref{bc}). The operator of dimension $\Delta=(d-1)/2$ is just $\phi$ restricted to the boundary, while the one of dimension $\Delta=(d+1)/2$ is the operator $\frac{\delta \cL_{int}}{\delta \phi}$ (this is a cubic operator in the $O(N)$ models we discuss below), which is related to $\phi$ by the boundary condition.

We can gain further insight on these protected operators by considering the bulk two-point function. Corresponding to two different OPE limits, 
there are two different ways to decompose the bulk two point function (see e.g. \cite{Mazac:2018biw, Liendo:2012hy,Billo:2016cpy}). We could do the usual OPE in the bulk and then do the boundary OPE of the fields that appear in the bulk OPE, or do the boundary OPE first and then do the usual OPE on the boundary. Correspondingly, a bulk two-point function can be expanded into either a set of boundary conformal blocks or a set of bulk conformal blocks, and the two expansions must be equal. Let us define the following cross-ratios
\begin{equation}
\xi \equiv  \frac{(\textbf{x}_1 - \textbf{x}_2)^2 + (y_1 - y_2)^2}{4 y_1 y_2}, \ \ \ z \equiv \frac{1}{1 + \xi}
\end{equation}
so that $ \xi \rightarrow \infty, \  z \rightarrow 0$ in the boundary OPE limit and $ \xi \rightarrow 0, \  z \rightarrow 1$ in the bulk OPE limit. We can then express the bulk two-point function of a scalar operator of dimension $\Delta_O$ as 
\begin{equation}
\begin{split}
\langle O(x_1) O(x_2) \rangle &= \frac{C_O}{(4 y_1 y_2)^{\Delta_O}} \cG(z) \\
\cG(z) &=  \frac{z^{\Delta_O}}{(1 - z)^{\Delta_O}} \sum_{k} \lambda_k f_{\textrm{bulk}} (\Delta_k; 1 - z) = \sum_l \mu_l^2 f_{\textrm{bdy}} (\hat{\Delta}_l; z)  
\label{2pt-block}
\end{split}
\end{equation}  
where $\lambda_k$ is the product of the bulk OPE coefficient and one point function of the operator, and $C_O \mu_l^2 = (B_O^{\hat{O}})^2 \hat{C}_{\hat{O}}$. The bulk and boundary blocks can be determined to be \cite{McAvity:1995zd}
\begin{equation}
\begin{split}
f_{\textrm{bulk}} (\Delta_k; z) &= z^{\frac{\Delta_k}{2}} \ {}_2 F_1 (\frac{\Delta_k + 1 -d}{2}, \frac{\Delta_k}{2}; \Delta_k + \frac{1 - d}{2}; z) \\ 
f_{\textrm{bdy}} (\hat{\Delta}_l; z)  &=  z^{\hat{\Delta}_l} \ {}_2 F_1 \bigg(\hat{\Delta}_l, \hat{\Delta}_l + \frac{1 - d}{2} ; 2 \hat{\Delta}_l + 1 - d;z \bigg).
\label{bulk-bdy-blocks}
\end{split}
\end{equation}
In the case of a bulk free field $\phi$, the equation of motion for the bulk two-point function $\langle\phi(x)  \phi(x')\rangle$ has two solutions corresponding to Neumann and Dirichlet boundary conditions
\begin{eqnarray} \label{FreeTwoPointNeumann}
G^{N/D}_{\phi} (x,x') &=& \frac{\Gamma(\frac{d + 1}{2})}{(d-1) 2 \pi^{\frac{d + 1}{2}}} \bigg( \frac{1}{((\textbf{x} - \textbf{x}')^2 + (y - y')^2)^{\frac{d - 1}{2}} } \pm  \frac{1}{((\textbf{x} - \textbf{x}')^2 + (y + y')^2)^{\frac{d - 1}{2}}} \bigg) \cr
&=& \frac{\Gamma(\frac{d + 1}{2})}{(d-1) 2 \pi^{\frac{d + 1}{2}} (4 y_1 y_2)^{\frac{d -1}{2}}} \bigg( \bigg( \frac{z}{1 - z} \bigg)^{\frac{d -1}{2}} \pm z^{\frac{d -1}{2}}   \bigg).
\end{eqnarray} 
In general, the bulk two-point function can then be a linear combination of these two solutions
\begin{equation}
\langle\phi(x)  \phi(x')\rangle = \frac{\Gamma(\frac{d + 1}{2})}{(d-1) 2 \pi^{\frac{d + 1}{2}} (4 y_1 y_2)^{\frac{d -1}{2}}} \bigg( \bigg( \frac{z}{1 - z} \bigg)^{\frac{d -1}{2}} + \lambda_{\phi^2} \  z^{\frac{d -1}{2}}   \bigg),
\label{2pt-gen-lc}
\end{equation}
where the $\lambda_{\phi^2}$ coefficient is related to the bulk one-point function of the $\phi^2$ operator. To see this, note that the bulk OPE expansion of the two-point function of $\phi$ contains, in addition to the identity block, a single block corresponding to the operator $\phi^2$ with $\Delta_{\phi^2} = d - 1$. The coefficient of the identity is just fixed by the normalization of the field $\phi$. Comparing with (\ref{2pt-block})-(\ref{bulk-bdy-blocks}), we see that the second term in (\ref{2pt-gen-lc}) indeed correspond to the $\phi^2$ operator. The coefficient $\lambda_{\phi^2}$ is equal to $\pm 1$ for Neumann or Dirichlet boundary conditions, but is arbitrary for generalized Neumann case. On the boundary, there are two possible blocks corresponding to operators with dimensions $(d - 1)/2$ and $(d + 1)/2$, as shown above, with OPE coefficients say $\mu_N^2$ and $\mu_D^2$. The blocks simplify for these values of conformal dimensions and the crossing equation relating the bulk and boundary OPE coefficients simply becomes 
\begin{equation}
1 + \lambda_{\phi^2} \ (1 - z)^{\frac{d -1}{2}} = \frac{\mu_N^2}{2} (1 + (1 - z)^{\frac{d -1}{2}} ) \ + \ \frac{2\mu_D^2}{d-1} (1 - (1 - z)^{\frac{d -1}{2}})\,.
\end{equation}
Equating the coefficients gives
\begin{equation}
\frac{\mu_N^2}{2} + \frac{2\mu_D^2}{d-1} = 1, \ \ \ \frac{\mu_N^2}{2} - \frac{2\mu_D^2}{d-1} = \lambda_{\phi^2}.
\end{equation}
As we expect, $\lambda_{\phi^2} = 1$ corresponds to Neumann and gives $\mu_D^2 = 0$, while $\lambda_{\phi^2} = -1$ corresponds to Dirichlet and gives $\mu_N^2 = 0$. The case of generic $\lambda_{\phi^2}$ has both operators present in the boundary spectrum and corresponds to the case of interacting theory on the boundary. 
\subsection{Displacement operator and its higher spin cousins} \label{HSDisplacementGeneral}
This section uses several results from \cite{McAvity:1993ue} about curved manifolds with a boundary. We refer the reader to \cite{McAvity:1993ue, Billo:2016cpy} for more detailed derivations. The action for the kind of theories we consider can be written in curved space as 
\begin{equation}
S = \int_{\mathcal{M}} d^{d + 1} x \sqrt{g} \bigg(  \frac{g^{\mu \nu}}{2} \partial_{\mu} \phi^I \partial_{\nu} \phi^I + \frac{\tau}{2} R \phi^I \phi^I \ \bigg) + \int_{\mathcal{\partial M}} d^{d} \hat{x} \sqrt{\gamma} \bigg(  \mathcal{L}_{\textrm{int}} + \frac{\rho}{2} K \phi^I \phi^I \ \bigg)
\end{equation}
where the boundary (or defect) is located at $x^{\mu} = X^{\mu} (\hat{x}^i)$, $K = \gamma^{i j} K_{ij}$ is the trace of the extrinsic curvature, and the boundary metric is defined by
\begin{equation}
\gamma_{i j} = e^{\mu}_{i} e^{\nu}_j g_{\mu \nu}, \ \ \ \ e^{\mu}_{i} = \frac{\partial X^{\mu}}{\partial \hat{x}^i}.
\end{equation}
By the usual variational principle, we can determine the following equation of motion and the boundary condition 
\begin{equation}
\nabla^2 \phi^I - \tau R \phi^I = 0, \ \ \ (\partial_n \phi^I - \rho K \phi^I - \mathcal{L}_{\textrm{int}}')|_{\partial \mathcal{M}} = 0.
\end{equation}
It can be shown \cite{McAvity:1993ue} that for Weyl invariance, we need $\rho = 2 \tau = \frac{d - 1}{2 d}$. From the variation of the above action with respect to the metric, we can determine the stress energy tensor, which in flat space with a flat defect reduces to  
\begin{equation}
\begin{split}
T_{\mu \nu}^{\textrm{tot}} &= T_{\mu \nu} + \delta_D(y) \delta^i_{\mu} \delta^j_{\nu} \delta_{i j} (- \mathcal{L}_{\textrm{int}}(\phi^I) + 2 \tau \mathcal{L}_{\textrm{int}}'(\phi^I) \phi^I ) \\
T_{\mu \nu} &= \partial_{\mu} \phi^I \partial_{\nu} \phi^I - \frac{\delta_{\mu \nu}}{2} (\partial_{\rho} \phi^I)^2 - \frac{d - 1}{4 d} (\partial_{\mu} \partial_{\nu} - \delta_{\mu \nu} \partial^2) \phi^I \phi^I.
\end{split}
\end{equation}
In a similar fashion, we can derive the displacement operator which can be defined by the variation of action with respect to the embedding coordinate $X^{\mu} (\hat{x}^i)$. Let $n^{\mu}$ be the normal to the defect. We shift the boundary along the normal as $\delta_t X^{\mu} (\hat{x}^i) = - n^{\mu} \delta t  $ and we let $\delta t$ be a function of the boundary coordinates here. Under this variation, the trace of the extrinsic curvature changes as \cite{McAvity:1993ue} 
\begin{equation}
\delta_t K = 3 \delta t K_{ij} K^{ij} - \hat{\gamma}^{i j} \hat{\nabla}_i \partial_j \delta t - \hat{\gamma}^{i j} R_{n j n i} \delta t. 
\label{K-var}
\end{equation}
Using this, one can see that, specializing to flat space with a flat defect, the variation of action is given by
\begin{equation}
\delta_t S = \int d^d \textbf{x} \left[ \delta t \left( \frac{1}{2} (\partial_y \phi^I)^2 + \frac{1}{2} (\partial_i \phi^I)^2 - \frac{\delta \mathcal{L}_{\textrm{int} }}{\delta \phi^I} \partial_y \phi^I \right) - \frac{\rho}{2} \phi^I \phi^I (\partial_i \partial^i \delta t) \right].
\label{S-var-t}
\end{equation}
The first two terms in the above equation come from the bulk piece of the action. The third term comes from the $\mathcal{L}_{\textrm{int}} $ piece of the boundary action. Since $\mathcal{L}_{\textrm{int}}$ is a function of boundary fields, which are just the bulk fields restricted to the boundary, its variation when we move the boundary should be given by $- \partial_y \mathcal{L}_{\textrm{int}} \ \delta t $, which simplifies to what we wrote above. The variation of $K$, as written in (\ref{K-var}), has three pieces, but only one of them survives in the flat space case, yielding the last term in (\ref{S-var-t}). After using the boundary condition and integration by parts, we get the displacement operator
\begin{equation}
D(\textbf{x}) = n^{\mu} \frac{\delta S}{\delta X^{\mu}} = \bigg[ \frac{1}{2} (\partial_y \phi^I)^2 - \frac{1}{2 d } (\partial_i \phi^I)^2 + \frac{d - 1}{2 d} \phi^I \partial_i^2 \phi^I \bigg] \bigg|_{y \rightarrow 0} = T_{y y}|_{y \rightarrow 0}.
\end{equation}
Another way to define the same operator is through its appearance in the divergence of stress tensor, as reviewed in the introduction
\begin{equation}
\partial_{\mu} T^{\mu i} = 0, \ \ \ \partial_{\mu} T^{\mu y} = D(\textbf{x})\delta(y)
\end{equation} 
By doing a volume integral over a Gaussian pill box located at the boundary, we can get the following relation
\begin{equation}
T^{y y}|_{y \rightarrow 0} = D(\textbf{x}). 
\end{equation}
which agrees with what we get from the other definition above. Since the stress tensor is conserved, the displacement operator must be protected on the boundary. Now, if the bulk theory is free, as in the models we study in this paper, we will have a tower of exactly conserved higher spin currents.  These are then expected to imply a tower of spinning protected operators on the boundary, which we may view as higher-spin ``cousins" of the displacement operator
\begin{equation}
\partial_{\mu} J^{\mu \mu_1 ... \mu_{s} y} = D^{\mu_1....\mu_{s-2}} (\textbf{x})\delta(y), \ \ \implies  J^{y \mu_1 ... \mu_{s - 2} y}|_{y \rightarrow 0} = D^{\mu_1....\mu_{s-2}}(\textbf{x}) .
\end{equation}
From the point of view of the theory on the boundary, the operator $D^{\mu_1....\mu_{s-2}}$ contains operators of all spins between $0$ and $s-2$, with $0$ being the case when all the $\mu's$ are equal to $y$ while $s - 2$ being the case when none of the $\mu's$ are equal to $y$. So we expect to see protected boundary operators of dimension $d + 1 + s - 2$ (same as the dimension of bulk spin $s$ current) and a spin between $0$ and $s-2$. In the boundary theory, these will be bilinears in the boundary operator\footnote{We use the same letter $\phi$ for the bulk field $\phi(\textbf{x}, y)$ and its boundary value $\phi(\textbf{x})$. It will be clear which one we mean from the context. This will make the expressions less messy by reducing the appearance of ``hats".} $\phi$ schematically of the form $\phi \vec{\partial}^{2 n} \partial_{\nu_1} \partial_{\nu_2}.. \partial_{\nu_l} \phi$ with dimensions $d -1 + 2 n + l$ and spin $l$. In section \ref{HSDisplacement}, we will give several pieces of evidence, within perturbation theory, for the fact that these boundary operators are protected.  

\section{$O(N)$ BCFT in $1<d<4$} \label{Models}
In this section, we describe perturbative fixed points of $O(N)$ invariant field theories with boundary localized interactions in boundary dimensions $1 < d < 4$. We calculate anomalous dimensions of various boundary operators and two point function of the bulk fundamental field at these fixed points and perform appropriate checks wherever different perturbative expansions are expected to match. 
\subsection{$\phi^4$ theory in $d=2-\epsilon$}
\label{$s=1$Quartic}
Let us first consider $N$ scalar fields on $d+1$ dimensional flat space with  a $d$ dimensional flat boundary, and a quartic $O(N)$ invariant interaction localized at the boundary:\footnote{We thank Igor Klebanov for useful suggestions and initial collaboration on the calculations presented in this Section.}
\begin{equation}
S = \int d^{d + 1} x \frac{1}{2} \partial_{\mu} \phi^I \partial^{\mu} \phi^I + \int d^{d} x \frac{g}{4} (\phi^I \phi^I)^2.
\label{phi4-ON}
\end{equation}
The coupling becomes marginal in $d = 2$, and it is relevant for $d<2$, so we will study this model in $d = 2 - \epsilon$. To do the calculation in momentum space, we can Fourier transform the free propagator along the boundary directions to get
\begin{equation}
\begin{split}
\langle \phi^I (- \textbf{p}, y) \phi^J (\textbf{p},y) \rangle =  \delta^{I J}\tilde{G}^0_{\phi} (p) &= \delta^{I J}\int_{\partial \mathcal{M}} d^{d} \textbf{x} e^{- i \textbf{p}\cdot ( \textbf{x}_1 - \textbf{x}_2)} G^0_{\phi} (y_1, \textbf{x}_1;y_2,\textbf{x}_2) \\
&= \delta^{I J}\frac{ e^{- p |y_1 - y_2|} + e^{- p (y_1 + y_2)}}{2 p} 
\end{split}
\end{equation}
which becomes $1/p$ on the boundary where $y_1, y_2 \rightarrow 0$.

To look for a fixed point, we compute the $\beta$ function up to two loops by first evaluating the following four point function and then requiring that it satisfies the Callan-Symanzik equation:

\begin{equation*}
\begin{split}
&G^4 = \vcenter{\hbox{\includegraphics[scale=1]{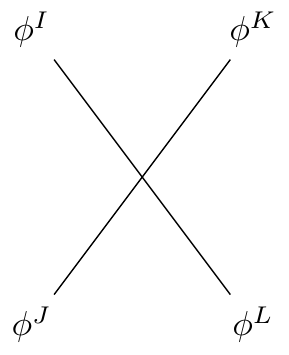}}} \ \   + \ \ \vcenter{\hbox{\includegraphics[scale=1]{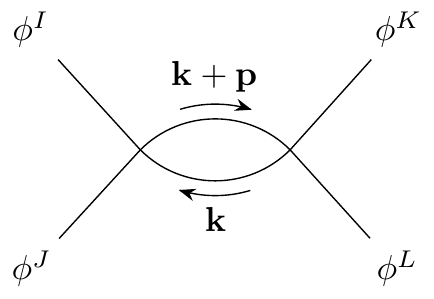}}} \ \   + \ \ \vcenter{\hbox{\includegraphics[scale=1]{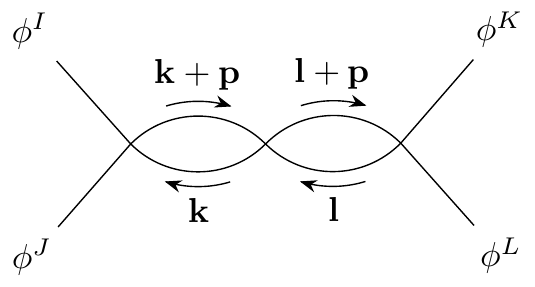}}} \\  &+ \vcenter{\hbox{\includegraphics[scale=1]{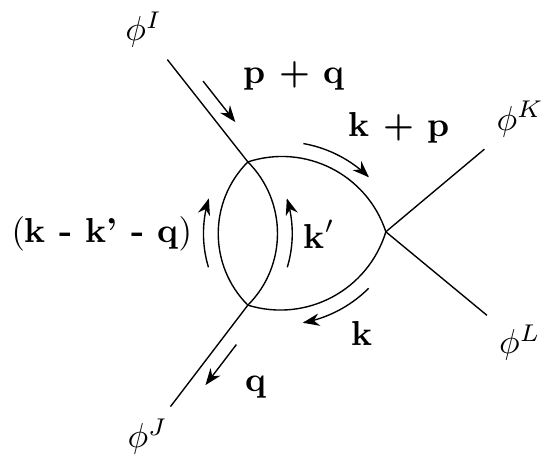}}} \\
\end{split}
\end{equation*}
\begin{equation}
\begin{split}
&=2 \delta^{IJ} \delta^{KL}  \bigg[ - (g + \delta_g) \ + \ (g + \delta_g)^2 (N + 8) \int \frac{d^{d} \textbf{k}}{(2 \pi)^{d}} \frac{1}{ |\textbf{k} + \textbf{p}|  | \textbf{k}|} -  g^3 (N^2 + 6 N + 20) \\
&\times \bigg( \int \frac{d^{d} \textbf{k}}{(2 \pi)^{d}} \frac{1}{ |\textbf{k} + \textbf{p}|  | \textbf{k}|} \bigg)^2 - 4 g^3 (5 N + 22)  \int \frac{d^d \textbf{k}}{(2 \pi)^d} \frac{d^d \textbf{k}'}{(2 \pi)^d} \frac{1}{|\textbf{k}| |\textbf{k + p}|} \frac{1}{|\textbf{k}'| |\textbf{k - k'- q}|}    \bigg]   \\
&= 2 \delta^{IJ} \delta^{KL} \bigg[ -(g + \delta_g) +  \frac{ (g + \delta_g)^2 (N + 8) \Gamma(\frac{d -1}{2})^2 \Gamma(1 - \frac{d}{2})}{(4 \pi)^{\frac{d}{2}}\pi \Gamma(d-1) (p^2)^{1 - \frac{d}{2}}}  \\ &- \frac{ g^3 (N^2 + 6 N + 20) \Gamma(\frac{d -1}{2})^4 \Gamma(1 - \frac{d}{2})^2}{(4 \pi)^{d} \ \pi^2 \Gamma(d-1)^2 (p^2)^{2 - d}} - \frac{4 g^3 (5 N + 22) \Gamma(\frac{d -1}{2})^3 \Gamma(1 - \frac{d}{2}) \Gamma(d - \frac{3}{2}) \Gamma(2 - d)}{(4 \pi)^{d} \pi^{3/2} \Gamma(d-1) \Gamma(\frac{3 -d}{2}) \Gamma(\frac{3 d}{2} - 2) (p^2)^{2 - d} }  \bigg].
\end{split}
\end{equation}
where we used an integral given in appendix \ref{Integrals} and evaluated the fourth diagram at $\textbf{q} = 0$. Expanding this in $d = 2 - \epsilon$ and demanding that the divergent terms cancel gives 
\begin{equation}
\delta_g = \frac{g^2 (N + 8)}{2 \pi \epsilon} - \frac{g^3 (5 N + 22) \log 2}{\pi^2 \epsilon} + \frac{g^3 (N + 8)^2}{4 \pi^2 \epsilon^2}.
\end{equation}
After canceling the divergent parts, the remaining finite parts need to satisfy Callan-Symanzik equation. Noting that in $2 - \epsilon$ dimensions, the bare coupling has a factor of $\mu^{\epsilon}$ on dimensional grounds, and then applying following equation
\begin{equation}
\bigg( \mu \frac{\partial}{\partial \mu} +  \beta \frac{\partial }{ \partial g} \bigg) G^4 = 0
\end{equation}
gives us 
\begin{equation}
\beta = - \epsilon g + \frac{g^2 (N + 8)}{2 \pi} - \frac{2 g^3 (5 N + 22) \log 2}{\pi^2}.
\end{equation}
There is a unitary IR fixed point at 
\begin{equation}
g_* = \frac{2 \pi \epsilon}{N + 8} + \frac{16 \pi ( 5 N + 22) \epsilon^2  \log 2 }{(N + 8)^3}.
\end{equation}

We can compute the anomalous dimensions of various operators at this fixed point. The simplest operator that gets an anomalous dimension is the $O(N)$ singlet on the boundary, $\phi^I \phi^I$. Its anomalous dimensions up to two loops can be determined from the following contributions to the boundary correlation function $\langle \phi^I \phi^I (\textbf{x}) \phi^J(\textbf{y}) \phi^K(\textbf{z})\rangle $ 

\begin{equation*}
\begin{split}
&G^{ 2,1} = \vcenter{\hbox{\includegraphics[scale=1]{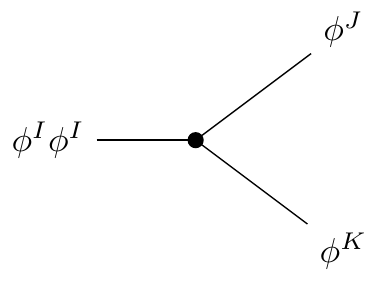}}} \ \  + \ \  \vcenter{\hbox{\includegraphics[scale=1]{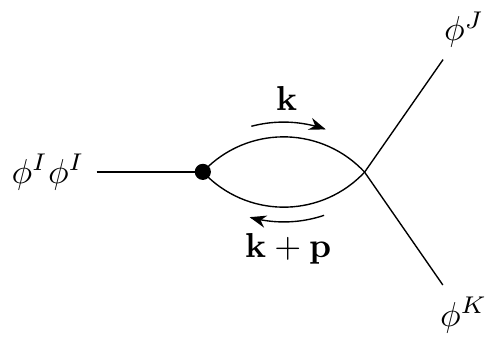}}} \\  &+ \vcenter{\hbox{\includegraphics[scale=1]{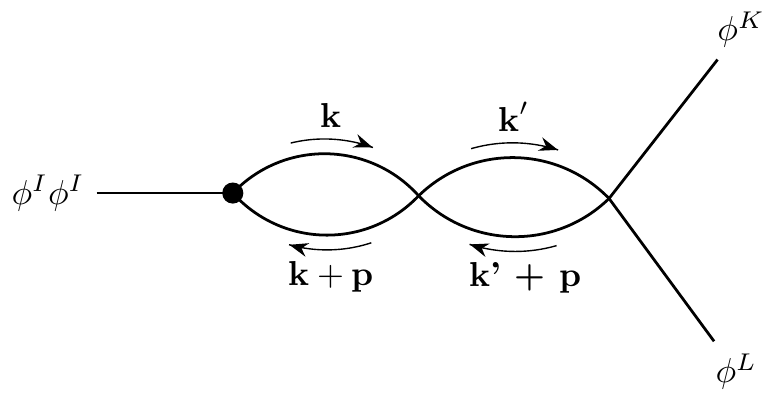}}} \ \ + \ \  \vcenter{\hbox{\includegraphics[scale=1]{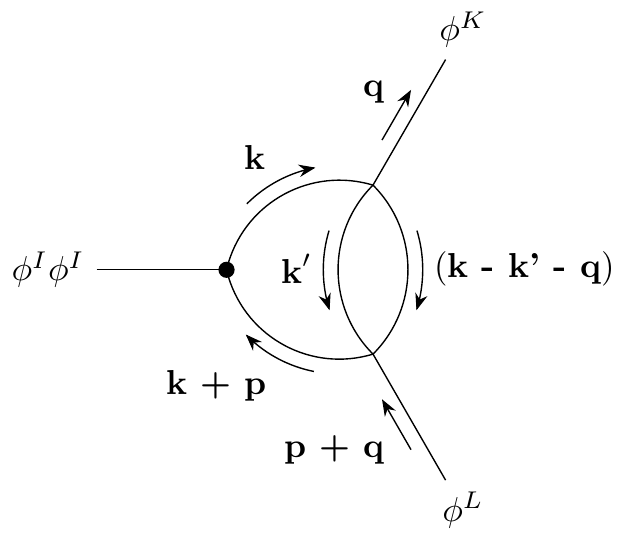}}}
\end{split}
\end{equation*}
\begin{equation}
\begin{split}
&=  2 \delta^{J K} \bigg[ 1 + \delta_{\phi^2} \ - \ (1 + \delta_{\phi^2}) (g +\delta_g) (N + 2) \int \frac{d^{d} \mathbf{k}}{(2 \pi)^{d}} \frac{1}{|\mathbf{k}| |\mathbf{k} + \mathbf{p}|} \\
&+ g^2 (N + 2)^2  \bigg(\int \frac{d^{d} \mathbf{k}}{(2 \pi)^{d}} \frac{1}{|\mathbf{k}| |\mathbf{k} + \mathbf{p}|} \bigg)^2 \ + \ 6 g^2 (N + 2) \int \frac{d^d \textbf{k}}{(2 \pi)^d} \frac{d^d \textbf{k}'}{(2 \pi)^d} \frac{1}{|\textbf{k}| |\textbf{k + p}|} \frac{1}{|\textbf{k}'| |\textbf{k - k'- q}|}   \bigg] \\
&= 2 \delta^{J K} \bigg[ 1 + \delta_{\phi^2} \ - \  (1 + \delta_{\phi^2})\frac{ (g + \delta_g) (N + 2) \Gamma(\frac{d -1}{2})^2 \Gamma(1 - \frac{d}{2})}{(4 \pi)^{\frac{d}{2}}\pi \Gamma(d-1) (p^2)^{1 - \frac{d}{2}}} \\
&+ \frac{ g^2 (N + 2)^2 \Gamma(\frac{d -1}{2})^2 \Gamma(1 - \frac{d}{2})^2}{(4 \pi)^{d}\pi^2 \Gamma(d-1)^2 (p^2)^{2 - d}} \ + \  \frac{6 g^2 (N + 2) \Gamma(\frac{d -1}{2})^3 \Gamma(1 - \frac{d}{2}) \Gamma(d - \frac{3}{2}) \Gamma(2 - d)}{(4 \pi)^{d} \pi^{3/2} \Gamma(d-1) \Gamma(\frac{3 -d}{2}) \Gamma(\frac{3 d}{2} - 2) (p^2)^{2 - d} }     \bigg].
\end{split}
\end{equation} 
where we evaluated the last diagram at $\textbf{q = 0}$ in this case as well. Again, expanding in $d = 2 - \epsilon$ and requiring that the divergent terms cancel gives 
\begin{equation}
\delta_{\phi^2} = \frac{ g (N + 2)}{2 \pi \epsilon}  - \frac{3 g^2 (N + 2) \log 2}{2 \pi^2 \epsilon} + \frac{g^2 (N + 2) (N + 5)}{4 \pi^2 \epsilon^2}
\end{equation}
Then applying Callan-Symanzik equation to the correlation function 
\begin{equation}
\bigg( \mu \frac{\partial}{\partial \mu} + \beta(g) \frac{\partial}{\partial g} + \hat{\gamma}_{\phi^2} \bigg) G^{2, 1} = 0\\
\end{equation}
gives us the anomalous dimension 
\begin{equation} \label{DimensionPhi^2Quartic}
\begin{split}
 \hat{\gamma}_{\phi^2} &= \frac{g_* (N + 2)}{2 \pi} - \frac{12 g_*^2 (N + 2) \log 2}{4 \pi^2} = \frac{N + 2}{N + 8} \epsilon + \frac{4 (N + 2) (7 N + 20) \log 2}{(N + 8)^3} \epsilon^2\\
\hat{\Delta}_{\phi^2} & =   d - 1 + \hat{\gamma}_{\phi^2} = 1 - \frac{6 \epsilon}{N + 8} + \frac{4 (N + 2) (7 N + 20) \log 2}{(N + 8)^3} \epsilon^2
\end{split}
\end{equation}

Another interesting operator to look at on the boundary is the $(\phi^I \phi^I) \phi^J$ operator which we dub as $\phi^3$ operator. For that we compute the following one loop contributions to the boundary correlator $\langle (\phi^I \phi^I) \phi^J (\textbf{x}) \phi^K(\textbf{y}) \phi^L(\textbf{z}) \phi^M (\textbf{w}) \rangle $ 

\begin{equation*}
G^{ 3,1} = \includegraphics[scale=0.9, valign = m]{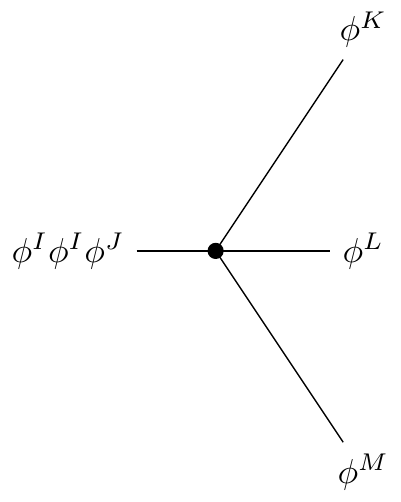} \ \ \ + \ \ \  \includegraphics[scale=1, valign = m]{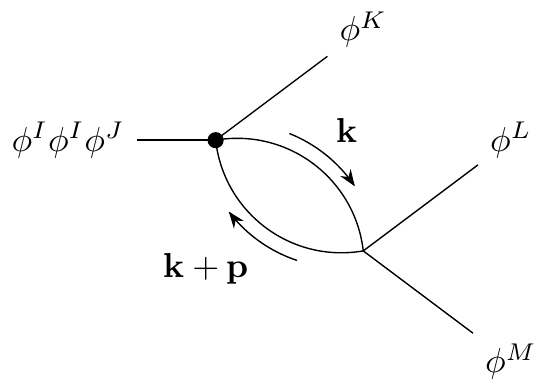} \\ 
\end{equation*}

\begin{equation}
\begin{split}
&=  2(\delta^{K L} \delta^{M J} + \delta^{K M} \delta^{L J} +  \delta^{L M} \delta^{K J} ) \bigg( 1 +  \delta_{\phi^3} - g (N + 8)  \int \frac{d^{d} \mathbf{k}}{(2 \pi)^{d}} \frac{1}{|\mathbf{k}| |\mathbf{k} + \mathbf{p}|} \bigg) \\
& =  2(\delta^{K L} \delta^{M J} + \delta^{K M} \delta^{L J} +  \delta^{L M} \delta^{K J} ) \bigg(1 +  \delta_{\phi^3} -  g (N + 8)  \frac{\Gamma(\frac{2 -d}{2})}{(4 \pi )^{\frac{d}{2}} (p^2)^{\frac{2 -d}{2}}}    \bigg).
\end{split}
\end{equation} 
To cancel the divergence we impose the condition that the order $g$ term vanish at momentum scale $\mu$ which implies 
\begin{equation}
\begin{split}
&\delta_{\phi^3} = \frac{ g (N + 8) \Gamma(\frac{2 -d}{2})}{(4 \pi )^{\frac{d}{2}} (\mu^2)^{\frac{2 -d}{2}}}, \ \ \ \  \hat{\gamma}_{\phi^3} = - \mu \frac{\partial}{\partial \mu} \delta_{\phi^3} =  \epsilon \\ 
&\hat{\Delta}_{\phi^3} = \frac{3 (d - 1)}{2} + \epsilon = \frac{3 - \epsilon}{2} = \frac{d + 1}{2}
\end{split}
\end{equation}
which agrees with our expectation since the boundary condition fixes $\phi^3 \sim \partial_{y} \phi$, so it must have dimension $\Delta_{\phi} + 1$. 

\begin{figure}[!ht]
\centering
\includegraphics[scale=1.5]{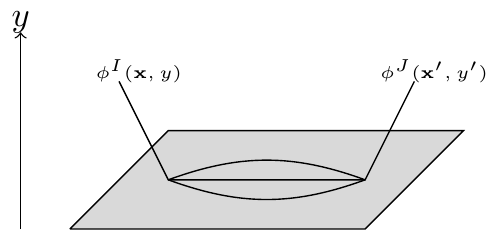}
\caption{Bulk two-point function at leading non-trivial order with $\phi^4$ interaction on the boundary}
\label{BulkTwoPoint}
\end{figure}

We will next compute the bulk two point of $\phi$ at this fixed point. In the free theory, it is still given by eq. \eqref{FreeTwoPointNeumann} but this will receive corrections because of interactions starting at order $g^2$. The leading perturbative correction is depicted in Figure \ref{BulkTwoPoint}. 
The computation of the corresponding Feynman diagram yields
\begin{equation*}
\tilde{G}^{I J}_{\phi}(p) = \includegraphics[scale=1,valign = b]{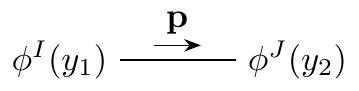} \ \ \  + \ \ \ \vcenter{\hbox{\includegraphics[scale=1]{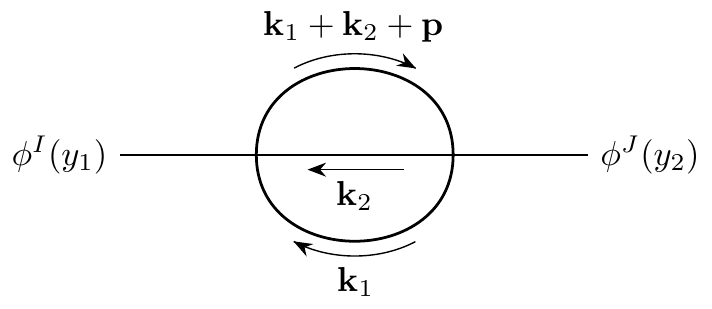}}} \\
\end{equation*}
\begin{equation}
\begin{split} 
&= \frac{\delta^{I J}(e^{- p |y_1 - y_2|} + e^{- p (y_1 + y_2)})}{2 p} \ + \ \frac{\delta^{I J} 2 g^2 (N + 2) e^{- p (y _1 + y_2)}}{p^2} \int \frac{d^{d} \mathbf{k_1}}{(2 \pi)^{d}} \frac{d^{d} \mathbf{k_2}}{(2 
\pi)^{d}}  \frac{1}{|k_1| |k_2| | \mathbf{k_1} + \mathbf{k_2} + \mathbf{p}|} \\
&= \frac{\delta^{I J} (e^{- p |y_1 - y_2|} + e^{- p (y_1 + y_2)}) }{ 2 p} \ + \ \frac{\delta^{I J} 2 g^2 (N + 2) e^{- p (y _1 + y_2)} \Gamma(\frac{2-d}{2}) \Gamma(\frac{d - 1}{2})^2}{(4 \pi)^{\frac{d}{2}} \pi \Gamma( d- 1) p^2} \int \frac{d^{d} \mathbf{k_2}}{(2 \pi)^{d}}  \frac{1}{ |k_2| | \mathbf{k_2} + \mathbf{p}|^{2-d}} \\
&= \frac{\delta^{I J} (e^{- p |y_1 - y_2|} + e^{- p (y_1 + y_2)}) }{2 p} \ + \ \frac{ \delta^{I J} 2 g^2 (N + 2) e^{- p (y _1 + y_2)} \Gamma(\frac{3-2d}{2}) \Gamma(\frac{d - 1}{2})^3 (p^2)^{d - \frac{5}{2}} }{(4 \pi)^{d} \pi^{\frac{3}{2}} \Gamma( \frac{3 d- 3}{2})} . 
\end{split}
\end{equation}
This doesn't have a divergence, in accordance with the fact that $\phi^I$ is a free field and does not get anomalous dimension. We can transform it back to position space and at the fixed point, this gives
\begin{equation}
G^{I J}_{\phi} (x_1, x_2) = \delta^{I J} G^0_{\phi} (x_1, x_2) - \frac{  \delta^{I J}   \epsilon^2 (N + 2)}{\pi (N + 8)^2 \sqrt{(\textbf{x}_1 - \textbf{x}_2)^2 + (y_1 + y_2)^2}} 
\end{equation}
This in particular gives corrections to the one point function of $\phi^I \phi^I$
\begin{equation}
\langle \phi^I \phi^I (\textbf{x},y) \rangle = \frac{N}{2 \pi y} \bigg( \frac{1}{4} - \frac{\epsilon^2 (N + 2)}{(N + 8)^2} \bigg)
\end{equation}

\subsection{Large $N$ description for general $d$}
\label{LargeN $s = 1$}
We can rewrite the quartic model introduced in the previous section in terms of a Hubbard-Stratonovich auxiliary field that lives only at the $d$-dimensional boundary:
\begin{equation} \label{LargeNActionBoundary}
S = \int d^{d + 1} x \frac{1}{2} \partial_{\mu} \phi^I \partial^{\mu} \phi^I + \int d^{d} x \bigg( \frac{\sigma \phi^I \phi^I}{2} -\frac{\sigma^2}{4 g} \bigg) .
\end{equation}
The equation of motion of $\sigma$ sets it equal to $g \phi^I \phi^I$ and plugging this in gives us back the original action. On the boundary, this is analogous to the usual O(N) model except for the fact that the propagator for $\phi$ is different. We can integrate out $\phi^I$ on the boundary to get a boundary effective action for $\sigma$
\begin{equation}
\begin{split}
e^{- S^{\textrm{eff}}_{\textrm{bdry}} [\sigma]} &= \int D \phi \ e^{- \int d^{d + 1} x \frac{1}{2} \partial_{\mu} \phi^I \partial^{\mu} \phi^I -  \int d^d x (\frac{\sigma \phi^I \phi^I}{2} -\frac{\sigma^2}{4 g} )} \\ 
&= e^{\frac{1}{8} \int d^{d} x_1 d^{d}  x_2 \ \sigma(x_1) \sigma(x_2) \langle \phi^I \phi^I (x_1) \phi^J \phi^J (x_2) \rangle_0 + \int d^{d} x \frac{\sigma^2}{4 g} + O(\sigma^3) }
\end{split}
\end{equation}
where 
\begin{equation}
\langle \phi^I \phi^I (\textbf{x}_1) \phi^J \phi^J (\textbf{x}_2) \rangle_0 = 2 N  [G_{\phi} (\textbf{x}_1-\textbf{x}_2)]^2
\end{equation}
with 
\begin{equation}
\begin{split}
[G_{\phi} (\textbf{x}_1 - \textbf{x}_2)]^2 &= \int \frac{d^{d} \mathbf{k_1}}{(2 \pi)^{d}} \int \frac{d^{d} \mathbf{k_2}}{(2 \pi)^{d}} \frac{e^{i (\mathbf{k_1} + \mathbf{k_2})\cdot(\mathbf{x_1} - \mathbf{x_2})}}{k_1 k_2} = \int \frac{d^{d} \mathbf{p}}{(2 \pi)^{d}}e^{i \mathbf{p}\cdot(\mathbf{x_1} - \mathbf{x_2})}  \int \frac{d^{d} \mathbf{q}} {(2 \pi)^{d }} \frac{1}{q |\mathbf{p} - \mathbf{q}|} \\
&= -\int \frac{d^{d} \mathbf{p}}{(2 \pi)^{d}}e^{i \mathbf{p}\cdot(\mathbf{x_1} - \mathbf{x_2})} \frac{2}{\tilde{C_{\sigma}}} (p)^{d-2}
\end{split}
\end{equation}
where 
\begin{equation}
\tilde{C_{\sigma}} = -\frac{2 \pi (4 \pi)^{\frac{d}{2}} \Gamma(d-1) }{\Gamma (\frac{2-d}{2}) \Gamma(\frac{d - 1}{2})^2}.
\end{equation}
This gives the quadratic part of the boundary effective action for sigma to be 
\begin{equation}
S_2 = \int \frac{d^{d} \mathbf{p}}{(2 \pi)^{d}} \frac{\sigma(\mathbf{p}) \sigma(-\mathbf{p})}{2} \bigg( \frac{N}{\tilde{C_{\sigma}}} (p)^{d-2} - \frac{1}{2 g}  \bigg).
\end{equation}
From here, it is clear that for $d<2$, the second term in the quadratic action can be dropped in the IR limit, while for $d>2$, it can be dropped in the UV limit.  This only leaves the induced kinetic term in the quadratic action and leads to the following two point function for $\sigma$
\begin{equation}
\langle \sigma(\mathbf{p}) \sigma(-\mathbf{p}) \rangle = \frac{\tilde{C_{\sigma}}}{N} (p^2)^{\frac{2 - d}{2}}
\end{equation} 
which gives in position space
\begin{equation}
\langle \sigma(\textbf{x}_1) \sigma(\textbf{x}_2) \rangle = \frac{C_{\sigma}}{ |\textbf{x}_1 - \textbf{x}_2|^2},  \ \ \ C_{\sigma} = \tilde{C}_{\sigma}  \frac{ 4 }{ (4 \pi)^{\frac{d}{2}} \Gamma(\frac{d}{2} - 1)}
\end{equation}
which implies that the conformal dimension of sigma operator to this order is 1. The power law correlation suggests the existence of an IR fixed point in $d<2$ and a UV fixed point in $d>2$. 

We can also compute the anomalous dimension of $\sigma$ to order $1/N$. In general, it should be computed using the two loop correction to the $\sigma$ propagator, but in this case, since $\phi$ does not get an anomalous dimension, we can use the $1/N$ corrections to the following correlator
\begin{equation*}
\begin{split}
\langle\sigma(0) \phi^I(q) \phi^J(-q) \rangle &= 
\includegraphics[scale=1,valign = m]{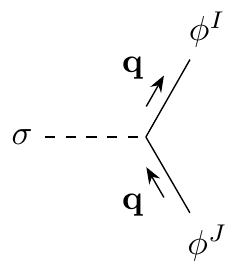} + \includegraphics[scale=1,valign = m]{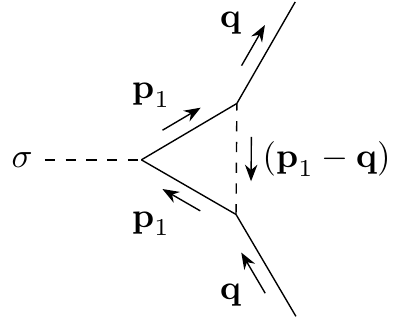} \  \\ \\
&+  \ \includegraphics[scale=1,valign = m]{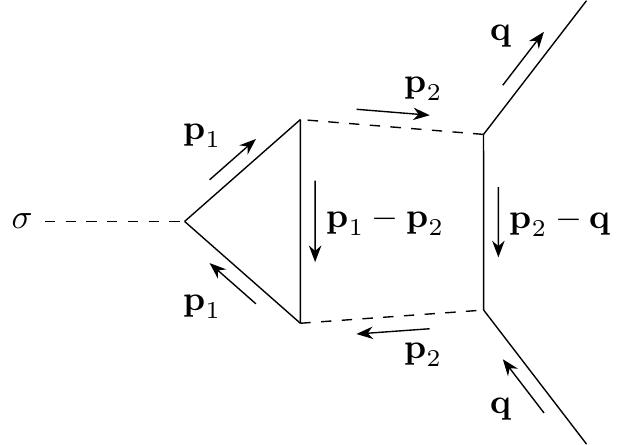} 
\end{split}
\end{equation*}

\begin{equation}
\begin{split}
&= \delta^{I J} + \frac{\tilde{C_{\sigma}} \delta^{I J}}{N} \int \frac{d^{d} p_1}{(2 \pi)^{d}} \frac{1}{|p_1|^2 |p_1 -q|^{d-2}}  + \frac{\tilde{C}_{\sigma}^2 \delta^{I J}}{N} \int \frac{d^{d} p_1}{(2 \pi)^{d}} \int \frac{d^{d} p_2}{(2 \pi)^{d}} \frac{1}{ |p_1|^2 |p_1 - p_2| |p_2 - q| |p_2|^ {2 (d - 2) }} \\
& = \delta^{I J} \bigg(1 - \frac{2 \log q   \ \tilde{C_{\sigma}}}{N (4 \pi)^{\frac{d}{2}} \Gamma(\frac{d}{2})}  - \frac{4 \log q \ \tilde{C_{\sigma}}^2 \ \Gamma(\frac{d - 1}{2}) \Gamma(\frac{3 - d}{2}) }{N (4 \pi)^{d} (d - 2) \sqrt{\pi} \Gamma( d- \frac{3}{2})} \bigg) \\
&= \delta^{I J} + \frac{\delta^{I J} \log( q^2/\mu^2) }{2 N} \bigg( \frac{2^{d} \sqrt{\pi}}{\Gamma(\frac{2 - 
d}{2}) \Gamma(\frac{d - 1}{2})} - \frac{2^{2 d -1} 
\sqrt{\pi} \Gamma( \frac{3 - d}{2}) 
\Gamma(\frac{d}{2}) \Gamma(\frac{d - 2}{2}) }
{\Gamma(d - \frac{3}{2}) \Gamma(\frac{d - 1}{2}) 
\Gamma(\frac{2 - d}{2})^2 }  \bigg).
\end{split}
\end{equation}
Applying Callan-Symanzik equation to it gives the anomalous dimension
\begin{equation}
\hat{\Delta}_{\sigma} = 1 + \hat{\gamma}_{\sigma} = 1 + \frac{1}{N} \bigg( \frac{2^{2 d -1} 
\sqrt{\pi} \Gamma( \frac{3 - d}{2}) 
\Gamma(\frac{d}{2}) \Gamma(\frac{d - 2}{2}) }
{\Gamma(d - \frac{3}{2}) \Gamma(\frac{d - 1}{2}) 
\Gamma(\frac{2 - d}{2})^2 } - \frac{2^{d} \sqrt{\pi}}{\Gamma(\frac{2 - d}{2}) \Gamma(\frac{d - 1}{2})}. \bigg)
\end{equation}
This can be expanded in $d = 2 - \epsilon$ 
\begin{equation}
\hat{\Delta}_{\sigma} = 1 - \frac{6 \epsilon}{N} + \frac{28 \epsilon^2 \log 2}{N}
\end{equation}
This precisely agrees with the dimension of $\phi^2$ operator in the $\epsilon$ expansion at large N in eq. \eqref{DimensionPhi^2Quartic}. This can also be expanded in $d = 1 + \epsilon$ 
\begin{equation}
\hat{\Delta}_{\sigma} = 1 - \frac{\epsilon^2}{N}
\end{equation}
and we will show that it agrees with the result obtained from non-linear sigma model in eq. \eqref{SigmaAnomNLSM} in the next subsection. Expanding in $d = 4 - \epsilon$  
\begin{equation}
\hat{\Delta}_{\sigma} = 1 - \frac{\epsilon^2}{N}
\end{equation}
which agrees with mixed $\sigma \phi$ theory described below in subsection \ref{MixedSigmaPhiBoundary}. 

The bulk propagator for $\phi$ now involves following contributions
\begin{equation}
\begin{split}
\langle \phi^I(-p,y_1) \phi^J(p,y_2) \rangle &= \includegraphics[scale=1, valign = b]{Phi4Prop} \ \ \ + \ \ \ 
\includegraphics[scale=1, valign = b]{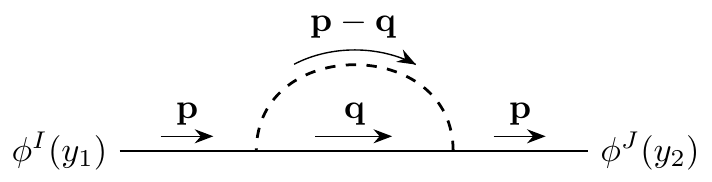} \\ \\
&= \frac{\delta^{IJ} (e^{- p |y_1 - y_2|} + e^{- p (y_1 + y_2)})}{|p|} + \frac{\tilde{C_{\sigma}} \delta^{I J} e^{- p (y_1 + y_2)} }{N |p|^2} \int \frac{d^{d} q}{(2 \pi)^{d}} \frac{1}{|q| ((p-q)^2)^{\frac{d-2}{2}}} \\
&= \frac{\delta^{IJ} (e^{- p |y_1 - y_2|} + e^{- p (y_1 + y_2)})}{|p|} + \frac{ \delta^{I J} 8 \pi \ \Gamma(d-1)e^{- p (y_1 + y_2)} }{N |p| (d-1) \Gamma(\frac{d-2}{2}) \Gamma(\frac{2-d}{2}) \Gamma(\frac{d - 1}{2})^2 } .
\end{split}
\end{equation}
We can Fourier transform it back to position space to get 
\begin{equation}
G^{I J}_{\phi} (x_1, x_2) = \delta^{I J} G^0_{\phi} (x_1, x_2) \ + \ \frac{ 4 \delta^{I J}  \ \Gamma(d-1) }{N \pi^{\frac{d - 1}{2}} (d-1) \Gamma(\frac{d-2}{2}) \Gamma(\frac{2-d}{2}) \Gamma(\frac{d - 1}{2}) } \frac{1}{((y_1 + y_2)^2 + (\textbf{x}_1 - \textbf{x}_2)^2)^{\frac{d - 1}{2}}} .
\end{equation}
The $1/N$ correction can be expanded in $d  = 2 -\epsilon$ and it matches with what we got in the previous subsection from the $\epsilon$ expansion. It can also be expanded in $d = 4 - \epsilon$ and it agrees with what we get from $\epsilon$ expansion in subsection \ref{MixedSigmaPhiBoundary}. 
\subsection{Non-linear sigma model in $d=1+\epsilon$} \label{NLSm $s = 1$}

Next model we will consider is related to the usual $O(N)$ non-linear sigma model, so let us first review the calculation of beta function for the usual case to set the notation. We define the model as 
\begin{equation}
S = \int d^d x \bigg( \frac{1}{2} \partial_{\mu} \phi^I \partial^{\mu} \phi^I + \sigma (\phi^I \phi^I - \frac{1}{t^2})  \bigg)
\end{equation}
where the Lagrange multiplier $\sigma$ imposes the constraint that $\phi^I \phi^I = \frac{1}{t^2}$. We can choose the following parametrization that solves the constraint
\begin{equation}
\phi^I = \psi^I, \ \ I = 1,...,N - 1; \ \ \ \  \phi^N = \frac{1}{t} \sqrt{1 - t^2 \psi^I \psi^I } = \frac{1}{t} - \frac{t}{2} \psi^I \psi^I + O(t^3).  
\end{equation} 
In terms of these variables, the action becomes 
\begin{equation}
S = \int d^d x \bigg( \frac{1}{2} \partial_{\mu} \psi^I \partial^{\mu} \psi^I + \frac{t^2}{2} \frac{(\psi^I \partial_{\mu}  \psi^I)^2}{1 - t^2 \psi^I \psi^I} \bigg) = \int d^d x \bigg( \frac{1}{2} \partial_{\mu} \psi^I  \partial^{\mu} \psi^I + \frac{t^2}{2} (\psi^I \partial_{\mu}\psi^I)^2 + O(t^4) \bigg) 
\end{equation}
We can then calculate the $\beta$ function by requiring that the correlation functions obey Callan-Symanzik equation
\begin{equation}
(\mu \frac{\partial}{\partial \mu} + \beta \frac{\partial}{\partial t} + n \gamma(t)) G^n = 0
\end{equation}
and the original O(N) symmetry forces the anomalous dimensions for all the $\phi^I$ to be the same. We can apply this to the two point function 

\begin{equation}
\begin{split}
\langle \psi^K (p) \psi^L (-p) \rangle & = \includegraphics[scale=1, valign = b]{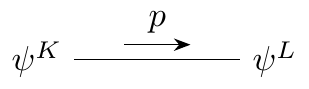} + \includegraphics[scale=1, valign = b]{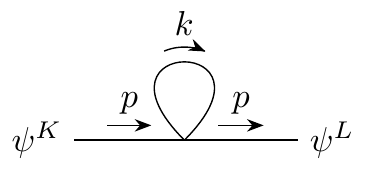} \\ \\
&= \frac{\delta^{K L}}{p^2} - \frac{t^2 \ \delta^{K L}}{(p^2)^2} \int \frac{d^d k }{(2 \pi)^d} \frac{p^2 + k^2}{k^2 + m^2} \\
&= \frac{\delta^{K L}}{p^2} - \frac{t^2 \delta^{K L} }{p^2}  \frac{\Gamma(1 - \frac{d}{2})}{(4 \pi)^{\frac{d}{2}} (m^2)^{1 - \frac{d}{2}}} - \frac{t^2 \delta^{K L} }{(p^2)^2}  \frac{ \frac{d}{2} \ \Gamma( - \frac{d}{2})}{(4 \pi)^{\frac{d}{2}} (m^2)^{- \frac{d}{2}}}
\end{split}
\end{equation}
where we have introduced an IR cutoff $m^2$. The last term vanishes as $m \rightarrow 0$ for all $d \ge 0$. The other two terms in $d = 2 + \epsilon$ give 
\begin{equation}
\langle \psi^K (p) \psi^L (-p) \rangle = \frac{\delta^{K L}}{p^2} \bigg( 1 - \frac{t^2}{4 \pi} \log \frac{\mu^2}{m^2} \bigg). 
\end{equation}
This satisfies Callan-Symanzik equation with 
\begin{equation}
\gamma_{\phi}(t) = \frac{t^2 }{4 \pi}.
\end{equation}
We next consider the one point function of $\phi^N$
\begin{equation}
\begin{split}
\langle \phi^N (0) \rangle  &= \frac{1}{t} - \frac{t}{2} \langle \psi^a \psi^a (0) \rangle - \frac{t^3}{8}\langle \psi^a \psi^a (0) \psi^b \psi^b (0) \rangle  \\
&=  \frac{1}{t} - \frac{t (N -1) }{2} G_0(0,0) + \frac{t^3 (N -1)}{2} \int d^{d} x G_0(x,x) (\partial_{\mu} G_0(0,x))^2 \\ 
&-\frac{t^3 ((N -1)^2 + 2 (N -1))}{8} (G_0(0,0))^2  \\
&= \frac{1}{t} - \frac{t (N - 1)}{2}  \int \frac{d^d k}{(2 \pi)^d} \frac{1}{k^2 + m^2} - \frac{t^3 ((N -1)^2 - 2 (N -1)) }{8} \bigg( \int \frac{d^d k}{(2 \pi)^d} \frac{1}{k^2 + m^2} \bigg)^2  \\ 
&=  \frac{1}{t} - \frac{t (N - 1)}{8 \pi} \log \frac{\mu^2}{m^2}  - \frac{t^3 (N -1)(N - 3)}{8 (4 \pi)^2} \bigg( \log \frac{\mu^2}{m^2} \bigg)^2
\end{split}
\end{equation}
where in the last line, we plugged in $d = 2 + \epsilon$. We can now apply Callan-Symanzik equation to it and we find
\begin{equation}
\beta(t) = \frac{\epsilon}{2} t -\frac{t^3 (N -2)}{4 \pi}
\end{equation}
where the first term is present because in $2 + \epsilon$ dimensions, $t$ has engineering dimensions $-\epsilon/2$. The sign of $\beta$ function suggests a UV fixed point in $2 + \epsilon$ dimensions at 
\begin{equation}
t^2 = t_*^2 = \frac{2 \pi \epsilon}{N - 2}.
\end{equation}
The anomalous dimensions of the field $\phi$ at the fixed point $\gamma_{\phi} = \frac{\epsilon}{2 (N - 2)}$ agrees with the known results. The anomalous dimensions of the Lagrange multiplier field $\sigma$ which is the analogue of the field $\sigma$ in the large N analysis, can be found by the following relation
\begin{equation}
\Delta_{\sigma} = d + \beta'(t_*) = d + \frac{\epsilon}{2} - \frac{3 t_*^2 (N - 2)}{4 \pi} = 2 + O(\epsilon^2). 
\end{equation}  

We will now consider a variant of the non-linear sigma model where the sphere constraint is only imposed on the $d$-dimensional boundary: 
\begin{equation}
S = \int d^{d + 1} x  \frac{1}{2} \partial_{\mu} \phi^I \partial^{\mu} \phi^I + \int d^{d} x \ \sigma (\phi^I \phi^I - \frac{1}{t^2})\,.  
\end{equation} 
As in the case of the local models, the auxiliary field $\sigma$ is related to the Hubbard-Stratonovich field introduced in the large $N$ treatment. The fact that $\hat{\Delta}_{\sigma}=1+O(1/N)$, as shown in the previous section, suggests that the lower critical dimension is $d=1$, and we should look for UV fixed points of the above model in $d=1+\epsilon$ boundary dimensions.  

As in previous sections, the bulk propagator induces a $1 / |\textbf{p}|$ propagator on the boundary, which in the position space looks like a non-local kinetic term
\begin{equation}
S_{\textrm{bdry}} = - \frac{\Gamma(\frac{d + 1}{2})}{\pi^{\frac{d + 1}{2}}}  \int d^{d} x \ d^{d} y \ \frac{ \phi^I (x)  \phi^I(y)}{|x-y|^{d + 1}} + \int d^{d} x \ \sigma (\phi^I \phi^I - \frac{1}{t^2})
\end{equation}
We can now solve the constraint on the boundary in terms of the variables $\psi^{a}$ as before to get 
\begin{equation}
S_{\textrm{bdry}} = - \frac{ \Gamma(\frac{d + 1}{2})}{\pi^{\frac{d + 1}{2}}} \int d^{d} x \ d^{d} y \ \frac{ \psi^a (x)  \psi^a(y)}{|x-y|^{d + 1}}  -  \frac{ \Gamma(\frac{d + 1}{2})}{\pi^{\frac{d + 1}{2}}}  \frac{t^2}{4} \int d^{d} x \ d^{d} y \ \frac{ \psi^a \psi^a (x)  \psi^b\psi^b(y)}{|x-y|^{d + 1}} + ...
\end{equation}
where we dropped a constant unimportant shift, as well as corrections at higher orders in $t^2$. So, for the purpose of computing boundary correlation functions, this action gives a  propagator for the $\psi^a$ field that goes like $1/|p|$, and we can use this to develop perturbation theory with the interaction term from above expression. Let us first try to compute the diagram that would give us the anomalous dimension of the field $\psi^a$. We will show that it vanishes in accord with the expectation  since $\phi^I$ is a free field in the bulk. The two point function of the field $\psi^a$ goes like
\begin{equation}
\begin{split}
\langle \psi^a(x) \psi^b(y) \rangle &= \delta^{ab} G_0(x,y) -  \frac{ \Gamma(\frac{d + 1}{2})}{\pi^{\frac{d + 1}{2}}} t^2 \delta^{ab} \int d^{d} z \ d^{d} w \frac{ G_0(x,w) G_0(y,z) G_0(z,w)}{|z-w|^{d + 1}}  \\
& -  \frac{ \Gamma(\frac{d + 1}{2})}{\pi^{\frac{d + 1}{2}}}  t^2 \delta^{ab} \int d^{d} z \ d^{d} w \frac{ (N -1) G_0(x,w) G_0(y,w) G_0(z,z)}{|z-w|^{d + 1 }}.
\end{split}
\end{equation}
The term in the second line vanishes when we do the integral over z. We can now go to momentum space to get 
\begin{equation}
\langle \psi^a(-p) \psi^b(p) \rangle = \frac{\delta^{ab} }{|p|} + \frac{t^2 }{|p|^2 } \int \frac{d^{d} q}{(2 \pi)^{d}} \frac{|p - q|}{|q|}. 
\end{equation}
The integral can be evaluated in dimensional regularization by adding a small mass and then expanding in mass in $d = 1 + \epsilon $ to get
\begin{equation}
\begin{split}
\langle \psi^a(-p) \psi^b(p) \rangle &= \frac{\delta^{ab} }{|p|} - \frac{t^2 }{|p|^2 } \frac{(m^2)^{\frac{d}{2}} \Gamma(-\frac{d}{2}) \Gamma(\frac{d + 1}{2}) \Gamma(-\frac{1}{2}) \ _2F_1 (-\frac{d}{2},-\frac{1}{2}, \frac{d}{2} ,-\frac{k^2}{m^2} ) }{2^{d + 1} \pi^{\frac{d + 2}{2}} \Gamma(\frac{d}{2})} \\
&= \frac{\delta^{ab} }{|p|} - \frac{t^2 }{|p|} \frac{(2  + \log \frac{m^2}{4 p^2} )}{2 \pi} + O(m^2).
\end{split}
\end{equation}
Since there is no $1/\epsilon$ pole, this implies that the field $\psi^a$ does not get an anomalous dimension. We next go on to compute the beta function for the coupling $t$. For that, we will apply the Callan-Symanzik equation to the one point function of the field $\phi^N(0)$ as before

\begin{equation*}
\langle \phi^N (0) \rangle = \includegraphics[scale=2, valign = b]{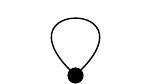} + \includegraphics[scale=0.50, valign = b]{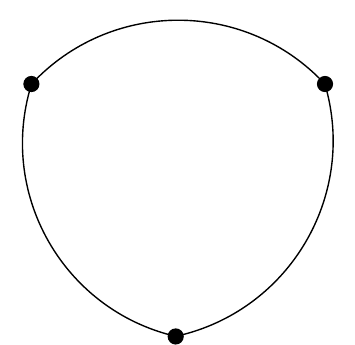} + \includegraphics[scale=2, valign = b]{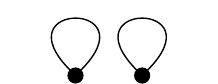} + \ \ \includegraphics[scale=2, valign = m]{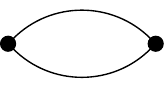}  
\end{equation*}
\begin{equation}
\begin{split}
&= \frac{1}{t} - \frac{t}{2} \langle \psi^a \psi^a (0) \rangle - \frac{t^3}{8}\langle \psi^a \psi^a (0) \psi^b \psi^b (0) \rangle \\
&= \frac{1}{t}  -\frac{t (N - 1)}{2} G_0(0,0) -  \frac{ \Gamma(\frac{d + 1}{2})}{\pi^{\frac{d + 1}{2}}}  \frac{4 (N - 1)t^3}{8} \int d^{d} z \ d^{d} w \frac{ G_0(0,w) G_0(0,z) G_0(z,w)}{|z-w|^{d + 1}} \\  &-\frac{t^3 ((N-1)^2 + 2 (N - 1))}{8} G_0(0,0)^2 \\ 
&= \frac{1}{t} - \frac{t (N - 1) }{2} \int \frac{d^{d} k}{(2 \pi)^{d}} \frac{1}{|k|} + \frac{ (N - 1) t^3}{2}  \int \frac{d^{d} k}{(2 \pi)^{d}} \frac{1}{|k|^2}  \int \frac{d^{d} l}{(2 \pi)^{d}} \frac{|k - l|}{|l|} \\
& - \frac{t^3  ((N-1)^2 + 2 (N - 1))}{8 } \int \frac{d^{d} k}{(2 \pi)^{d}} \frac{1}{|k|} \int \frac{d^{d} l}{(2 \pi)^{d}} \frac{1}{|l|}.    
\end{split}
\end{equation}
The integrals in the second and fourth term are straightforward. However, the integral in the third term is a bit subtle. Let us introduce an IR regulator mass, and perform the integral over l first, which gives in d dimensions
\begin{equation}
\int \frac{d^{d} l}{(2 \pi)^{d}} \frac{|k-l|}{\sqrt{l^2 + m^2}} = -\frac{(m^2)^{\frac{d}{2}} \Gamma(\frac{-d}{2}) \Gamma(\frac{d + 1}{2}) \Gamma(-\frac{1}{2}) \ _2F_1 (\frac{-d}{2},-\frac{1}{2}, \frac{d}{2} ,-\frac{k^2}{m^2} ) }{2^{d + 1} \pi^{\frac{d + 2}{2}} \Gamma(\frac{d}{2})}.
\end{equation}
Fortunately, it is possible to do the integral over $k$ now, and doing that and then taking $d = 1 + \epsilon$, gives, to leading order in $\epsilon$
\begin{equation}
\int \frac{d^{d} k}{(2 \pi)^{d}} \frac{1}{|k|^2}  \int \frac{d^{d} l}{(2 \pi)^{d}} \frac{|k - l|}{|l|} = \frac{1}{8 \pi^2} \bigg( \frac{4}{\epsilon^2} + \frac{4(\gamma + \log m^2 - \log 4 \pi)}{\epsilon} \bigg) - \frac{1}{4 \pi^2} \bigg( \frac{-2}{\epsilon} \bigg) + O(\epsilon^0)
\end{equation}
The other two integrals can be evaluated by usual means, and overall it gives 
\begin{equation}
\begin{split}
\langle \phi^N (0) \rangle &=  \frac{1}{t} \ + \ \frac{t (N - 1) }{2 \pi} \bigg( \frac{1}{\epsilon} + \frac{\gamma + \log m^2 - \log 4 \pi}{2} \bigg) 
+ \frac{(N -1) t^3}{4 \pi^2 \epsilon}  \\
& -\frac{t^3 (N -1)^2}{8 \pi^2}\bigg( \frac{1}{\epsilon^2} + \frac{\gamma + \log m^2 - \log 4 \pi}{\epsilon} \bigg)
\end{split}
\end{equation} 
We can now introduce the counterterms to cancel the divergences by redefining $t \rightarrow t_0 = t + \delta_t$ to get
\begin{equation}
\begin{split}
\langle \phi^N (0) \rangle &=  \frac{1}{t} \ - \frac{\delta_t}{t^2} + \frac{\delta_t^2}{t^3} \ +  \frac{(t + \delta_t)(N - 1) }{2 \pi} \bigg( \frac{1}{\epsilon} + \frac{\gamma + \log m^2 - \log 4 \pi}{2} \bigg) 
+ \frac{(N -1) t^3}{4 \pi^2 \epsilon}  \\
& -\frac{t^3 (N -1)^2}{8 \pi^2}\bigg( \frac{1}{\epsilon^2} + \frac{\gamma + \log m^2 - \log 4 \pi}{\epsilon} \bigg).
\end{split}
\end{equation} 
The counterterm is fixed by the requirement that it should cancel all the divergent terms which gives the original bare coupling in terms of renormalized coupling
\begin{equation}
t_0 = \mu^{-\epsilon/2} \bigg(t + \frac{(N - 1) t^3}{2 \pi \epsilon} +  \frac{(N - 1) t^5}{4 \pi^2 \epsilon} + \frac{3 (N - 1)^2 t^5}{8 \pi^2 \epsilon^2}  \bigg).
\end{equation} 
This gives the $\beta$ function
\begin{equation}
\beta(t) = \frac{\epsilon}{2} t - \frac{t^3 (N -1)}{2 \pi} - \frac{t^5(N-1)}{2 \pi^2}.
\end{equation} 
Notice that the $\beta$ function here is proportional to $N - 1$ as opposed to $N - 2$ in the usual local case. This tells us that the $N = 1$ case has to be treated separately, similar to what happens for $ N = 2$ case in the usual $O(N)$ model in two dimensions \cite{Kosterlitz_1973, Kosterlitz_1974}. This beta function gives a fixed point at 
\begin{equation}
t_*^2 = \frac{\epsilon \pi }{(N-1)} -\frac{\epsilon^2 \pi}{ (N - 1)^2}
\end{equation}
This gives the dimension of the field $\sigma$
\begin{equation} \label{SigmaAnomNLSM}
\hat{\Delta}_{\sigma} = d + \beta'(t_*) = 1 - \frac{\epsilon^2}{(N - 1)}
\end{equation}
in exact agreement with the prediction of the large $N$ expansion. 

\subsection{Mixed $\sigma\phi$ theory in $d=4-\epsilon$} \label{MixedSigmaPhiBoundary}
The large $N$ analysis described in subsection \ref{LargeN $s = 1$} applies for general $d$, and in particular it can be formally pushed to $d>2$. In $d=2+\epsilon$, one finds formal UV fixed points of the quartic model (\ref{phi4-ON}). The fact that at large $N$ the dimension of $\sigma$ is near 1 suggests that it becomes a free propagating field in $d=4$ boundary dimensions. Then, in close analogy with the situation for local $O(N)$ models \cite{Fei:2014yja}, one expects that a UV completion of the formal UV fixed point of the quartic model in $d>2$ is provided by the following model
\begin{equation} \label{MixedSigmaPhiAction}
S = \int d^{d + 1} x \frac{1}{2} (\partial_{\mu} \phi^I)^2 + \int d^{d} x \bigg( \frac{1}{2} (\partial \sigma )^2 + \frac{g_1}{2} \sigma \phi^I \phi^I  + \frac{g_2}{4!} \sigma^4 \bigg).
\end{equation}
where $\sigma$ propagates only on the boundary. The couplings $g_1$ and $g_2$ are classically marginal in $d=4$, and we can look for perturbative IR fixed points in $d=4-\epsilon$. 

The leading correction to $\sigma$ propagator is given by the one-loop diagram

\begin{equation}
\begin{split}
G^{ 2,0}  &= \includegraphics[scale=1, valign = m]{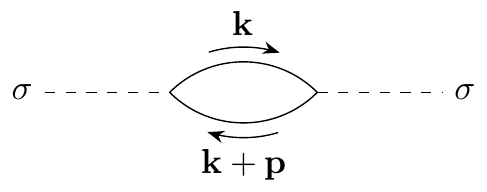} + \includegraphics[scale=1, valign = m]{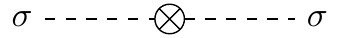} \\
&=\frac{N (-g_1)^2}{2} \int \frac{d^{d} \textbf{k}}{(2 \pi)^{d}} \frac{1}{|\textbf{p} +  \textbf{k}| \  k} - p^2 \delta_{\sigma} \\
&= \frac{N g_1^2 \Gamma(\frac{2 -d}{2})  (p^2)^{\frac{d-2}{2}} \Gamma(\frac{d - 1}{2})^2  }{2 (4 \pi)^{\frac{d}{2}} \Gamma(d - 1) \pi} - p^2 \delta_{\sigma} .
\end{split}
\end{equation}
We then take a derivative with $p^2$ at $p^2 = \mu^2$ and set the divergent part to 0. This gives
\begin{equation}
\delta_{\sigma} = -\frac{N g_1^2 \Gamma(\frac{4 -d}{2}) \Gamma(\frac{d - 1}{2})^2   }{2 (4 \pi)^{\frac{d}{2}} \Gamma(d - 1) \pi (\mu^2)^{\frac{4 -d}{2}} } = - \frac{N g_1^2}{8 \epsilon (4 \pi)^2} .
\end{equation}
Next, we can compute the corrections to the vertex $g_1$

\begin{equation}
\begin{split}
G^{ 1, 2} &= \includegraphics[scale=1, valign = m]{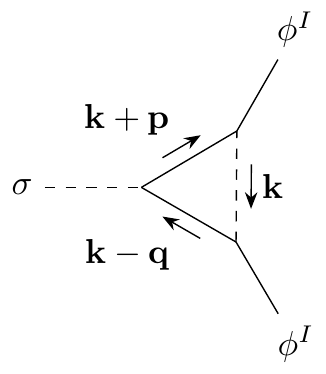} + \includegraphics[scale=1, valign = m]{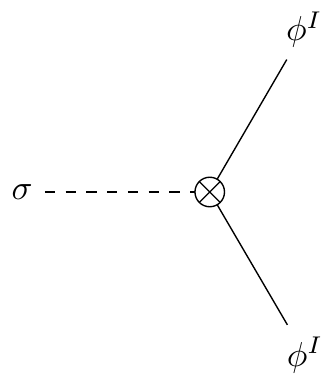} \\
&(-g_1)^3 \int \frac{d^{d} \textbf{k}}{(2 \pi)^{d}} \frac{1}{|\textbf{k} - \textbf{q}| |\textbf{k} + \textbf{p}| k^2} - \delta_{g_1} \\
&= -g_1^3 \frac{\Gamma(\frac{4-d}{2})}{(4 \pi)^{\frac{d}{2}}  (\mu^2)^{\frac{4-d}{2}}}  - \delta_{g_1}
\end{split}
\end{equation}
which in $d = 4-\epsilon$ gives 
\begin{equation}
\delta_{g_1} = -g_1^3 \frac{\Gamma(\frac{4-d}{2})}{(4 \pi)^{\frac{d}{2}}  (\mu^2)^{\frac{4-d}{2}}}  = -\frac{g_1^3}{8 \pi^2 \epsilon}.
\end{equation}
Similarly, the one loop correction to $g_2$ is given by the following diagrams (we are evaluating these at all external momenta $= \mu^2$)
\begin{equation*}
G^{ 4,0 } = \includegraphics[scale=1, valign = m]{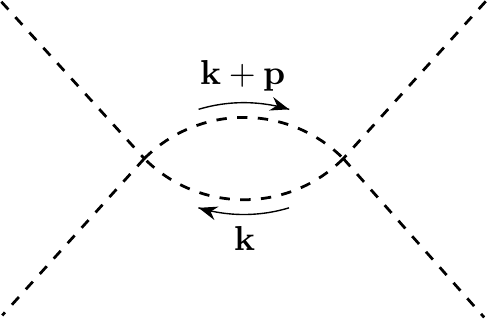} \ \ + \ \ \includegraphics[scale=1, valign = m]{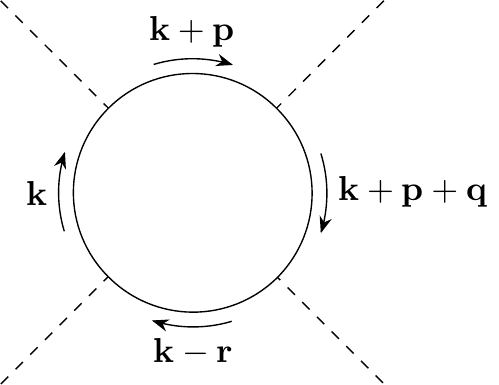} \ \ + \ \  \includegraphics[scale=1, valign = m]{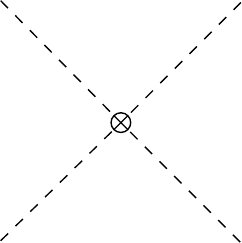}   
\end{equation*}
\begin{equation}
\begin{split}
& = \frac{3 (-g_2)^2}{2} \int \frac{d^{d} \textbf{k}}{(2 \pi)^{d}}  \frac{1}{k^2 (\textbf{k} + \textbf{p})^2} + 3 (-g_1)^4 N \int \frac{d^{d} \textbf{k}}{(2 \pi)^{d}}  \frac{1}{k |\mathbf{k} - \mathbf{r}| |\mathbf{k} + \mathbf{p}| |\mathbf{k} + \mathbf{p} + \mathbf{q}|}  - \delta_{g_2} \\
&= \frac{3 g_2^2 \Gamma(\frac{4-d}{2})} {2 (4 \pi)^{\frac{d}{2}}  (\mu^2)^{\frac{4-d}{2}}} + \frac{ 3 g_1^4 N \Gamma(\frac{4-d}{2})} {(4 \pi)^{\frac{d}{2}}  (\mu^2)^{\frac{4-d}{2}}} -\delta_{g_2} 
\end{split}
\end{equation}
which implies 
\begin{equation}
\delta_{g_2} = \frac{3 g_2^2 \Gamma(\frac{4-d}{2})} {2 (4 \pi)^{\frac{d}{2}}  (\mu^2)^{\frac{4-d}{2}}} + \frac{3 g_1^4 N \Gamma(\frac{4-d}{2})} {(4 \pi)^{\frac{d}{2}}  (\mu^2)^{\frac{4-d}{2}}} = \frac{3 g_2^2 + 6 g_1^4 N}{16 \pi^2 \epsilon}.
\end{equation}
Using these counterterms, we can calculate the $\beta$ function. The Callan-Symanzik equation for a correlation function with m external $\sigma$ lines and n external $\phi$ lines is 
\begin{equation}
(\mu \frac{\partial}{\partial \mu} + \beta_1 \frac{\partial}{\partial g_1} + \beta_2 \frac{\partial}{\partial g_2} + m \gamma_{\sigma} + n \gamma_{\phi} ) G^{m,n} = 0.
\end{equation}
Applying this to $G^{1,2}$ gives
\begin{equation}
\beta_1 = -\frac{\epsilon}{2} g_1 + \mu \frac{\partial}{\partial \mu} (-\delta_{g_1} + \frac{g_1}{2} (2 \delta_{\phi} + \delta_{\sigma})) = -\frac{\epsilon}{2} g_1 + \frac{(N-32) g_1^3}{16 (4 \pi)^2} 
\end{equation}
Applying Callan-Symanzik equation to $G^{4,0}$ gives
\begin{equation}
\beta_2 = -\epsilon g_2 +  \mu \frac{\partial}{\partial \mu} (-\delta_{g_2} + \frac{g_2}{2} (4 \delta_{\sigma}) ) = -\epsilon g_2 + \frac{12 g_2^2 + 24 g_1^4 N + g_1^2 g_2 N}{4 (4 \pi)^2}.
\end{equation}

It is possible to find two unitary fixed point at $N > N_{\rm crit}= 4544$ with coupling constants given by 
\begin{equation}
(g_1^*)^2 = \frac{8 (4 \pi)^2 \epsilon}{N-32}, \ \ \ (g_2^*)_{\pm} = \frac{12288 N \pi^2 \epsilon}{(N - 32) ( \pm \sqrt{1024 + N (N - 4544)} - (N + 32))}.
\end{equation}
Since we find two fixed points here, we should look at their IR stability by looking at the eigenvalues of the following matrix for the positive and negative sign root
\begin{equation}
M_{i j} = \frac{\partial \beta_i}{\partial g _j},
\ \ M = \begin{bmatrix}
\frac{-\epsilon}{2} + \frac{3 (N - 32)}{16 (4 \pi)^2} (g_1^*)^2 & 0 \\ \\
\frac{48 N (g_1^*)^3 + g_1^* g_2^* N}{ 2 (4 \pi)^2} & - \epsilon + \frac{24 g_2^* + N (g_1^*)^2}{4 (4 \pi)^2}
\end{bmatrix}
\end{equation}
For IR stability, we want both the eigenvalues of this matrix to be positive, and that only happens when we choose the negative root $(g_2^*)_-$ (sign of $g_1^*$ does not actually affect the eigenvalues). So the fixed point with $(g_2^*)_-$ is the IR stable fixed point and should be the one that matches the large $N$ fixed point near four dimensions. Note that the value of $g_2^*$ is negative for both the fixed points, indicating that this fixed point is non-perturbatively unstable, in the sense that the vacuum is not stable. For sufficiently large $N$, we may regard it as a metastable BCFT, similarly to the local $O(N)$ models in $4<d<6$ \cite{Giombi:2019upv}. 

We can also compute the anomalous dimensions at the fixed point. The field $\phi$ does not get any anomalous dimensions, while the anomalous dimension of the field $\sigma$ can be computed from $\delta_{\sigma}$
\begin{equation}
\hat{\gamma}_{\sigma} = \frac{\mu}{2} \frac{\partial }{\partial \mu} \log Z_{\sigma} = \frac{\epsilon N }{2 (N - 32)}
\end{equation}
which gives 
\begin{equation}
\hat{\Delta}_{\sigma} = 1 + \frac{16 \epsilon}{N - 32}
\end{equation}
in precise agreement with the large $N$ prediction, expanded near $d=4$.  

The correction to bulk propagator of the field $\phi$ is given by 
\begin{equation}
\begin{split}
&\langle \phi^I(-p,y_1) \phi^J(p,y_2) \rangle = \includegraphics[scale=1, valign = b]{Phi4Prop} + \includegraphics[scale=1, valign = b]{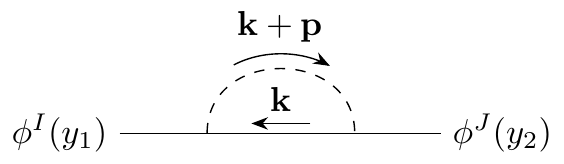} \\ \\
&= \frac{\delta^{IJ} (e^{- p |y_1 - y_2|} + e^{- p (y_1 + y_2)})}{|p|} +  \frac{ \delta^{I J} g_1^2 e^{- p (y_1 + y_2)}}{|p|^2} \int \frac{d^{d} \textbf{k}}{(2 \pi)^{d }} \frac{1}{(\textbf{p} +  \textbf{k})^2 \  k} \\
&= \frac{\delta^{IJ} (e^{- p |y_1 - y_2|} + e^{- p (y_1 + y_2)})}{|p|} + \frac{ \delta^{I J} g_1^2 e^{- p (y_1 + y_2)}  \Gamma(\frac{3 - d}{2}) \Gamma(\frac{d}{2} - 1) \Gamma(\frac{d - 1}{2}) (p^2)^{\frac{d - 5}{2}} }{ ( 4\pi )^{\frac{d}{2}} \sqrt{\pi} \Gamma(d - \frac{3}{2})}  .
\end{split}
\end{equation}
We can again Fourier transform back to position space to get 
\begin{equation}
G^{I J}_{\phi} (x_1, x_2) = \delta^{I J} G^0_{\phi} (x_1, x_2) - \frac{ 8 \delta^{I J}   \epsilon}{ 3 \pi^2 (N - 32) ((\textbf{x}_1 - \textbf{x}_2)^2 + (y_1 + y_2)^2)^{\frac{3}{2}}}. 
\end{equation}
At large $N$, this agrees with the result obtained from large $N$ expansion expanded in $d = 4 - \epsilon$.
\subsubsection{Boundary instanton}
The mixed $\sigma \phi$ theory described in eq. \eqref{MixedSigmaPhiAction} can be written on the boundary as 
\begin{equation}
S_{\textrm{bdry}} = \frac{2 \Gamma(\frac{d + 1}{2})}{\pi^{\frac{d}{2}} \Gamma(-\frac{1}{2})} \int d^d x d^d y \frac{\phi^I (x) \phi^I(y)}{|\textbf{x} - \textbf{y}|^{d + 1}} +  \int d^{d} x \bigg( \frac{1}{2} (\partial \sigma )^2 + \frac{g_1}{2} \sigma \phi^I \phi^I  + \frac{g_2}{4!} \sigma^4 \bigg)
\end{equation}
Since the coupling $g_2$ is negative at the fixed point, the vacuum $\sigma = \phi^I = 0$ can only be metastable and must tunnel to large absolute values of $\sigma$. Indeed for negative $g_2$, there is a real instanton solution responsible for this tunneling found in \cite{ Brezin:1976wa, Mckane:1978me, McKane:1978md} in the context of usual $\phi^4$ interaction in four dimensions 
\begin{equation}
\phi^I = 0 , \ \ \ \sigma = \sqrt{\frac{- 48}{g_2}} \frac{\lambda}{1 + \lambda^2 (\textbf{x} - \textbf{a})^2}.
\end{equation}
This instanton solution is expected to give non-perturbatively small imaginary parts to critical exponents \cite{McKane:1984eq}. This is because the fluctuations of $\sigma$ about the instanton background include a negative mode which yields an imaginary contribution to the partition function. 

We can perform a conformal mapping of the boundary to $S^4$, which will result in a $\sigma^2$ conformal coupling term in the action, and the solution just changes by a Weyl factor
\begin{equation}
\sigma = \sqrt{\frac{- 12}{g_2}} \frac{\lambda (1 + \textbf{x}^2)}{1 + \lambda^2 (\textbf{x} - \textbf{a})^2}.
\end{equation}
For $\lambda = 1$ and $\textbf{a} = 0$, it just becomes a constant VEV on the sphere, and the action evaluated on the solution turns out to be 
\begin{equation}
S^{\textrm{inst}}_{\textrm{bdry}} = -\frac{16 \pi^2}{g_2}. 
\end{equation}
This can be evaluated at the fixed point and then we can take the large $N$ limit to compare with the result from large $N$ calculation 
\begin{equation}
S^{\textrm{inst}}_{\textrm{bdry}} = -\frac{16 \pi^2}{(g_2^*)_-} =\frac{(N - 32) ( \sqrt{1024 + N (N - 4544)} + (N + 32))}{768 N  \epsilon} +O(\epsilon^0) \stackrel{N\gg 1}{\approx}  \frac{N}{384 \epsilon} . 
\label{inst-eps}
\end{equation}

The same result can be derived in the large $N$ theory by writing eq. \eqref{LargeNActionBoundary} as an action on the boundary
\begin{equation} 
S = \frac{2 \Gamma(\frac{d + 1}{2})}{\pi^{\frac{d}{2}} \Gamma(-\frac{1}{2})} \int d^d x d^d y \frac{\phi^I (x) \phi^I(y)}{|\textbf{x} - \textbf{y}|^{d + 1}}  + \int d^{d} x  \frac{\sigma \phi^I \phi^I}{2} .
\end{equation}
We can conformally map it to a sphere 
\begin{equation}
S = \frac{2 \Gamma(\frac{d + 1}{2})}{\pi^{\frac{d}{2}} \Gamma(-\frac{1}{2})} \int d^d x d^d y \sqrt{g(x)} \sqrt{g(y)} \frac{\phi^I (x) \phi^I(y)}{s(x,y)^{d + 1}}  + \int d^{d} x \sqrt{g(x)}  \frac{\sigma \phi^I \phi^I}{2}
\end{equation}
We will again look for the classical solution with a constant $\sigma$ on the sphere and compute the instanton action by integrating out $\phi^I$
\begin{equation}
S^{\textrm{inst}}_{\textrm{bdry}} (\sigma) = \frac{N}{2} \log \det \bigg( \frac{2 \Gamma(\frac{d + 1}{2})}{\pi^{\frac{d}{2}} \Gamma(-\frac{1}{2})} \frac{1}{s(x,y)^{d + 1}}  + \frac{\sigma}{2} \delta (\textbf{x} - \textbf{y})   \bigg).
\end{equation}
In general, the chordal distance on the sphere can be decomposed into spherical harmonics as follows \cite{Giombi:2019upv}
\begin{equation}
\frac{1}{s(x,y)^{2 \Delta}} = \sum_{n = 0}^{\infty} k_n(\Delta) Y^*_{n, \vec{m}} (x) Y_{n,\vec{m}} (y), \ \ \ k_n(\Delta) = \pi^{d/2} 2^{d - 2 \Delta} \frac{\Gamma(\frac{d}{2} - \Delta) \Gamma(n + \Delta)}{\Gamma(\Delta) \Gamma(d + n  - \Delta)}
\end{equation}
These spherical harmonics form a complete set of eigenfunctions with the following eigenvalue equation 
\begin{equation}
\int d^d y \sqrt{g(y)} \frac{1}{s(x,y)^{2 \Delta}} Y_{n,\vec{m}}(y) = k_n (\Delta) Y_{n, \vec{m}} (x). 
\end{equation} 
Using this, the required determinant becomes 
\begin{equation}
S^{\textrm{inst}}_{\textrm{bdry}} (\sigma) = \frac{N}{2} \sum_n D_n \log \bigg( \frac{2 \Gamma(\frac{d + 1}{2})}{\pi^{\frac{d}{2}} \Gamma(-\frac{1}{2})} k_n \bigg(\frac{d + 1}{2} \bigg)  + \frac{\sigma}{2} \bigg), \ \ \ D_n = \frac{(2 n + d -1) \Gamma(n + d -1)}{n! \Gamma(d)}
\end{equation}
where $D_n$ is the degeneracy of the eigenvalue $k_n$ with all the degenerate states labeled by $\vec{m}$ above. The constant value of $\sigma$ which extremizes this action can be found by solving 
\begin{equation}
\frac{\partial S^{\textrm{inst}}_{\textrm{bdry}}}{\partial \sigma} = 0 = \frac{N}{4} \sum_n \frac{ D_n}{ \frac{\Gamma(n  + (d + 1)/2)}{\Gamma(n  + (d - 1)/2)} + \frac{\sigma}{2} } = \frac{N \sigma \Gamma(1 - d) \Gamma(\frac{d - 1 + \sigma}{2})}{ 4 \Gamma(\frac{3 - d + \sigma}{2})}.
\end{equation}
So apart from the usual vacuum $\sigma = 0$, we also have other saddles
\begin{equation} \label{SigmaSaddle}
\sigma = d - 3 - 2 n
\end{equation} 
for positive integer $n$. The saddle point value of $\sigma$ is effectively the mass of field $\phi^I$ at large $N$. We want it to be positive for stability of $\phi^I = 0$ vacuum. Hence for $d<3$, $\sigma = 0$ is the only allowed saddle, while for $3 < d < 4$, the $n = 0$ saddle in eq. \ref{SigmaSaddle} is also allowed. So we expect the $n = 0$ instanton configuration to match the classical solution found above in $4 - \epsilon$ dimensions. Instanton action for this configuration is
\begin{equation}
S^{\textrm{inst}}_{\textrm{bdry}} (\sigma) - S_{\textrm{bdry}} (0) = \int_0^{d - 3} d \sigma \frac{\partial S^{\textrm{inst}}_{\textrm{bdry}}}{\partial \sigma}.
\end{equation} 
This clearly vanishes in $d = 3$. We can perform this integral in $d = 4 - \epsilon$ and compare with the result of the $\epsilon$ expansion in the previous section. We find
\begin{equation}
S^{\textrm{inst}}_{\textrm{bdry}} (\sigma) - S_{\textrm{bdry}} (0) = \frac{N}{384 \epsilon}+O(\epsilon^0)
\end{equation}
which precisely matches the $\epsilon$ expansion result (\ref{inst-eps}). 

\section{Higher-spin displacement operators} \label{HSDisplacement}
As discussed in section \ref{HSDisplacementGeneral}, a spin $s$ conserved current in the bulk induces a tower of protected operators on the boundary with dimension $d + 1 + s -2$ and spin ranging between $0$ and $s -2$. They are bilinears in the boundary operator $\phi$ and have the schematic form $\sim \phi \vec{\partial}^{2 n} \partial_{\nu_1} \partial_{\nu_2}.. \partial_{\nu_l} \phi$ with $n \ge 1$.  They appear in the conformal block decomposition of the four point function of the boundary field $\phi$. The scalar ones with boundary spin $0$ also appear in the boundary channel conformal block decomposition of two point function of the bulk scalar $\phi^I \phi^I$. In the following subsections, we will see that these operators have protected dimensions in perturbation theory using their appearance in both these conformal block decompositions. Then we will go on to calculate the anomalous dimensions of the first few of these operators using Feynman diagrams and verify that they vanish.  
\subsection{$\phi^4$ theory in $d=2 - \epsilon$}
\subsubsection{Decomposition of boundary four-point function}
Let us compute the four-point function of the leading boundary operator $\phi^I$ in the quartic theory of subsection \ref{$s=1$Quartic}. In the free theory, the four-point function just comes from the Wick contractions 
\begin{equation}
\langle \phi^I(\textbf{x}_1)\phi^J(\textbf{x}_2) \phi^K(\textbf{x}_3) \phi^L(\textbf{x}_4)\rangle_0 =  \hat{C}_{\phi \phi}^2 \bigg(\frac{\delta^{I J} \delta^{K L}}{(\textbf{x}_{12}^2)^{\hat{\Delta}} (\textbf{x}_{34}^2)^{\hat{\Delta}} }  + \frac{\delta^{I K} \delta^{J L}}{(\textbf{x}_{13}^2)^{\hat{\Delta}} (\textbf{x}_{24}^2)^{\hat{\Delta}} } + \frac{\delta^{I L} \delta^{J K}}{(\textbf{x}_{14}^2)^{\hat{\Delta}} (\textbf{x}_{23}^2)^{\hat{\Delta}} }   \bigg).
\end{equation}
In the s-channel, $12 \rightarrow 34$, the leading term just comes from the identity operator, while the other two come from the double trace operators of dimensions $2 \hat{\Delta} + 2 n + l$ \cite{Fitzpatrick:2011dm}
\begin{equation}
\frac{1}{(\textbf{x}_{13}^2)^{\hat{\Delta}} (\textbf{x}_{24}^2)^{\hat{\Delta}} } = \frac{(-1)^l}{(\textbf{x}_{14}^2)^{\hat{\Delta}} (\textbf{x}_{23}^2)^{\hat{\Delta}}} = \frac{1}{(\textbf{x}_{12}^2)^{\hat{\Delta}} (\textbf{x}_{34}^2)^{\hat{\Delta}} } \sum_{l,n} a_{\tau = 2 \hat{\Delta} + 2 n,l} u^{\hat{\Delta} + n} g_{\tau = 2 \hat{\Delta} + 2 n , l } (u,v)
\end{equation} 
where 
\begin{equation}
 a_{\tau = 2 \hat{\Delta} + 2 n,l} = \frac{ (-1)^l[(\hat{\Delta} - \frac{d}{2} + 1)_n (\hat{\Delta})_{l + n}]^2}{l! n! (l + \frac{d}{2})_n (2 \hat{\Delta} + n - d + 1)_n (2 \hat{\Delta} + 2 n + l -1)_l (2 \hat{\Delta} + n + l - \frac{d}{2})_n}
\end{equation}
and 
\begin{equation}
u = \frac{\textbf{x}_{12}^2 \textbf{x}_{34}^2}{\textbf{x}_{13}^2 \textbf{x}_{24}^2}, \ \ \ v = \frac{\textbf{x}_{14}^2 \textbf{x}_{23}^2}{\textbf{x}_{13}^2 \textbf{x}_{24}^2}.
\end{equation}
In our case $\hat{\Delta} = \frac{d - 1}{2}$ and $g_{\tau, l} (u,v)$ is the $d$ dimensional conformal block for four-point function. At first order in the coupling, we have the following connected contribution to the four-point function
\begin{equation}
\langle \phi^I(\textbf{x}_1) \phi^J(\textbf{x}_2) \phi^K(\textbf{x}_3) \phi^L(\textbf{x}_4)\rangle_1 = -2 g (\delta^{IJ} \delta^{KL} + \delta^{IK} \delta^{JL} + \delta^{IL} \delta^{JK}) \int d^d \textbf{x}_0  \frac{\hat{C}_{\phi \phi}^4 }{(\textbf{x}_{10}^2)^{\hat{\Delta}} (\textbf{x}_{20}^2)^{\hat{\Delta}}(\textbf{x}_{30}^2)^{\hat{\Delta}} (\textbf{x}_{40}^2)^{\hat{\Delta}}}. \\ 
\end{equation}
To make life simpler, we are going to evaluate this integral in $d = 2$ so that $\hat{\Delta}= 1/2$. In that case, the integral can be computed in terms of the $\bar{D}$ function  
\begin{equation}
\langle \phi^I(\textbf{x}_1) \phi^J(\textbf{x}_2) \phi^K(\textbf{x}_3) \phi^L(\textbf{x}_4)\rangle_1 = -\frac{2g}{\pi} (\delta^{IJ} \delta^{KL} + \delta^{IK} \delta^{JL} + \delta^{IL} \delta^{JK}) \frac{\hat{C}_{\phi \phi}^4}{(\textbf{x}_{12}^2 \textbf{x}_{34}^2)^{\frac{1}{2}}} u^{\frac{1}{2}} \bar{D}_{\frac{1}{2}, \frac{1}{2},\frac{1}{2},\frac{1}{2}} (u,v) 
\end{equation} 
This particular $\bar{D}$ function can be expressed in terms of the $H$ function, which can then be expanded in a power series in $u$ and $1-v$ \cite{Dolan:2000ut,Dolan:2000uw, Giombi:2018vtc}
\begin{equation} 
\begin{split}
\bar{D}_{\frac{1}{2}, \frac{1}{2},\frac{1}{2},\frac{1}{2}} (u,v) &=   -\pi^2 \log u \ G (\frac{1}{2}, \frac{1}{2},1,1;u,1-v ) + \sum \limits_{m,n = 0}^{\infty} \frac{\Gamma(\frac{1}{2} + m)^2 \Gamma(\frac{1}{2} + m + n)^2}{ (m!)^2 \  n! \ \Gamma(1 + 2 m + n) } f_{mn} u^m (1 - v)^n,    \\
f_{m n} &= 2 \psi(1 + m) + 2 \psi(1 + 2 m + n)  - 2\psi(\frac{1}{2} + m) - 2 \psi(\frac{1}{2} + m + n) . 
\end{split}
\end{equation}
The $G$ function appearing above can also be expanded in to powers 
\begin{equation}
G(\alpha, \beta, \gamma, \delta; u, 1-v) = \sum_{m,n = 0}^{\infty} \frac{(\delta - \alpha)_m (\delta - \beta)_m}{m! (\gamma)_m} \frac{(\alpha)_{m + n} (\beta)_{m + n} }{n! (\delta)_{2 m + n}} u^m (1 - v)^n
\end{equation}
and in particular,
\begin{equation}
\begin{split}
G (\frac{1}{2}, \frac{1}{2},1,1;u,1-v ) &= \sum \limits_{m,n = 0}^{\infty} \frac{\Gamma(\frac{1}{2} + m)^2 \Gamma(\frac{1}{2} + m + n)^2}{\pi^2 (m!)^2 \  n! \ \Gamma(1 + 2 m + n) } u^m (1-v)^n \\
& = \sum \limits_{m = 0}^{\infty} \frac{\Gamma(\frac{1}{2} + m)^4}{\pi^2 (m!)^2 \  \Gamma(1 + 2 m )} u^m \ {}_2F_1 (\frac{1}{2} + m, \frac{1}{2} + m, 1 + 2 m, 1- v).
\end{split}
\end{equation}
The $\log u$ term appearing above in the four-point function directly gives the anomalous dimensions as we now discuss. On general grounds, we can decompose the four point function as follows
\begin{equation} \label{FourPointDecomSTA}
\begin{split}
\langle \phi^I(\textbf{x}_1) \phi^J(\textbf{x}_2) \phi^K(\textbf{x}_3) \phi^L(\textbf{x}_4)\rangle &= \delta^{I J } \delta^{K L} \mathcal{G}_S +   \bigg( \frac{\delta^{I K } \delta^{J L} + \delta^{I L } \delta^{J K} }{2} - \frac{\delta^{I J } \delta^{K L}}{N}  \bigg) \mathcal{G}_T  \\
&  +   \frac{\delta^{I K } \delta^{J L} - \delta^{I L } \delta^{J K} }{2}  \mathcal{G}_A 
\end{split}
\end{equation}
where $S, T, A$ refer to singlet, traceless symmetric and anti-symmetric representations of $O(N)$. For each of these representations, we can have a decomposition into conformal blocks 
\begin{equation}
\mathcal{G} = \frac{\hat{C}_{\phi \phi}^2}{(\textbf{x}_{12}^2 \textbf{x}_{34}^2)^{\frac{1}{2}}} \mathcal{F} (u,v), \ \ \ \mathcal{F} (u,v) = \sum_{\tau , l} a_{\tau,l} u^{\frac{\tau}{2}} g_{\tau,l} (u,v).
\end{equation} 
From our discussion above, we have 
\begin{equation}
\begin{split}
\mathcal{F}_S (u,v) &= 1 + \sum \limits_{\substack{l, n \\ l: \mathrm{even} }} a_{S \ n , l}^0 u^{\frac{1}{2} + n} g_{\tau_n^0, l} -  \frac{2 g (N + 2) \hat{C}_{\phi \phi}^2 }{\pi N} u^{\frac{1}{2}} \bar{D}_{\frac{1}{2}, \frac{1}{2},\frac{1}{2},\frac{1}{2}} (u,v)\\
\mathcal{F}_T (u,v) &= \sum \limits_{\substack{l, n \\ l: \mathrm{even} }} a_{T \ n , l}^0 u^{\frac{1}{2} + n} g_{\tau_n^0, l} -  \frac{4 g  \hat{C}_{\phi \phi}^2 }{\pi} u^{\frac{1}{2}} \bar{D}_{\frac{1}{2}, \frac{1}{2},\frac{1}{2},\frac{1}{2}} (u,v) \\
\mathcal{F}_A (u,v) &= \sum \limits_{\substack{l, n \\ l: \mathrm{odd} }} a_{A \ n , l}^0 u^{\frac{1}{2} + n} g_{\tau_n^0, l}
\end{split} 
\end{equation}
where $\tau_n^0 = 1 + 2 n$ and 
\begin{equation}
a_{S \ n , l}^0 = \frac{1}{N} a_{T \ n , l}^0 = \frac{1}{N} a_{A \ n , l}^0 =  \frac{2}{N} \frac{ (-1)^l[( \frac{1}{2})_n (\frac{1}{2})_{l + n}]^2}{l! n! (l + 1)_n ( n )_n (2 n + l)_l ( n + l)_n}.
\end{equation}
Leading corrections to $\mathcal{F}$ can also be expressed in terms of anomalous dimensions and corrections to OPE coefficients: using $\tau_{n,l} = \tau_{n}^0 +  \hat{\gamma}_{n,l}$ and $a_{n,l} = a_{n,l}^0 + \delta a_{n,l}$ we have 
\begin{equation}
\delta \mathcal{F} (u,v) = u^{\frac{1}{2}}\sum_{n = 0}^{\infty} u^n \sum_{l : \mathrm{even}} \bigg( \frac{1}{2} a^0_{n,l}  \hat{\gamma}_{n,l} \log u + \delta a_{n,l} + \frac{1}{2} a_{n,l}^0 \hat {\gamma}_{n,l} \partial_n  \bigg) g_{\tau_n^0,l} (u,v).
\end{equation}
It is clear that the operators in the anti-symmetric representation do not get anomalous dimension or corrections to OPE coefficient to leading order in $g$. For the singlet representation, comparing the terms proportional to $\log u$, we have the following equation which implicitly determines the anomalous dimensions 
\begin{equation} \label{ImplicitAnomalous}
\begin{split}
\sum \limits_{\substack{l, n = 0 \\ l: \mathrm{even} }}^{\infty} u^n  \frac{1}{2} a^0_{S \ n,l}  \hat{\gamma}^S_{n,l} g_{\tau_n^0,l} (u,v) &= \frac{2 \pi g (N + 2) \hat{C}_{\phi \phi}^2}{N} \sum \limits_{m = 0}^{\infty} \frac{\Gamma(\frac{1}{2} + m)^4}{\pi^2 (m!)^2 \  \Gamma(1 + 2 m )} u^m  \\ 
&\times {}_2F_1 (\frac{1}{2} + m, \frac{1}{2} + m, 1 + 2 m, 1- v).
\end{split}
\end{equation}
A similar equation can be obtained for symmetric traceless case. For small values of $u$, in two dimensions and for even spins, the conformal block on the LHS has the following expansion \cite{Dolan:2000ut} to leading order in $u$
\begin{equation}
g_{\tau_n^0,l} (u,v) = (1 - v)^l {}_2 F_1 (\frac{1}{2} + n + l, \frac{1}{2} + n + l, 1 + 2 n + 2 l, 1- v ) + O(u). 
\end{equation}
Also, for $l = 0$, we have the following expansion to all orders in $u$
\begin{equation}
g_{\tau_n^0,l = 0} =  \sum _{m = 0}^{\infty} u^m \frac{\Gamma(\frac{1}{2} + m + n)^4 \Gamma(1 + 2 n)^2}{\Gamma(\frac{1}{2} + n)^4 \ m! (m + 2 n )! (2 m + 2 n)! } \  {}_2 F_1 (\frac{1}{2} + m + n, \frac{1}{2} + m + n, 1 + 2 m + 2 n, 1 - v ).
\end{equation}
We can use these expansions to compare coefficients of different powers of $u$ in eq. \eqref{ImplicitAnomalous}. At zeroth order in u, this implies
\begin{equation}
 \sum_{l : \mathrm{even}} \frac{1}{2} a^0_{S \ 0,l} \hat{ \gamma}^S_{0,l} x^l F_{\frac{1}{2} + l}(x)  =  \frac{2 g \pi (N + 2) \hat{C}_{\phi \phi}^2}{N} F_{\frac{1}{2}} (x)
\end{equation}
where $F_{\beta}(x)$ is defined by
\begin{equation}
F_{\beta} (x) \equiv {}_2F_1 (\beta, \beta, 2 \beta, x), \ \ \ x \equiv 1 - v
\end{equation}
and it obeys an orthogonality relation 
\begin{equation}
\frac{1}{2 \pi i} \oint_{x = 0} x^{\beta - \beta' - 1} F_{\beta} (x) F_{1 - \beta'} (x) = \delta_{\beta, \beta'}.
\end{equation}
Using this and $\hat{C}_{\phi \phi} = 1/2 \pi$, we get
\begin{equation}
 \hat{\gamma}^S_{0,l} = \delta_{0l} \frac{g (N + 2)}{2 \pi}.
\end{equation}
For $ l = 0 $, it agrees with the anomalous dimension of the boundary operator $\phi^2$ found in eq. \ref{DimensionPhi^2Quartic}. It vanishes for all other spins, which is perhaps not so surprising given that in the usual $O(N)$ model, the anomalous dimensions of leading twist bilinear operators (weakly broken higher spin currents) start at $O(\epsilon^2)$ in $4 - \epsilon$ dimensions. Similarly for the symmetric traceless case
\begin{equation}
\hat{\gamma}^T_{0,l} = \delta_{0l} \frac{g }{\pi}.
\end{equation}
At next order in $u$, equation \eqref{ImplicitAnomalous} implies
\begin{equation}
\sum_{l : \mathrm{even}} \frac{1}{2} a^0_{S \ 1,l} \hat{\gamma}^S_{1,l} x^l F_{\frac{3}{2} + l}(x) + \frac{1}{64} a^0_{S \ 0,0} \hat{\gamma}^S_{0,0} F_{\frac{3}{2}} (x)  =  \frac{2 g \pi (N + 2) \hat{C}_{\phi \phi}^2}{ 32 N} F_{\frac{3}{2}} (x)
\end{equation}
which just gives
\begin{equation}
\sum_{l : \mathrm{even}} \frac{1}{2} a^0_{S \ 1,l} \hat{ \gamma} ^S_{1,l} x^l F_{\frac{3}{2} + l}(x) = 0
\end{equation}
which implies
\begin{equation}
\hat{ \gamma}^S_{1,l} = 0
\end{equation}
for all values of $l$. For $l = 0$, this is just the displacement operator. We could use this result to go to next subleading twist and so on, since we know the conformal block for $l = 0$ to all orders in $u$. In general, it follows that if the anomalous dimensions of operators with all spins vanish from level $1$ through level $n - 1$, then at level n, we have the following equation 
\begin{equation}
\begin{split}
&\sum_{l : \mathrm{even}} \frac{1}{2} a^0_{S \ n,l} \hat{ \gamma} ^S_{n,l} x^l F_{n + \frac{1}{2} + l}(x) + \frac{1}{2} a^0_{S \ 0,0} \hat {\gamma}^S_{0,0} \frac{\Gamma(\frac{1}{2} + n)^4 }{\pi^2 (\ n!)^2  (2 n)! }  F_{n + \frac{1}{2}} (x)  \\
&=  \frac{2 g \pi (N + 2) \hat{C}_{\phi \phi}^2}{ N} \frac{\Gamma(\frac{1}{2} + n)^4 }{\pi^2 (\ n!)^2  (2 n)! }  F_{n + \frac{1}{2}} (x)
\end{split}
\end{equation} 
which gives 
\begin{equation}
\hat{ \gamma}^S_{n,l} = 0.
\end{equation}
In this way we can extend this result to all values of twist. Note that it was important that the leading twist anomalous dimensions vanish for all spins other than $l = 0$. These subleading twist operators with free dimension $d-1 + 2 n + l, n \ge 1$ and spin $l$ are exactly the operators we called higher-spin ``cousins" of displacement and we have just shown that their anomalous dimension vanishes to leading order in $g$. Similar reasoning goes through for the symmetric traceless case.

\subsubsection{Decomposition of bulk two point function}
Let us now discuss the conformal block decomposition of the bulk two-point function of the $\phi^I \phi^I$ operator.
In the case of free theory, using the cross-ratio $z$ defined in section \ref{GeneralRemarks}, we can write
\begin{equation}
\begin{split}
\langle \phi^I \phi^I (\textbf{x}_1,y_1) \phi^J \phi^J (\textbf{x}_2, y_2) \rangle_0 &= N^2 (G^0(0,0))^2 + 2 N (G^0_{\phi} (x_1,x_2))^2  \\
& = \frac{ N \Gamma(\frac{d -1}{2})^2} {16 \pi^{d + 1} (4 y_1 y_2)^{d - 1}} \bigg[ N + 2 \  \bigg(\frac{z}{1 - z}\bigg)^{d -1} + 2 \ z^{d - 1} + 4 \ \frac{z^{d -1}}{(1 - z)^{\frac{d -1}{2}}}  \bigg] \\
& = \frac{ N \Gamma(\frac{d -1}{2})^2} {16 \pi^{d + 1} (4 y_1 y_2)^{d - 1}} \cG(z)  .
\end{split}
\end{equation} 
We can determine the coefficients of the blocks using Euclidean inversion formulae \cite{Mazac:2018biw, Hogervorst:2017kbj}. On the boundary, we can define the coefficient function
\begin{equation}
\hat{I}_{\hat{\Delta}} = \frac{1}{\Gamma(\frac{d + 1}{2})} \int_0^1 d z \ z^{- (d + 1)} \ (1 - z)^{\frac{d -1}{2}} \ {}_2 F_1 \bigg( \hat{\Delta}, d - \hat{\Delta}; \frac{d + 1}{2}; \frac{z - 1}{z} \bigg) \cG (z) 
\end{equation}
and its residues are related to the coefficients of conformal block expansion as
\begin{equation}
\hat{I}_{\hat{\Delta}} \ \frac{\Gamma(\hat{\Delta}) \Gamma( \hat{\Delta} + \frac{1 - d}{2} ) }{2 \Gamma(2 \hat{\Delta} - d) } \sim \ - \frac{\mu_{\hat{O}}^2}{\hat{\Delta} - \hat{\Delta}_{\hat{O}} }. 
\end{equation}
Doing this procedure tells us that we have the identity block on the boundary, with coefficient $\mu_0^2 = N$, and a tower of blocks with dimensions $d -1 + 2 n$ and coefficients 
\begin{equation}
\begin{split}
\mu_{d - 1 + 2 n}^2 &= \frac{2}{{\Gamma (2 n+1)}} \bigg[ 2 \ \delta_{n,0} \ + \ \frac{\Gamma \left(\frac{3 - d}{2}\right) \, {}_2F_1(1-2 n,-2 n;-d-4 n+3;1)}{\Gamma \left(\frac{3 - d}{2} - 2 n \right)} +  \Gamma \left(\frac{d+1}{2}\right) 
\\ &\times  \Gamma (-d-4 n+3) \, {}_3\tilde{F}_2\left(\frac{3 - d}{2} -2 n ,-2 n,1-2 n;-d-4 n+3,\frac{d + 1}{2} -2 n ;1\right)\bigg]
\end{split}
\end{equation}
where regularized Hypergeometric function is defined by 
\begin{equation}
{}_3 \tilde{F}_2(a1,a2,a2;b1,b2;z) = \frac{{}_3 {F}_2(a1,a2,a2;b1,b2;z)}{\Gamma(b1) \Gamma(b2)}.
\end{equation} 
Similarly in the bulk, we have the coefficient function 
\begin{equation}
I_{\Delta} =  \int_0^1 d y \ y^{\frac{d - 5}{2}} (1 - y)^{-d + 1} {}_2 F_1 \bigg( \frac{\Delta}{2}, \frac{d + 1 - \Delta}{2}, 1, 1 - \frac{1}{y} \bigg) \cG (1 - y)
\end{equation}
and then the bulk data is determined using
\begin{equation}
I_{\Delta} \ \frac{\Gamma(\frac{\Delta}{2}) \Gamma(\frac{\Delta + 1 -d}{2})}{2 \Gamma (\Delta - \frac{d + 1}{2})} \sim \ -\frac{\lambda_O}{\Delta - \Delta_O}. 
\end{equation}
Using this, it can be seen that in the bulk channel, the two-point function contains identity, $\phi^2$ ( with dimension $d -1$), and a tower of primaries $\phi^2 \partial^{2 n} \phi^2$ with dimensions $2 d - 2 + 2 n$ with following OPE coefficients
\begin{equation*}
\lambda_0 = 2, \  \ \  \lambda_{d -1} = 4,
\end{equation*}
\begin{equation}
\begin{split}
\lambda_{2 d - 2 + 2 n} &= (-1)^n N \Gamma (1-d)
\bigg(\frac{\pi  (-2 d^2-3 d n+4 d-2 n^2+5 n-2 ) 
\sec (\frac{3 \pi  d}{2}) \Gamma (1-d)}{ 2 \ \Gamma 
(-d-n+2)^2 \ \Gamma (n+1) \Gamma (-\frac{3 d}{2}-n+\frac{7}
{2}) \Gamma (\frac{3 d}{2}+2 n-\frac{5}{2})} \\
& + \frac{{}_2F_1 (1-n,\frac{-d-2 n+3}{2};\frac{-3 d-4 n+7}
{2} ;1) }{ \Gamma (-d-n
+2) \Gamma (n)} \bigg) \\
&-\frac{ (2 \pi  (-1)^n \sec 
(\frac{3 \pi  d}{2}) \, _3\tilde{F}_2(-d-n+2,
\frac{-d-2 n + 3}{2},-n;\frac{- 3 d - 4 n + 7}{2},1-n;
1)}{\Gamma (\frac{1}{2} (3 d-5)+2 n) \
\Gamma (n+1)  }.
\end{split}
\end{equation}
When we add boundary interactions to the theory, the dimensions of the operators in the bulk channel will remain the same since the theory is free in the bulk, but the OPE coefficients $\lambda_O$ can receive corrections which will depend on the interaction strength. 

Note that the operators appearing in the boundary channel are scalars with dimensions $d - 1 + 2 n$. We will now show by an explicit perturbative calculation in the interacting theory, that for $n \ge 1$, they don't acquire anomalous dimensions, which is consistent with the fact that they are induced by bulk conserved higher spin currents. At leading order, we have   
\begin{equation}
\begin{split}
\langle \phi^I \phi^I (\textbf{x}_1,y_1) \phi^J \phi^J (\textbf{x}_2, y_2) \rangle_1 &= - 2 g N (N + 2) \int d^d \textbf{x}_0  (G^0_{\phi} (\textbf{x}_1,y_1;\textbf{x}_0,0))^2(G^0_{\phi} (\textbf{x}_0,0;\textbf{x}_2,y_2))^2.
\end{split}
\end{equation} 
This requires computing the following integral, which can be done, for example, by using Feynman parameters
\begin{equation}
\int d^d \textbf{x}_0 \frac{1}{(\textbf{x}_{1 0}^2 + y_1^2)^{d - 1} (\textbf{x}_{2 0}^2 + y_2^2)^{d - 1} } = \frac{\pi z}{2  y_1 y_2 \sqrt{1 - z}} \tanh^{-1} \bigg( \frac{2 \sqrt{1 - z} }{ 2 - z} \bigg)   
\end{equation}
where we already set $d = 2$ for the integral since we are computing the leading correction in $d=2-\epsilon$. This gives the two point function as 
\begin{equation}
\begin{split}
\langle \phi^I \phi^I (\textbf{x}_1,y_1) \phi^J \phi^J (\textbf{x}_2, y_2) \rangle &= \frac{ N \Gamma(\frac{d -1}{2})^2} {16 \pi^{d + 1} (4 y_1 y_2)^{d - 1}} \bigg[ N + 2  \  \bigg(\frac{z}{1 - z}\bigg)^{d -1} + 2 \ z^{d - 1} + 4 \ \frac{z^{d -1}}{(1 - z)^{\frac{d -1}{2}}}  \bigg] \\
& - \frac{ g N (N + 2) z}{ 16 \pi^3 y_1 y_2 \sqrt{1 - z}} \tanh^{-1} \bigg( \frac{2 \sqrt{1 - z} }{ 2 - z} \bigg). 
\end{split}
\end{equation} 
We can compute the anomalous dimensions of the operators appearing in boundary channel decomposition by extracting $\log z$ from our two point function. In the boundary channel, $\log z$ comes from the $z^{\hat{\Delta}}$ present in the boundary conformal block. So in the following, we only keep track of the $\epsilon \log z$ term of the leading order perturbation to the free propagator. Then using the decomposition from above, we have at the fixed point  
\begin{equation}
\begin{split}
\langle \phi^I \phi^I (\textbf{x}_1,y_1) \phi^J \phi^J (\textbf{x}_2, y_2) \rangle &\ni \frac{ N \Gamma(\frac{d -1}{2})^2} {16 \pi^{d + 1} (4 y_1 y_2)^{d - 1}} \bigg[ N + \sum_{n = 0}^{\infty} \mu^2_{d - 1 + 2 n} f_{\textrm{bdry}} (d - 1 + 2n; z) \bigg] \\
&+ \frac{N}{ 64 \pi^2 y_1 y_2} \bigg(8 \log z\frac{N + 2}{N + 8} \epsilon \bigg)
\end{split}
\end{equation}
where there will be other order $\epsilon$ terms which will contribute to the corrections to OPE coefficients, but we have only kept $\log z$ terms. Noting that the boundary block for $\hat{\Delta} = d -1$ simplifies, this again precisely gives the value of anomalous dimension of the boundary operator $\phi^2$ found in \eqref{DimensionPhi^2Quartic} and tells us that none of the other operators get anomalous dimensions. This is consistent since the operators with $n \ge 1$ correspond to higher spin displacements with boundary spin $0$ and are equal to the boundary value of conserved currents with all $2 n$ indices being $y$, $J^{y y... y}$. 
\subsubsection{Direct Computation}
It is possible to compute these anomalous dimensions more directly as well, by explicitly writing down the operator induced by conserved currents on the boundary and computing their anomalous dimensions. For the displacement, the operator is 
\begin{equation}
\begin{split}
D = T_{y y} &= \frac{d -1}{4 d} ( \phi^I (\partial_i^2 \phi^I ) + \phi^I (\partial_i^2 \phi^I )) - \frac{1}{2 d} \partial_i \phi^I  \partial^i \phi^I + \frac{1}{2} \partial_y \phi^I \partial_y \phi^I \\
&= \frac{d -1}{4 d} ( \phi^I (\partial_i^2 \phi^I ) + \phi^I (\partial_i^2 \phi^I )) - \frac{1}{2 d} \partial_i \phi^I  \partial^i \phi^I + \frac{g^2}{2} ((\phi^K \phi^K)) \phi^I (\phi^L \phi^L) \phi^I
\end{split}
\end{equation}  
where we used modified Neumann boundary condition $\partial_y \phi^I =  g (\phi^J \phi^J) \phi^I$. We will calculate its anomalous dimension to order $g^2$. To this order, the last term in the above expression will not contribute and it will start contributing at order $g^3$. This is actually a primary operator in the boundary theory as it matches up to a coefficient to a ``double trace"\footnote{The operators we discuss here are bilinears in the fundamental fields $\phi^I$ and hence should be thought of as single trace operators. However, we will sometimes loosely use the terminology ``double trace" to make contact with some of the literature on the subject.} operator. We will denote by $O^{IJ}_{n,l}$, the operator with dimensions $2 \hat{\Delta} + 2 n + l$ and spin $l$. For $n = 1$ and $l = 0$, the ``double trace" primary operator takes the form \cite{Bekaert:2015tva} 
\begin{equation}
O^{IJ}_{1,0} = \frac{d - 1}{2} ((\partial_i^2 \phi^I) \phi^J + \phi^I (\partial_i^2 \phi^J)) - \partial_{i} \phi^I \partial^{i} \phi^J. 
\end{equation}
We want to show that the anomalous dimension of this operator vanishes by computing its three point function with two other $\phi$. To two loop order, following are the non trivial diagrams that will contribute, and we want to show that these do not have any logarithmic divergence.

\begin{equation*}
G^{1,2} = \includegraphics[scale=1,valign = m]{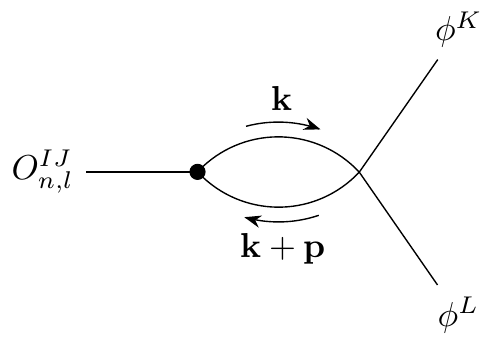} \ +\  \ \includegraphics[scale=1,valign = m]{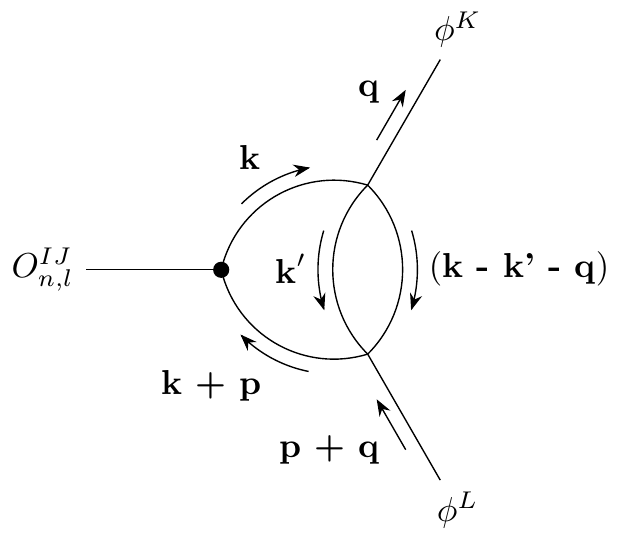}
\end{equation*}
\begin{equation}
\begin{split}
&= 2 g (\delta^{I J} \delta^{K L}  + \delta^{I K} \delta^{J L} + \delta^{I L} \delta^{J K }) \int \frac{d^d \textbf{k}}{(2 \pi)^d} \frac{1}{|\textbf{k}| |\textbf{k + p}|} \tilde{O}_{1,0} (\textbf{k}, \textbf{p}) \\
&+ g^2 (  8 \delta^{I J} \delta^{K L}  + 2 (N + 6)( \delta^{I K} \delta^{J L} + \delta^{I L} \delta^{J K }) \mathcal{I}_1
\end{split} 
\end{equation}
where 
\begin{equation}
\tilde{O}_{1,0} (\textbf{k}, \textbf{p}) = \frac{d }{2} (\textbf{k}^2 + (\textbf{k + p})^2 ) -\frac{\textbf{p}^2}{2}.  
\end{equation}
It is easy to see that the first one loop diagram vanishes identically, which is why we do not need to consider other two loop diagrams which contain this diagram as a subdiagram. Now for the second two loop diagram, we have to perform the integral
\begin{equation}
\begin{split}
\mathcal{I}_1 &= \int \frac{d^d \textbf{k}}{(2 \pi)^d} \frac{d^d \textbf{k}'}{(2 \pi)^d} \frac{1}{|\textbf{k}| |\textbf{k + p}||\textbf{k}'| |\textbf{k - k'- q}|} \tilde{O}_{1,0} (\textbf{k}, \textbf{p})   \\
&= \frac{\Gamma(\frac{d -1}{2})^2 \Gamma(1 - \frac{d}{2}) }{(4 \pi)^{\frac{d}{2}}\pi \Gamma(d -1)} \int \frac{d^d \textbf{k}}{(2 \pi)^d} \frac{1}{|\textbf{k}| |\textbf{k + p}| |\textbf{q - k}|^{2 - d}} \tilde{O}_{1,0} (\textbf{k}, \textbf{p}) \\
&=  -\frac{2^{1-d} \sec \left(\frac{\pi  d}{2}\right) \Gamma \left(2-\frac{d}{2}\right) \Gamma \left(d-\frac{3}{2}\right)  \Gamma(\frac{d -1}{2})^2 \Gamma(2 - \frac{d}{2}) }{ \Gamma \left(\frac{3}{2}-\frac{d}{2}\right) \Gamma \left(\frac{3 d}{2}-1\right) (4 \pi)^{d}\pi \Gamma(d -1) (p^2)^{1 - d}} 
\end{split}
\end{equation}
where we computed the integral at $\textbf{q} = 0$, since we are just using this diagram to calculate the anomalous dimension. This is finite in $ d = 2 - \epsilon$ which implies that to this order, the operator $O^{I J}_{1,0}$ does not get anomalous dimensions.

Let us now talk about the operators induced by the bulk spin $4$ current on the boundary. If the bulk is 3 dimensional (which will be sufficient for our perturbative calculation), it can be explicitly constructed using the generating function 
\begin{equation}
O^{IJ}(x,\epsilon) = \sum_{s = 0}^{\infty} J^{IJ}_{\mu_1 .... \mu_s} (x) \epsilon^{\mu_1} \cdots \epsilon^{\mu_s}.
\end{equation}
This generating function can be calculated by using the conditions of current conservation and tracelessness and it turns out to be \cite{Giombi:2009wh} 
\begin{equation}
O^{IJ}(x,\epsilon) = \phi^I(x - \epsilon) \sum_{n = 0}^{\infty} \frac{(2 \epsilon^2 \overset{\leftarrow}{\partial_x} \cdot \overset{\rightarrow}{\partial_x} - 4 (\epsilon \cdot \overset{\leftarrow} {\partial_x} )( \epsilon \cdot \overset{\rightarrow}{\partial}_x )  )^n}{(2 n)!} \phi^J (x+\epsilon).
\end{equation}
This can be expanded to fourth order in $\epsilon$, which gives the spin 4 current
\begin{equation}
\begin{split}
J^{I J}_{\mu \nu \rho \sigma} &=  \frac{1}{4!} \bigg[ \frac{1}{24} \partial_{(\mu} \partial_{\nu} \partial_{\rho} \partial_{\sigma)}\phi^I \phi^J  - \frac{7}{6} \partial_{(\mu} \partial_{\nu} \partial_{\rho} \phi^I  \partial_{\sigma)} \phi^J + \frac{1}{2} \delta_{(\mu \nu} \partial_{\rho}\partial_{\sigma)}\partial_{\alpha} \phi^I \partial^{\alpha} \phi^J + (I \leftrightarrow J) \\
&+ \frac{1}{6} \delta_{(\mu \nu} \delta_{\rho \sigma)} \partial_{\alpha} \partial_{\beta} \phi^I \partial^{\alpha} \partial^{\beta} \phi^J - \frac{5}{3} \delta_{(\mu \nu} \partial_{\alpha} \partial_{\rho} \phi^I \partial^{\alpha} \partial_{\sigma)} \phi^J + \frac{35}{12} \partial_{(\mu} \partial_{\nu} \phi^I \partial_{\rho} \partial_{\sigma)} \phi^J     \bigg]
\end{split}
\end{equation}
where the symmetrization sign means that we add all the terms related by exchange of indices. Now, we can take all its components to be transverse to the boundary and obtain an operator on the boundary, which with Neumann boundary condition looks like 
\begin{equation}
\begin{split}
J^{I J}_{y y y y } &=  \bigg[ \frac{1}{24} ((\partial_i^2)^2 \phi^I) \phi^J - \frac{1}{2} (\partial_i^2 \partial_{j} \phi^I) \partial^{j} \phi^J + (I \leftrightarrow J) \\
&+ \frac{1}{6}( \partial_{i} \partial_{j} \phi^I \partial^{i} \partial^{j} \phi^J ) + \frac{17}{12} (\partial_{i}^2 \phi^I)( \partial_{i}^2 \phi^J)    \bigg] + O(g^2).
\end{split}
\end{equation}
From the boundary point of view, this is an operator with dimensions $2 \Delta_{\phi} + 4$ and spin $0$. Using recursion relations from \cite{Bekaert:2015tva}, we can write down the form of a primary of the same dimension and spin in $d$ dimensions 
\begin{equation}
\begin{split}
O^{I J}_{2,0} &=  \bigg(\partial_i \partial_j \phi^I \partial^i \partial^j \phi^J + \frac{(d + 1)(d + 3) + 2}{2} (\partial_{i}^2 \phi^I)( \partial_{i}^2 \phi^J) \bigg) - (d + 1)( (\partial_i^2 \partial_{j} \phi^I) \partial^{j} \phi^J + (I \leftrightarrow J) ) \\
& + \frac{(d + 1) (d -1)}{12} (((\partial_i^2)^2 \phi^I) \phi^J + (I \leftrightarrow J) ).
\end{split}
\end{equation} 
The relative coefficients of various terms in this operator indeed match what we get from the operator that the spin 4 current defines on the boundary. So they are the same operator up to a constant. We can now try to compute its anomalous dimensions using the following correlation function, which involves the same set of diagrams as the displacement operator $O^{I J}_{1,0}$ case but with different factors of external momentum
\begin{equation}
\begin{split}
\langle O^{I J}_{2,0} (- \textbf{p}) \phi^K( - \textbf{q}) \phi^L(\textbf{p + q}) \rangle &= 2 g (\delta^{I J} \delta^{K L}  + \delta^{I K} \delta^{J L} + \delta^{I L} \delta^{J K }) \int \frac{d^d \textbf{k}}{(2 \pi)^d} \frac{1}{|\textbf{k}| |\textbf{k + p}|} \tilde O_{2 ,0 }(\textbf{k}, \textbf{p})\\
 &+ g^2 (  8 \delta^{I J} \delta^{K L}  + 2 (N + 6)( \delta^{I K} \delta^{J L} + \delta^{I L} \delta^{J K }) \mathcal{I}_2
\end{split}
\end{equation}
where
\begin{equation}
\begin{split}
\tilde{O}_{2,0}(\textbf{k}, \textbf{p}) &=    \frac{(d + 2) (d + 4)}{12} (|\textbf{k}|^4 + |\textbf{k + p}|^4 ) + \frac{(d + 2) (d + 4)}{2}|\textbf{k}|^2 |\textbf{k + p}|^2 \\
&- \frac{d + 2}{2} |\textbf{p}|^2 (|\textbf{k}|^2 + |\textbf{k + p}|^2 ) + \frac{1}{4} |\textbf{p}|^4 .
\end{split} 
\end{equation}
The one loop diagram again vanishes identically and the two loop diagram requires the following integral, which we again evaluate at $\textbf{q} = 0$
\begin{equation}
\begin{split}
\mathcal{I}_2 &= \int \frac{d^d \textbf{k}}{(2 \pi)^d} \frac{d^d \textbf{k}'}{(2 \pi)^d} \frac{1}{|\textbf{k}| |\textbf{k + p}|}  \frac{1}{|\textbf{k}'| |\textbf{k} - \textbf{k}'- \textbf{q}|} \tilde{O}_{2,0}(\textbf{k}, \textbf{p}) \\
&= \frac{\Gamma(\frac{d -1}{2})^2 \Gamma(1 - \frac{d}{2}) }{(4 \pi)^{\frac{d}{2}}\pi \Gamma(d -1)} \int \frac{d^d \textbf{k}}{(2 \pi)^d} \frac{1}{|\textbf{k}| |\textbf{k + p}| |\textbf{q - k}|^{2 - d}} \tilde{O}_{2,0} (\textbf{k}, \textbf{p}) \\
& = \frac{\Gamma(\frac{d -1}{2})^2 \Gamma(1 - \frac{d}{2})^2 2^{-d-4} \Gamma \left(d-\frac{3}{2}\right) }{(4 \pi)^{d}\pi \Gamma(d -1) (p^2)^{ - d}}    \bigg(\frac{(d^5 + 8 d^4 + 39 d^3 - 80 d^2 - 4 d + 48) \Gamma \left(\frac{d-1}{2}\right)}{\pi  \Gamma \left(\frac{3 d}{2}+1\right)}\\ 
&-\frac{16 (d+2) \sec \left(\frac{\pi  d}{2}\right)}{\Gamma \left(\frac{1}{2}-\frac{d}{2}\right) \Gamma \left(\frac{3 d}{2}-1\right)}\bigg)
\end{split}
\end{equation}
and this is finite in $ d = 2 - \epsilon$. This implies that to this order, the operator $O^{I J}_{2,0}$ does not get anomalous dimensions.

The next operator we consider is the spin 2 operator on the boundary induced by the spin 4 current in the bulk. It can be obtained by taking two of the components of the current to be in the normal direction and it gives 
\begin{equation}
\begin{split}
J^{IJ}_{y y i j} &= \frac{1}{12} \bigg[ -\frac{1}{2} \partial_k^2 \partial_i \partial_j \phi^I \phi^J + \frac{7}{2} \partial_k^2 \partial_{(i} \phi^I \partial_{j)} \phi^J + \partial_i \partial_j \partial_k \phi^I \partial^k \phi^J - \frac{35}{6} \partial_k^2 \phi^I \partial_i \partial_j \phi^J + (I \leftrightarrow J) \\
 & - \frac{5}{3} \partial_k \partial_{( i} \phi^I \partial^k \partial_{j)} \phi^J -\frac{8}{3} \delta_{i j} \partial_k^2 \phi^I \partial_k^2 \phi^J + \frac{2}{3} \delta_{i j} \partial_k \partial_l \phi^I \partial^k \partial^l \phi^J -\delta_{ij} (\partial_k^2 \partial_l \phi^I \partial^l \phi^J + I \leftrightarrow J) \bigg]
\end{split}
\end{equation}
where $\partial_{(i} \phi^I \partial_{j)} \phi^J = \partial_{i} \phi^I \partial_{j} \phi^J + \partial_{j} \phi^I \partial_{i} \phi^J$. This is symmetric in $i,j$ indices and we can project it onto a symmetric traceless part
\begin{equation}
\begin{split}
&J^{IJ}_{y y i j (T)} = \bigg( \delta_{ik} \delta_{jl} - \frac{\delta_{ij} \delta_{kl}}{d} \bigg) V_{kl} =
\frac{1}{12} \bigg[ - \frac{5}{3} \partial_k \partial_{( i} \phi^I \partial^k \partial_{j)} \phi^J  \\& + \bigg(-\frac{1}{2} \partial_k^2 \partial_i \partial_j \phi^I \phi^J + \frac{7}{2} \partial_k^2 \partial_{(i} \phi^I \partial_{j)} \phi^J + \partial_i \partial_j \partial_k \phi^I \partial^k \phi^J - \frac{35}{6} \partial_k^2 \phi^I \partial_i \partial_j \phi^J + (I \leftrightarrow J) \bigg) \\
& + \delta_{i j}\bigg( \bigg(\frac{1}{4}  \partial_k^2  \partial_l^2 \phi^I  \phi^J -4 \partial_k^2 \partial_l \phi^I \partial^l \phi^J + I \leftrightarrow J \bigg) + \frac{35 }{6} \partial_k^2 \phi^I \partial_l^2 \phi^J + \frac{5 }{3} \partial_k \partial_l \phi^I \partial^k \partial^l \phi^J  \bigg)\bigg].
\end{split}
\end{equation}
As is probably familiar by now, we can write the ``double trace" primary with spin $2$ and dimensions $2 \Delta_{\phi} + 4$ using results from \cite{Bekaert:2015tva}
\begin{equation}
\begin{split}
&O^{IJ}_{1,2 \ i j (T)} = \\
& \bigg[\frac{1- d}{2} \partial_k^2 \partial_i \partial_j \phi^I \phi^J + \frac{d + 5}{2} \partial_k^2 \partial_{(i} \phi^I \partial_{j)} \phi^J + \partial_i \partial_j \partial_k \phi^I \partial^k \phi^J - \frac{(3 + d) (5 + d)}{2 (1 + d)} \partial_k^2 \phi^I \partial_i \partial_j \phi^J + (I \leftrightarrow J) \bigg] \\& - \frac{d + 3}{d + 1} \partial_k \partial_{( i} \phi^I \partial^k \partial_{j)} \phi^J  + \delta_{i j}\bigg[ \bigg( \frac{d - 1}{2 d}  \partial_k^2  \partial_l^2 \phi^I  \phi^J -\frac{d + 6}{d} \partial_k^2 \partial_l \phi^I \partial^l \phi^J + I \leftrightarrow J \bigg) \\
&+ \frac{(d + 3) (d + 5)}{d (1 + d)} \partial_k^2 \phi^I \partial_l^2 \phi^J + \frac{2(d + 3) }{d (d + 1)} \partial_k \partial_l \phi^I \partial^k \partial^l \phi^J  \bigg]
\end{split}
\end{equation}
which matches, up to an overall constant, to the operator we need. Repeating the same procedure as other operators
\begin{equation}
\begin{split}
&\langle O^{I J}_{1,2 i j (T)} (- \textbf{p}) \phi^K( - \textbf{q}) \phi^L(\textbf{p + q}) \rangle = \\
& 2 g (\delta^{I J} \delta^{K L}  + \delta^{I K} \delta^{J L} + \delta^{I L} \delta^{J K }) \int \frac{d^d \textbf{k}}{(2 \pi)^d} \frac{1}{|\textbf{k}| |\textbf{k + p}|} \tilde{O}_{1, 2 , T} (\textbf{k}, \textbf{p}) \\
&+ g^2 (  8 \delta^{I J} \delta^{K L}  + 2 (N + 6)( \delta^{I K} \delta^{J L} + \delta^{I L} \delta^{J K }) \mathcal{I}_3
\end{split}  
\end{equation}
where
\begin{equation}
\begin{split}
\tilde{O}_{1, 2, T}( \textbf{k}, \textbf{p} ) &= \bigg[ k_i k_j \bigg( -\frac{2 (d + 4) (d + 2)}{d + 1} (|\textbf{k}|^2 + |\textbf{k + p}|^2) +  \frac{2 (d + 2)}{(d + 1)} |\textbf{p}|^2 \bigg) \\ 
& + k_{(i} p_{j)} \bigg( -\frac{(d + 6) (d + 2)}{d + 1} |\textbf{k}|^2 -\frac{(d + 2)^2}{d + 1} |\textbf{k + p}|^2 + \frac{ (d + 2)}{(d + 1)} |\textbf{p}|^2  \bigg) \\
&+ p_i p_j \bigg( -\frac{d}{2} |\textbf{k + p}|^2 - \frac{d^2 + 9 d + 16}{2 (d + 1)} |\textbf{k}|^2  + \frac{|\textbf{p}|^2}{2} \bigg) \\
&+ \delta_{i j} \bigg( \frac{(d + 2)^2}{d(d + 1)} (|\textbf{k}|^4 + |\textbf{k + p}|^4) + \frac{2 (d + 6)(d + 2)}{d (d + 1)} |\textbf{k}|^2|\textbf{k + p}|^2 \\
&- \frac{d^2 + 9 d + 16}{2 d (d + 1)} (|\textbf{k}|^2 + |\textbf{k + p}|^2) |\textbf{p}|^2  + \frac{d + 3}{2 d (1 + d)} |\textbf{p}|^4  \bigg) \bigg].
\end{split} 
\end{equation}
The one loop contribution vanishes, and we can use some integrals from the appendix \ref{Integrals} to evaluate the integral appearing in the two loop diagram
\begin{equation}
\begin{split}
\mathcal{I}_3 &= \int \frac{d^d \textbf{k}}{(2 \pi)^d} \frac{d^d \textbf{k}'}{(2 \pi)^d} \frac{1}{|\textbf{k}| |\textbf{k + p}|}  \frac{1}{|\textbf{k}'| |\textbf{k} - \textbf{k}'- \textbf{q}|} \tilde{O}_{1, 2, T}( \textbf{k}, \textbf{p} ) \\
&= \frac{\Gamma(\frac{d -1}{2})^2 \Gamma(1 - \frac{d}{2}) }{(4 \pi)^{d}\pi \Gamma(d -1) (p^2)^{-d}} \bigg[ \frac{p_i p_j}{ 2^{d+2} p^2  \Gamma (\frac{3 d}{2})} \bigg( \frac{ (5 d^3+26 d^2-40 d-16) \Gamma (1-\frac{d}{2}) \Gamma (d-\frac{3}{2}) \Gamma (\frac{d-1}{2})}{\pi } \\
&-\frac{8 (d+2) (d+4) \sec (\frac{\pi  d}{2}) \Gamma (-\frac{d}{2}) \Gamma (d+\frac{1}{2})}{3 \Gamma (\frac{3}{2}-\frac{d}{2})} \bigg) + \delta_{i j} \frac{(d-2) \sec (\frac{\pi  d}{2}) \Gamma (4-\frac{d}{2}) \Gamma (d-\frac{3}{2})}{ 2^{d} d \Gamma (\frac{3}{2}-\frac{d}{2}) \Gamma (\frac{3 d}{2}+1)} \bigg].
\end{split}
\end{equation}
As anticipated, this is finite in $ d = 2 - \epsilon$  which implies that to this order, the operator $O^{I J}_{1,2 ij (T)} $ also does not get anomalous dimensions.

\subsection{Large $N$ expansion}
We will now do the calculation of anomalous dimensions of the same operators in the large N model of subsection \ref{LargeN $s = 1$} using the Feynman diagrams. Starting with the displacement, we have the following contributions

\begin{equation*}
\begin{split}
\langle O^{I J}_{1,0}(0) \phi^K(q) \phi^L(-q) \rangle &= 
\includegraphics[scale=1,valign = m]{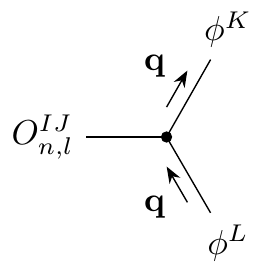} + \  \includegraphics[scale=1,valign = m]{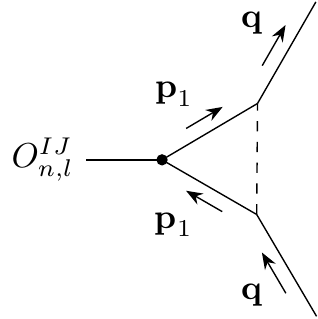} \   \\ \\
&+ \ \includegraphics[scale=1,valign = m]{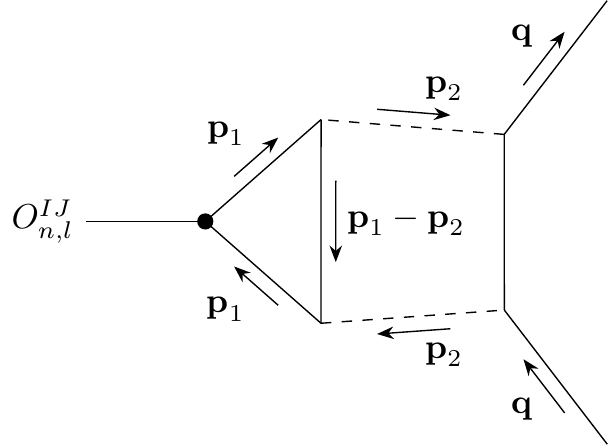}
\end{split}
\end{equation*}
\begin{equation}
\begin{split}
&= (\delta^{I  K} \delta^{J L} + \delta^{I  L} \delta^{J K} ) \bigg( - d q^2 + \frac{\tilde{C_{\sigma}}}{N} \int \frac{d^{d} p_1}{(2 \pi)^{d}} \frac{(- d p_1^2)}{|p_1|^2 |p_1 -q|^{d-2}} \bigg)  \\
& + \frac{ 2 \tilde{C}_{\sigma}^2 \delta^{I J} \delta^{K L}}{N^2} \int \frac{d^{d} p_1}{(2 \pi)^{d}} \int \frac{d^{d} p_2}{(2 \pi)^{d}} \frac{(-d p_1^2)}{ |p_1|^2 |p_1 - p_2| |p_2 - q| |p_2|^ {2 (d - 2) }} \\
& =  q^2 \bigg[(\delta^{I  K} \delta^{J L} + \delta^{I  L} \delta^{J K} ) \bigg( - d  + \frac{2  d   \ \tilde{C_{\sigma}}}{N (4 \pi)^{\frac{d}{2}}  d \Gamma(\frac{d}{2} - 1 )} \bigg)  + \frac{ \delta^{I J} \delta^{K L} \ d \tilde{C_{\sigma}}^2 (d - 3) \ \Gamma(\frac{d - 1}{2}) \Gamma( \frac{1 - d}{2}) }{ 2 N^2 (4 \pi)^{d} \sqrt{\pi} d   \Gamma( d- \frac{1}{2})} \bigg]. 
\end{split}
\end{equation}
There is no $ \log q$ term which tells us that there is no anomalous dimension. Both the $1/N$ corrections start at $O(\epsilon^2)$ in $d = 2 - \epsilon$ which is consistent with the fact that the $O(g)$ contribution to this correlator vanish in the $\epsilon$ expansion. Similar computation can be done for the two operators induced by the spin 4 current on the boundary. For the boundary scalar, we have 

\begin{equation}
\begin{split}
&\langle O^{I J}_{2,0}(0) \phi^K(q) \phi^L(-q) \rangle = \frac{2 (d + 2)(d + 4) }{3} \bigg[ (\delta^{I  K} \delta^{J L} + \delta^{I  L} \delta^{J K} ) \bigg( q^4 + \frac{\tilde{C_{\sigma}}}{N} \int \frac{d^{d} p_1}{(2 \pi)^{d}} \frac{ p_1^4}{|p_1|^2 |p_1 -q|^{d-2}} \bigg)  \\
& + \frac{ 2 \tilde{C}_{\sigma}^2 \delta^{I J} \delta^{K L}}{N^2} \int \frac{d^{d} p_1}{(2 \pi)^{d}} \int \frac{d^{d} p_2}{(2 \pi)^{d}} \frac{p_1^4}{ |p_1|^2 |p_1 - p_2| |p_2 - q| |p_2|^ {2 (d - 2) }} \bigg] \\
& =  \frac{2 (d + 2)(d + 4) q^4  }{3}\bigg[(\delta^{I  K} \delta^{J L} + \delta^{I  L} \delta^{J K} ) \bigg(  1  - \frac{ \ \tilde{C_{\sigma}}}{N (4 \pi)^{\frac{d}{2}}  (d + 2) \Gamma(\frac{d}{2} - 1 )} \bigg)  \\
&- \frac{ \delta^{I J} \delta^{K L} \ 3 \tilde{C_{\sigma}}^2  \ \Gamma(\frac{d - 1}{2})^2 \Gamma( \frac{-1 - d}{2}) }{ 4  N^2 (4 \pi)^{d} \sqrt{\pi} (d + 2)   \Gamma( d+ \frac{1}{2}) \Gamma( \frac{d -5}{2})} \bigg]. 
\end{split}
\end{equation}
This also does not have any $\log q$ terms indicating no anomalous dimensions. The corrections here also start at $O(\epsilon^2)$ in $d = 2 - \epsilon$. Finally, for the spin two operator, we have 
\begin{equation}
\begin{split}
& \langle O^{I J}_{1,2 i j (T)}(0) \phi^K(q) \phi^L(-q) \rangle = \frac{4 (d + 2)(d + 4) }{d + 1} \bigg[ (\delta^{I  K} \delta^{J L} + \delta^{I  L} \delta^{J K} ) \\
&\times \bigg( - q_i q_j q^2 + \frac{\delta_{i j} q^4}{d} + \frac{\tilde{C_{\sigma}}}{N} \int \frac{d^{d} p_1}{(2 \pi)^{d}} \frac{1}{|p_1|^2 |p_1 -q|^{d-2}} (- p_{1i} p_{1j} p_1^2 + \frac{\delta_{i j} p_1^4}{d}) \bigg)  \\
&+ \frac{ 2 \tilde{C}_{\sigma}^2 \delta^{I J} \delta^{K L}}{N^2} \int \frac{d^{d} p_1}{(2 \pi)^{d}} \int \frac{d^{d} p_2}{(2 \pi)^{d}} \frac{1}{ |p_1|^2 |p_1 - p_2| |p_2 - q| |p_2|^ {2 (d - 2) }} (- p_{1i} p_{1j} p_1^2 + \frac{\delta_{i j} p_1^4}{d}) \bigg] \\
& =  \frac{4 (d + 2)(d + 4)   }{d + 1} ( - q_i q_j q^2 + \frac{\delta_{i j} q^4}{d})\bigg[(\delta^{I  K} \delta^{J L} + \delta^{I  L} \delta^{J K} )  \\
 &\times \bigg( 1 - \frac{ 2 \tilde{C_{\sigma}}}{N (4 \pi)^{\frac{d}{2}}  (d + 4) \Gamma(\frac{d}{2} - 1 )} \bigg) - \frac{ 15 \ \delta^{I J} \delta^{K L}  \tilde{C_{\sigma}}^2  \ (d - 3) \sqrt{\pi} \sec (\frac{d \pi}{2}) }{ 4 \ N^2 (4 \pi)^{d}  (d + 4) (d - 1)   \Gamma( d+ \frac{3}{2})} \bigg]
\end{split}
\end{equation}
which also does not contain $\log q$ implying that there is no anomalous dimension.

\section{Long Range $O(N)$ Models} \label{LongRange}

It is natural to generalize the analysis of the previous sections to general non local models in $d$-dimensional Euclidean space, where the free propagator takes the form $1/|p|^s$ in momentum space, and the kinetic term in position space is
\begin{equation}
\frac{2 ^s \Gamma(\frac{d + s}{2})}{\pi^{\frac{d}{2}} \Gamma(-\frac{s}{2})} \int d^d x d^d y \frac{\phi^I (x) \phi^I (y)}{|x - y|^{d+s}}, \ \ \ \Delta_{\phi} = \frac{d - s}{2}. 
\end{equation}
For the applications discussed below, $d$ is some fixed dimension (which can be taken to be integer), and $s$ is a free parameter that controls the power of the long range propagator. 

\subsection{Quartic interaction}
First we consider the following model with a quartic interaction 
\begin{equation}
S = \frac{2 ^s \Gamma(\frac{d + s}{2})}{\pi^{\frac{d}{2}} \Gamma(-\frac{s}{2})} \int d^d x d^d y \frac{\phi^I (x) \phi^I (y)}{|x - y|^{d+s}} + \frac{g}{4} \int d^d x (\phi^I \phi^I)^2.
\end{equation}
This coupling becomes marginal when $s = d/2$, so we will study this model perturbatively in $s = \frac{d + \epsilon}{2}$ when g has dimensions equal to $\epsilon$. For $s = 1$ this is equivalent to the boundary model we studied in subsection \ref{$s=1$Quartic} and all the diagrams remain the same with modified propagators. So we will not give all the details here and just sketch out the main points. 

The computation of the four point function now requires the following integrals
\begin{equation}
\begin{split} 
G^4 &= 2 \delta^{IJ} \delta^{KL}  \bigg[ - (g + \delta_g) \ + \ (g + \delta_g)^2 (N + 8) \int \frac{d^{d} \textbf{k}}{(2 \pi)^{d}} \frac{1}{ |\textbf{k} + \textbf{p}|^s  | \textbf{k}|^s} -  g^3 (N^2 + 6 N + 20) \\& \times \bigg( \int \frac{d^{d} \textbf{k}}{(2 \pi)^{d}} \frac{1}{ |\textbf{k} + \textbf{p}|^s  | \textbf{k}|^s} \bigg)^2 - 4 g^3 (5 N + 22)  \int \frac{d^d \textbf{k}}{(2 \pi)^d} \frac{d^d \textbf{k}'}{(2 \pi)^d} \frac{1}{|\textbf{k}|^s |\textbf{k + p}|^s} \frac{1}{|\textbf{k}'|^s |\textbf{k - k'- q}|^s}    \bigg] \\
&= 2 \delta^{IJ} \delta^{KL} \bigg[ -(g + \delta_g) +  \frac{ (g + \delta_g)^2 (N + 8) \Gamma(\frac{d -s}{2})^2 \Gamma(s - \frac{d}{2})}{(4 \pi)^{\frac{d}{2}} \Gamma(\frac{s}{2})^2 \Gamma(d-s) (p^2)^{s - \frac{d}{2}}}  \\ &- \frac{ g^3 (N^2 + 6 N + 20) \Gamma(\frac{d -s}{2})^4 \Gamma(s - \frac{d}{2})^2}{(4 \pi)^{d} \ \Gamma(\frac{s}{2})^4 \Gamma(d-s)^2 (p^2)^{2 s - d}} - \frac{4 g^3 (5 N + 22) \Gamma(\frac{d -s}{2})^3 \Gamma(s - \frac{d}{2}) \Gamma(d - \frac{3 s}{2}) \Gamma(2 s - d)}{(4 \pi)^{d} \Gamma(\frac{s}{2})^3 \Gamma(d-s) \Gamma(\frac{3 s -d}{2}) \Gamma(\frac{3 d}{2} - 2 s) (p^2)^{2 s - d} }  \bigg].
\end{split}
\end{equation}
Requiring that the divergent terms cancel when $s = \frac{d + \epsilon}{2}$ fixes $\delta_g$ and then applying Callan-Symanzik equation on the finite piece gives the $\beta$ function
\begin{equation}
\beta(g) = -\epsilon g + \frac{2 g^2 (N + 8) }{(4 \pi)^{\frac{d}{2}} \Gamma(\frac{d}{2}) } + \frac{8 g^3 (5 N + 22)}{(4 \pi)^d \Gamma(\frac{d}{2})^2} (\gamma + 2 \psi(d/4) - \psi(d/2)).
\end{equation}
This gives the fixed point at
\begin{equation}
g = g_* = \frac{(4 \pi)^{\frac{d}{2}} \Gamma(\frac{d}{2})}{2 (N + 8)} \ \epsilon + \frac{(4 \pi)^{\frac{d}{2}} \Gamma(\frac{d}{2}) (5 N + 22) (-\gamma - 2 \psi(d/4) + \psi(d/2))  }{ (N + 8)^3} \epsilon^2 .
\end{equation}
The computation of anomalous dimensions of the operator $\phi^I \phi^I$ at this fixed point also closely follows the boundary case and the result is 
\begin{equation} \label{DimensionPhi2squartic}
\begin{split}
\gamma_{\phi^2} &= \frac{2 g_* (N + 2)}{(4 \pi )^{\frac{d}{2}} \Gamma(\frac{d}{2})} + \frac{12 (N + 2) g_*^2 (\gamma + 2 \psi(d/4) - \psi(d/2))}{(4 \pi )^{d} \Gamma(\frac{d}{2})^2} \\
& = \frac{(N + 2)}{(N + 8)} \epsilon - \frac{ (N + 2) ( 7 N + 20) (\gamma + 2 \psi(d/4) - \psi(d/2))}{(N + 8)^3} \epsilon^2 \\
\Delta_{\phi^2} &= d - s + \gamma_{\phi^2} = \frac{d}{2} + \frac{(N - 4) \epsilon}{2 (N + 8)} - \frac{ (N + 2) ( 7 N + 20) (\gamma + 2 \psi(d/4) - \psi(d/2))}{(N + 8)^3} \epsilon^2 .
\end{split}
\end{equation}
This agrees with what was found in \cite{PhysRevLett.29.917}.

\subsection{Large $N$ description}
Similar to subsection \ref{LargeN $s = 1$} we can develop a complementary approach to study the fixed point studied above in continuous and arbitrary s and d, but in an expansion in $1/N$. For that, we consider the following action with an auxiliary field $\sigma$
\begin{equation}
S = \frac{2 ^s \Gamma(\frac{d + s}{2})}{\pi^{\frac{d}{2}} \Gamma(-\frac{s}{2})} \int d^d x d^d y \frac{\phi^I (x) \phi^I (y)}{|x - y|^{d+s}} + \int d^{d} x \bigg( \frac{\sigma \phi^I \phi^I}{2} -\frac{\sigma^2}{4 g} \bigg).
\end{equation}
As usual, we will integrate out the $\phi$ field to get an effective quadratic action in terms of $\sigma$
\begin{equation}
S_2 = \int \frac{d^{d} \mathbf{p}}{(2 \pi)^{d}} \frac{\sigma(\mathbf{p}) \sigma(-\mathbf{p})}{2} \bigg( \frac{N}{\tilde{C_{\sigma}}} (p^2)^{\frac{d}{2}-s} - \frac{1}{2 g}  \bigg).
\end{equation}
where 
\begin{equation}
\tilde{C_{\sigma}} = -\frac{2 (4 \pi)^{\frac{d}{2}} \Gamma(\frac{s}{2})^2 \Gamma(d-s) }{\Gamma (s - \frac{d}{2}) \Gamma(\frac{d-s}{2})^2}.
\end{equation}
From here, it is clear that for $s > \frac{d}{2}$, the second term in the quadratic action can be dropped in the IR limit, while for $s < \frac{d}{2}$, it can be dropped in the UV limit.  This only leaves the induced kinetic term in the quadratic action and leads to the following two point function for $\sigma$
\begin{equation}
\langle \sigma(x_1) \sigma(x_2) \rangle = \frac{C_{\sigma}}{ N  |x_1 - x_2|^{2 s}}, \ \ \ C_{\sigma} = \tilde{C}_{\sigma}  \frac{2^{2 s} \ \Gamma(s)}{ (4 \pi)^{\frac{d}{2}} \Gamma(\frac{d}{2} - s)}
\end{equation}
which implies that the conformal dimension of sigma operator, to this order, is $s$. The computation of its anomalous dimension involves same diagrams and similar integrals as the boundary case and the result is
\begin{equation}
\Delta_{\sigma} = s + \frac{1}{N} \bigg(\frac{8  \Gamma(\frac{d}{2} - s) \Gamma(\frac{ 3 s - d}{2}) \Gamma(\frac{s}{2})^3 \Gamma(d - s)^2 }{\Gamma(s - \frac{d}{2})^2 \Gamma(\frac{d - s}{2})^3 \Gamma(s) \Gamma( d - \frac{3 s}{2}) \Gamma(\frac{d}{2}) } - \frac{4 \Gamma(\frac{s}{2})^2 \Gamma(d - s)}{\Gamma(s - \frac{d}{2}) \Gamma(\frac{d -s}{2})^2 \Gamma(\frac{d}{2})}   . \bigg) 
\end{equation}
This agrees with what was found in \cite{PhysRevLett.29.917, Gubser:2017vgc}. We can expand it in an $\epsilon$ expansion with $s = \frac{d + \epsilon}{2}$
\begin{equation}
\Delta_{\sigma} = s + \frac{1}{N}(-6 \epsilon - 7 (\gamma + 2 \psi(d/4) - \psi(d/2) ) \epsilon^2 + O(\epsilon^3)).
\end{equation}
It agrees with the $\epsilon$ expansion result above in eq. \ref{DimensionPhi2squartic} when expanded at large $N$. We can also expand when $ s= d - \epsilon$ which gives 
\begin{equation}
\Delta_{\sigma} = s + \frac{1}{N}\bigg(- \epsilon^2 \frac{(\gamma + \psi(\frac{-d}{2}) - \psi(\frac{d}{2}) + \psi(d) )}{2} + O(\epsilon^3) \bigg).
\end{equation}
As we show below, this agrees with the result from the non-local non-linear sigma model in eq. \ref{DimensionsigmasNLSM} at large $N$.

\subsection{Non-local non-linear sigma model}
In line with subsection \ref{NLSm $s = 1$} we can also study this fixed point by an epsilon expansion at the other end, $s = d - \epsilon$ using a non-local non-linear sigma model (note that the scalar becomes dimensionless at $s = d$). A variant of this model, aiming at a more general target manifold, was considered in \cite{Gubser:2019uyf}. We restrict ourselves to $O(N)$, but it should be possible to generalize our approach to other homogeneous spaces. To do that, we consider the following action
\begin{equation}
S = \frac{2 ^s \Gamma(\frac{d + s}{2})}{\pi^{\frac{d}{2}} \Gamma(-\frac{s}{2})} \int d^d x d^d y \frac{\phi^I (x) \phi^I (y)}{|x - y|^{d+s}} + \int d^{d} x \sigma (\phi^I \phi^I - \frac{1}{t^2}). 
\end{equation}
The constraint can be solved using the same parametrization as the boundary case. The one-point function required for $\beta$ function computation now involves the following modified integrals

\begin{equation}
\begin{split}
\langle \phi^N (0) \rangle  &= \frac{1}{t} - \frac{t}{2} \langle \varphi^a \varphi^a (0) \rangle - \frac{t^3}{8}\langle \varphi^a \varphi^a (0) \varphi^b \varphi^b (0) \rangle  \\
&= \frac{1}{t} - \frac{t (N - 1) }{2} \int \frac{d^{d} k}{(2 \pi)^{d}} \frac{1}{|k|^s} + \frac{ (N - 1) t^3}{2 }  \int \frac{d^{d} k}{(2 \pi)^{d}} \frac{1}{|k|^{2 s}}  \int \frac{d^{d} l}{(2 \pi)^{d}} \frac{|k - l|^s}{|l|^s} \\
& - \frac{t^3  ((N-1)^2 + 2 (N - 1))}{8 } \int \frac{d^{d} k}{(2 \pi)^{d}} \frac{1}{|k|^s} \int \frac{d^{d} l}{(2 \pi)^{d}} \frac{1}{|l|^s} \\
&= \frac{1}{t} + \frac{t (N -1)}{2^d \pi^{\frac{d}{2}} \Gamma(\frac{d}{2})} \bigg(\frac{1}{\epsilon} + \frac{\gamma + \log m^2 + \psi(\frac{d}{2})}{2} \bigg) + \frac{t^3 (N - 1) (\gamma + \psi(-\frac{d}{2}) - \psi(\frac{d}{2}) + \psi(d) )}{ 2 \Gamma(\frac{d}{2})^2 (4 \pi)^d \epsilon} \\
&- \frac{t^3 (N -1)^2}{8} \bigg( \frac{1}{2^{2 d -2} \pi^d \Gamma(\frac{d}{2})^2 \epsilon^2 } + \frac{\gamma + \log m^2 + \psi(\frac{d}{2})}{\epsilon \Gamma(\frac{d}{2})^2 2^{2 d -2} \pi^d } \bigg)    
\end{split}
\end{equation}
where we used techniques similar to boundary case to perform the integrals and expanded in $s = d - \epsilon$. The $\beta$ function can be extracted from this one-point function
\begin{equation}
\beta(t) = \frac{\epsilon}{2} t - \frac{t^3 (N -1)}{(4 \pi)^{\frac{d}{2}} \Gamma(\frac{d}{2})} - \frac{t^5(N-1) (\gamma + \psi(-\frac{d}{2}) - \psi(\frac{d}{2}) + \psi(d) )}{(4 \pi)^{d} \Gamma(\frac{d}{2})^2}.
\end{equation} 
This beta function gives a fixed point at 
\begin{equation}
t_*^2 = \frac{\epsilon (4 \pi)^{\frac{d}{2}} \Gamma(\frac{d}{2}) }{2 (N-1)} -\frac{\epsilon^2 (4 \pi)^{\frac{d}{2}} \Gamma(\frac{d}{2}) (\gamma + \psi(-\frac{d}{2}) - \psi(\frac{d}{2}) + \psi(d) ) }{ 4 (N - 1)^2}
\end{equation}
and the dimension of the field $\sigma$ at this fixed point is
\begin{equation} \label{DimensionsigmasNLSM}
\Delta_{\sigma} = d  + \beta'(t_*) = s - \frac{\epsilon^2 (\gamma + \psi(-\frac{d}{2}) - \psi(\frac{d}{2}) + \psi(d)  )}{2(N - 1)}
\end{equation}
in agreement with the large $N$ result.

\subsection{Some Pad\'e estimates for the $d=1$ long range $O(N)$ model}
\begin{figure} [!ht] 
\centering
\includegraphics[scale=0.4]{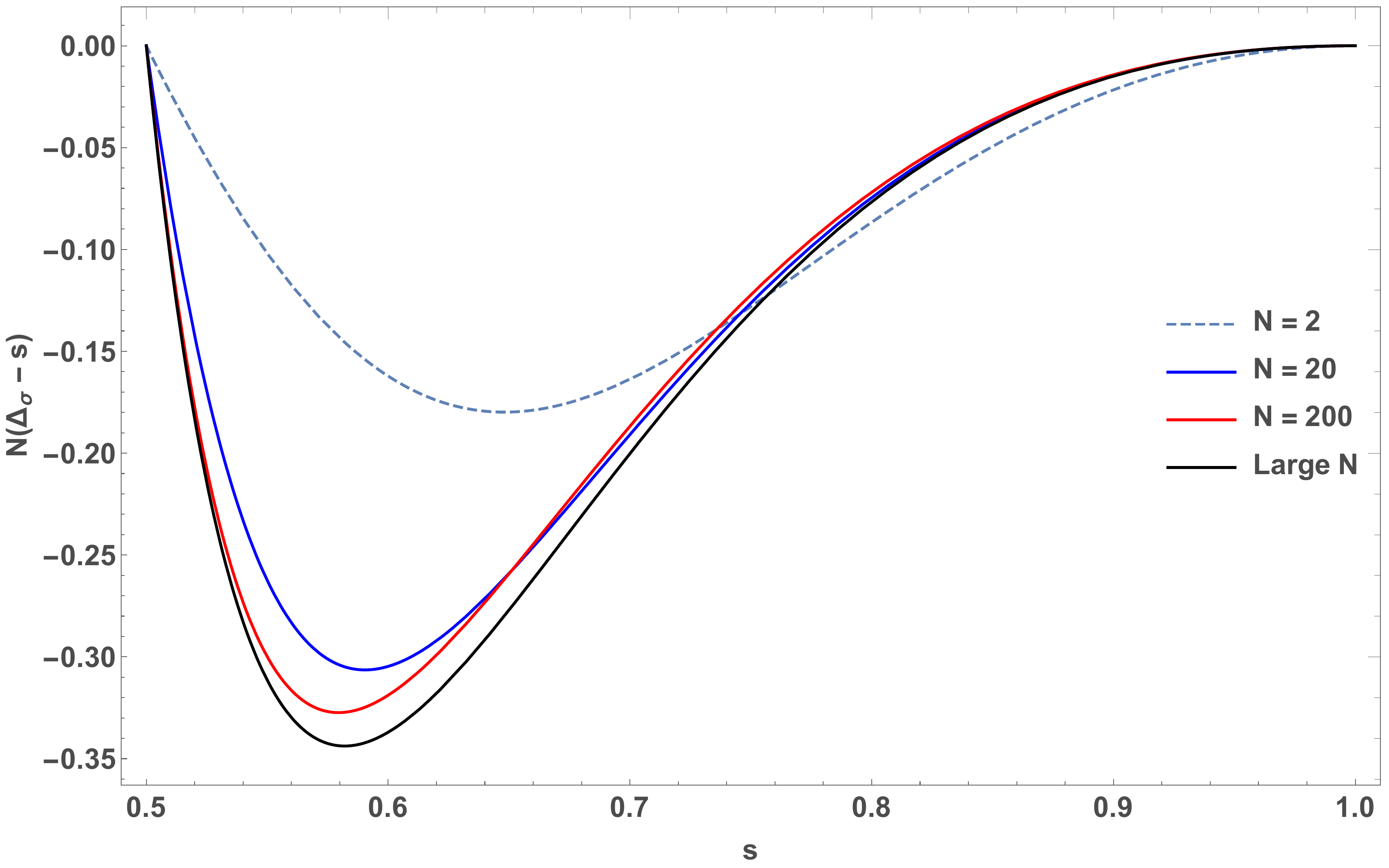}
\caption{Pad\'e result for $\Delta_{\sigma}$ for $N = 2, 20$ and $200$. We plot $N (\Delta_{\sigma} - s)$ against $s$ because that is easier to compare with the large $N$ result. The Pad\'e result approaches large N result as we  go to larger $N$.}
\label{PadeFig}
\end{figure}

The quartic model and the non-linear sigma model approximate the fixed point of one dimensional long range $O(N)$ model near the two ends in s, i.e. $s = \frac{d}{2} + \frac{\epsilon}{2}$ and $s = d - \epsilon$ respectively. The large $N$ model interpolates between the two ends, but we can also develop a two-sided Pad\'e approximant to interpolate the intermediate range of $s$ for finite $N$. By that, we mean that we consider an ansatz $\textrm{Pad\'e}_{m,n} = \frac{\sum_{i = 0}^m a_i s^i}{ 1 + \sum_{j = 1}^n b_j s^j}$ and equate its series expansion with the available perturbative series expansion. We do this for $\Delta_{\sigma}$ which is related to the critical exponent $\nu$ as $\Delta_{\sigma} = 1 - 1/\nu$ (this is the dimension of $\sigma$ in non-linear sigma model and of $\phi^2$ in the quartic theory). From the models anlayzed in the previous sections, we have the following series expansions for the anomalous dimension of $\sigma$ in $d = 1$ 
\begin{equation}
\begin{split}
\Delta_{\sigma} &= \frac{1}{2} + \frac{(N - 4) ( s - \frac{1}{2} )}{N + 8} + \frac{ 4 (N + 2) ( 7 N + 20) (\pi + 4 \log 2)}{(N + 8)^3} \bigg(s-\frac{1}{2}\bigg)^2 + O\bigg(s-\frac{1}{2}\bigg)^3, \   s \sim 1/2 
\\
\Delta_{\sigma} &= s - \frac{( 1 - s )^2}{N - 1}  + O( 1 - s)^3 , \ \ \  s \sim 1.
\end{split}
\end{equation}
We have six possible Pad\'e approximants corresponding to choices of $m,n$ such that $m + n = 5$. Only $\textrm{Pad\'e}_{2,3}$ and $\textrm{Pad\'e}_{3,2}$ are well behaved at all $s$ and $N$ and have a large $N$ behaviour close to our large $N$ result (i.e., they go as $s + 1/N$ at large $N$). We take their average and plot that to compare it with the large $N$ result in figure \ref{PadeFig}.

The non-linear sigma model description clearly breaks down for the Ising case $N = 1$, since the $\beta$ function vanishes and the anomalous dimension diverges in that case. But the dimension of $\sigma$ near $s = 1$ for the case of long range Ising was found in \cite{PhysRevLett.37.1577} to be 
\begin{equation} 
\Delta_{\sigma} = 1 - \sqrt{ 2 (1 - s )}  , \ \ \  s \sim 1 , \ \ \ N = 1.
\end{equation}
Since there is a square root, we will switch variables to $x = \sqrt{1 - s}$ and do a two sided Pad\'e between $0 < x < \frac{1}{\sqrt{2}} $ with the following two constraints
\begin{equation}
\begin{split}
\Delta_{\sigma} &= \frac{1}{2} + \frac{\sqrt{2}}{3} \bigg(x - \frac{1}{\sqrt{2}} \bigg) + \frac{ 3 +  8 (\pi + 4 \log 2)}{9} \bigg(x - \frac{1}{\sqrt{2}} \bigg)^2 + O\bigg(x - \frac{1}{\sqrt{2}} \bigg)^3, \  x \sim \frac{1}{\sqrt{2}} 
\\ 
\Delta_{\sigma} &= 1 - \sqrt{ 2} x + O(x^2) , \ \ \  x \sim 0.
\end{split}
\end{equation}
Again, there are five possibilities and $\textrm{Pad\'e}_{3,1}$, $\textrm{Pad\'e}_{1,3}$ and $\textrm{Pad\'e}_{2,2}$ are all close to each other. We take their average and tabulate the results in table \ref{Pade}, where we also include the Pad\'e estimates for higher values of $N$ obtained as described above. For $N=1$ our estimates are close to the available Monte Carlo results found in \cite{2014arXiv1401.6805C , LuijtenThesis}.
\begin{table}[!ht]
\centering
\begin{tabular}{|c|c|c|c|c|c|c|}
\hline
s & 0.6 & 0.65453 & 0.7 & 0.8 & 0.875 & 0.9 \\
\hline
N = 1 Pad\'e & 0.488 & 0.494 & 0.506 & 0.553 & 0.616 & 0.646 \\
\hline
N = 1 Monte Carlo \cite{2014arXiv1401.6805C}\ & - & 0.494(14)  & -  & - &0.5876(13) & - \\
\hline
N = 1 Monte Carlo \cite{LuijtenThesis}\ & 0.50(2)  & -  & 0.50(4)  & 0.54(5) &- & 0.63(7) \\
\hline
N = 2 & 0.519 & 0.565 & 0.618 & 0.757 & 0.858 & 0. 889 \\
\hline 
N = 3 & 0. 534 & 0.588 & 0. 643 & 0.774 & 0.865 & 0.894  \\
\hline
N = 4 & 0.544 & 0.601 & 0.656 & 0.781 & 0.868 & 0.896 \\
\hline
N = 5 & 0.552 & 0.610 & 0.664 & 0.785 & 0.870 & 0.897 \\
\hline
N = 10 & 0.572 & 0.630 & 0.681 & 0.792 & 0.872 & 0.898 \\
\hline
\end{tabular}
\caption{The numerical results for $\Delta_{\sigma} = 1 - 1/\nu$ from our Pad\'e approximants and the available Monte Carlo results for various values of $s$. As $N$ grows, the results approach the prediction of the large $N$ expansion, which gives  $\Delta_{\sigma} = s + O(1/N)$.}
\label{Pade}
\end{table}
\section*{Acknowledgments}

We thank Igor Klebanov for collaboration at the initial stages of this project, and for many useful discussions and comments. We also thank Christian Jepsen, Ziming Ji, Brian Trundy, Amos Yarom and Xinan Zhou for related discussions. We are grateful to the Perimeter Institute for hospitality during the workshop ``Boundaries and Defects in Quantum Field Theory", where preliminary results of this work were presented. This research was supported in part by the US NSF under Grants No.~PHY-1620542 and PHY-1914860.

\appendix

\section{Other Examples of BCFT with free fields in the bulk}

In this Apppendix we briefly discuss some other examples of BCFTs with free fields in the bulk and interactions localized on the boundary.   

\subsection{Scalar Yukawa like interaction in $d = 5 - \epsilon$ boundary dimensions}
Consider the following model of a free scalar field interacting with $N$ bosons on the boundary with an action 
\begin{equation}
S = \int d^{d + 1} x \ \frac{1}{2} (\partial \sigma)^2 + \int d^{d} x \bigg( \frac{1}{2} (\partial_{\mu} \phi^I \partial^{\mu} \phi^I) + \frac{g}{2} \sigma \phi^I \phi^I \bigg).
\end{equation}
where $I = 1,2... N$. The interaction becomes marginal in $d = 5$, and it is weakly coupled in $d = 5 - \epsilon$ dimensions. As usual, $\sigma$ does not get renormalized and has dimensions fixed at classical value. The one loop correction to the propagator of $\phi$ is  
\begin{equation}
\begin{split}
G^{0,2} &= (-g)^2 \int \frac{d^{d} \textbf{k}}{(2 \pi)^{d}} \frac{1}{(\textbf{p} +  \textbf{k})^2 \  |\textbf{k}|} - p^2 \delta_{\phi} \\
&= \frac{g^2 \Gamma(\frac{d - 2}{2}) \Gamma(\frac{d - 1}{2}) \Gamma(\frac{3 - d}{2}) }{ (4 \pi)^{\frac{d}{2}} \sqrt{\pi} \Gamma(d - \frac{3}{2}) (p^2)^{\frac{3 - d}{2}} }- p^2 \delta_{\phi}
\end{split}
\end{equation}
which implies in $d = 5 - \epsilon$
\begin{equation}
Z_{\phi} =  1 - \frac{g^2}{60 \pi^3 \epsilon}. 
\end{equation}
The one loop correction to the vertex is 
\begin{equation}
\begin{split}
G^{1,2} &= (-g)^3 \int \frac{d^{d} \textbf{k}}{(2 \pi)^{d}} \frac{1}{(\textbf{p} +  \textbf{k})^2 (\textbf{k} -  \textbf{q})^2  \  | \textbf{k}|} - \delta_{g} \\
&= - \frac{4 g^3 \Gamma(\frac{5 - d}{2})}{3 \pi^{\frac{d + 1}{2}} 2^{d} (\mu^2)^{\frac{ 5 - d}{2} } } - \delta_{g}
\end{split}
\end{equation}
which implies 
\begin{equation}
Z_{g} = g + \delta_g = g - \frac{g^3}{12 \pi^3 \epsilon}.
\end{equation}
Using the relation $g_0 Z_{\sigma}^{1/2} Z_{\phi} = \mu^{\epsilon/2} Z_g $ gives the $\beta-$ function as 
\begin{equation}
\beta(g) = \mu \frac{\partial g}{\partial \mu} \bigg|_{g_0} = - \mu \frac{\partial_{\mu} g_0 |_{g}}{\partial_{g} g_0|_{\mu}} = - \frac{\epsilon g}{2} - \frac{g^3}{15 \pi^3}. 
\end{equation}
So there exists a non unitary fixed point at 
\begin{equation}
g_*^2 = -\frac{15 \pi^3 \epsilon}{2}.
\end{equation}
The boundary field $\phi$ acquires an anomalous dimension 
\begin{equation}
\hat{\gamma}_{\phi} = \mu \frac{\partial}{\partial \mu} \log Z_{\phi}^{1/2} = \beta (g) \frac{\partial}{\partial g} \log Z_{\phi}^{1/2} = \frac{g^2}{120 \pi^3 }
\end{equation}
which at the non unitary fixed point becomes $\hat{\gamma}_{\phi}|_{g^*} = -\epsilon /16 $.

\subsection{ $N + 1$ free scalars interacting on $d = 3 - \epsilon$ boundary dimensions}
Next model we consider is $N + 1$ free scalars in the bulk interacting only on the boundary
\begin{equation}
S = \int d^{d + 1} x \bigg( \frac{1}{2} (\partial \sigma)^2 + \frac{1}{2} \partial_{\mu} \phi^I \partial^{\mu} \phi^I \bigg)  + \int d^{d} x \bigg(\frac{g_1}{2} \sigma \phi^I \phi^I + \frac{g_2}{6} \sigma^3 \bigg).
\end{equation}
where $I = 1,2... N$. The couplings are marginal in $d = 4$ and the model becomes weakly coupled in $d = 3 -\epsilon$. Both $\sigma$ and $\phi^I$ are now free bulk fields and they don't get renormalized. The one loop correction to the $g_1$ vertex is 
\begin{equation}
\begin{split}
G^{1,2} &= ((-g_1)^3 + (-g_1)^2 (-g_2)) \int \frac{d^{d} \textbf{k}}{(2 \pi)^{d}} \frac{1}{ |\textbf{k} + \textbf{p}| |\textbf{k} - \textbf{q}|  |\textbf{k}|} - \delta_{g_1}\\
&= -\frac{(g_1^3 + g_1^2 g_2)}{2^{d-1} \pi^{\frac{d + 1}{2}}} \frac{\Gamma( \frac{3 - d}{2})}{(\mu^2)^{ \frac{3 - d}{2}}} - \delta_{g_1}
\end{split}
\end{equation}
which implies 
\begin{equation}
Z_{g_1} = g_1 +  \delta_{g_1} = g_1 - \frac{(g_1^3 + g_1^2 g_2)}{2 \pi^2 \epsilon}.
\end{equation}
The one loop correction to $g_2$ vertex is similarly
\begin{equation}
\begin{split}
G^{3, 0} & = (N(-g_1)^3 + (-g_2)^3) \int \frac{d^{d} \textbf{k}}{(2 \pi)^{d}} \frac{1}{ |\textbf{k} + \textbf{p}| |\textbf{k} - \textbf{q}|  |\textbf{k}|} - \delta_{g_2}\\
&= -\frac{(N g_1^3 +  g_2^3)}{2^{d-1} \pi^{\frac{d + 1}{2}}} \frac{\Gamma(\frac{3 - d}{2})}{(\mu^2)^{ \frac{3 - d}{2}}} - \delta_{g_2}
\end{split}
\end{equation}
which implies 
\begin{equation}
Z_{g_2} = g_2 +  \delta_{g_2} = g_2 - \frac{(N g_1^3 +  g_2^3)}{2 \pi^2 \epsilon}.
\end{equation}
The bare couplings are related to the renormalized couplings as 
\begin{equation}
\begin{split}
g_{1_0} Z_{\sigma}^{1/2} Z_{\phi} &= \mu^{\epsilon/2} (g_1 + \delta_{g_1}) \\
g_{2_0} Z_{\sigma}^{3/2} &= \mu^{\epsilon/2} (g_2 + \delta_{g_2}) 
\end{split}
\end{equation}
The $\beta$ functions can then be computed using following relations
\begin{equation}
\begin{split}
-\mu \partial_{\mu} g_{1_0}|_{g_1, g_2} &= \beta(g_1) \frac{\partial g_{1_0}}{\partial g_1} \bigg|_{\mu, g_2} + \beta(g_2) \frac{\partial g_{1_0}}{\partial g_2} \bigg|_{\mu, g_1} \\
-\mu \partial_{\mu} g_{2_0}|_{g_1, g_2} &= \beta(g_1) \frac{\partial g_{2_0}}{\partial g_1} \bigg|_{\mu, g_2} + \beta(g_2) \frac{\partial g_{2_0}}{\partial g_2} \bigg|_{\mu, g_1}.
\end{split}
\end{equation}
These give the following $\beta$ functions
\begin{equation}
\begin{split}
\beta(g_1) &= -\frac{\epsilon}{2} g_1 - \frac{g_1^3 + g_1^2 g_2}{2 \pi^2} \\
\beta(g_2) &= -\frac{\epsilon}{2} g_2 - \frac{N g_1^3 + g_2^3}{2 \pi^2}
\end{split}
\end{equation}
which give rise to non-unitary fixed points.
\subsection{Mixed Dimensional QED in $d=5$ boundary dimensions}
Another interesting model to consider is the following higher derivative variant of the mixed dimensional QED discussed in \cite{Herzog:2017xha}
\begin{equation}
S = \frac{1}{4} \int d^{d + 1} x F^{\mu \nu} (-\nabla^2) F_{\mu \nu} - \int d^{d} x \ \bar{\psi} \gamma^{\mu} (\partial_{\mu} + i g A_{\mu}) \psi.
\end{equation}
The engineering dimension of the gauge field here is $(d + 1)/2 -2$, hence the coupling is marginal in $d = 5$ dimensions. We will analyze this model in $d = 5-\epsilon$. The higher derivative term will give a $\frac{\eta_{A B}}{p^4}$ propagator in the bulk. We can Fourier transform back to position space in the direction perpendicular to the boundary and get the propagator on the boundary to be $\frac{\eta_{A B}}{4 |p|^3}$ . We have the standard propagator for the fermion $- i \frac{\slashed{p}}{p^2}$. The gauge field is free in the bulk, so it should not receive any anomalous dimensions. So to compute the $\beta$ function, we need to compute the one loop correction to the fermion propagator and the vertex. \\
The one loop correction to the fermion propagator is 
\begin{equation}
\begin{split}
G^{0,2} &= (i g)^2 \int \frac{d^{d} k}{(2 \pi)^{d}} \frac{\gamma^A (- i \slashed{k})\gamma^B \eta_{A B}}{4 | \textbf{p} - \textbf{k}|^3 k^2} - i\delta_{\psi} \slashed{p} \\
&= \frac{- i g^2 (d-2) \slashed{p} \Gamma(\frac{5 - d}{2}) }{5 \sqrt{\pi} (4 \pi)^{\frac{d}{2}}} - i\delta_{\psi} \slashed{p}.
\end{split}
\end{equation}
Requiring that the divergent part of the above expression vanish in $d = 5 - \epsilon$ gives us
\begin{equation}
\delta_{\psi} = -\frac{3 g^2}{80 \pi^3 \epsilon}.
\end{equation}
The one loop correction to the vertex is 
\begin{equation}
G^{1,2} = (i g)^3 \int \frac{d^{d} p}{(2 \pi)^{d}} \frac{\gamma^C (  - i (\slashed{p} + \slashed{q_1})) \gamma^A (- i  (\slashed{p} + \slashed{q_2})) \gamma^B \eta_{BC}}{(p + q_1)^2 (p + q_2)^2 4 |\textbf{p}|^3} + i \delta_g \gamma^A
\end{equation}
We can evaluate the divergent part of the first term in the above expression which must be cancelled by the counterterm which gives
\begin{equation}
\delta_g = -\frac{3 g^3}{80 \pi^3 \epsilon}.
\end{equation}
Using relation $g_0 Z_{\gamma}^{1/2} Z_{\psi} = (g + \delta_g) \mu^{\epsilon/2}$ this gives a finite value for $g_0$. This implies that the beta function actually vanishes in $5$ dimensions to this order.  
\section{Some useful integrals} \label{Integrals}
In this appendix, we mention some useful integrals which we use throughout the paper. The first one was performed in \cite{Fei:2014yja} 

\begin{equation}
\int \frac{d^d \textbf{k}}{(2 \pi)^d} \frac{1}{|\textbf{k}|^{2 \alpha} |\textbf{k + p}|^{2 \beta}} = \frac{1}{(4 \pi)^{\frac{d}{2}} |\textbf{p}|^{2 \alpha + 2 \beta - d }} \frac{\Gamma(\frac{d}{2}  - \alpha) \Gamma(\frac{d}{2}  - \beta) \Gamma( \alpha + \beta - \frac{d}{2})}{\Gamma(\alpha) \Gamma(\beta) \Gamma( d - \alpha - \beta)}.
\end{equation}
The following two variants of it can be performed by using very similar methods
\begin{equation}
\begin{split}
\int \frac{d^d \textbf{k}}{(2 \pi)^d} \frac{k_i k_j}{|\textbf{k}|^{2 \alpha} |\textbf{k + p}|^{2 \beta}} &= \frac{1}{(4 \pi)^{\frac{d}{2}} |\textbf{p}|^{2 \alpha + 2 \beta - d - 2}} \bigg( \frac{\delta_{ij}}{2} \frac{\Gamma(\frac{d}{2} + 1 - \alpha) \Gamma(\frac{d}{2} + 1 - \beta) \Gamma( \alpha + \beta - \frac{d}{2} - 1 ) }{\Gamma(\alpha) \Gamma(\beta) \Gamma(2 + d - \alpha - \beta)} \\
&+ \frac{p_i p_j}{|\textbf{p}|^{2}} \frac{\Gamma(\frac{d}{2} + 2 - \alpha) \Gamma(\frac{d}{2}  - \beta) \Gamma( \alpha + \beta - \frac{d}{2}) }{\Gamma(\alpha) \Gamma(\beta) \Gamma(2 + d - \alpha - \beta)} \bigg)
\end{split}
\end{equation}
and
\begin{equation}
\int \frac{d^d \textbf{k}}{(2 \pi)^d} \frac{k_i p_j}{|\textbf{k}|^{2 \alpha} |\textbf{k + p}|^{2 \beta}} = -\frac{p_i p_j}{(4 \pi)^{\frac{d}{2}} |\textbf{p}|^{2 \alpha + 2 \beta - d }} \frac{\Gamma(\frac{d}{2} + 1 - \alpha) \Gamma(\frac{d}{2}  - \beta) \Gamma( \alpha + \beta - \frac{d}{2})}{\Gamma(\alpha) \Gamma(\beta) \Gamma(1 + d - \alpha - \beta)}. 
\end{equation}

\bibliographystyle{ssg}
\bibliography{BCFT-Bib}

\end{document}